\documentclass[aps,pra,twocolumn,showpacs,prabib, superscriptaddress]{revtex4}
\usepackage{graphicx}
\usepackage{amsmath}
\usepackage{amssymb,wasysym}
\usepackage{amsfonts}
\usepackage{bm,verbatim}

\usepackage{color}

\newcommand{\bea}{\begin{eqnarray}}
\newcommand{\eea}{\end{eqnarray}}
\newcommand{\be}{\begin{equation}}
\newcommand{\ee}{\end{equation}}
\newcommand{\ds}{\displaystyle}
\newcommand{\rr}{\mathbf{r}}
\newcommand{\kk}{\mathbf{k}}
\newcommand{\KK}{\mathbf{K}}
\newcommand{\qq}{\mathbf{q}}
\newcommand{\xx}{\mathbf{x}}
\newcommand{\nn}{\mathbf{n}}
\newcommand{\uu}{\mathbf{u}}
\newcommand{\RR}{\mathbf{R}}
\newcommand{\CC}{\mathbf{C}}
\newcommand{\vn}{\mathbf{0}}

\newcommand{\ra}{\rangle}
\newcommand{\la}{\langle}
\newcommand{\si}{\sigma}

\newcommand{\ktyp}{k_{\rm typ}}
\newcommand{\up}{\uparrow}
\newcommand{\down}{\downarrow}

\newcommand{\epsk}{\epsilon_{\mathbf{k}}}
\newcommand{\epsq}{\epsilon_{\mathbf{q}}}

\newcounter{fnnumberter}
\newcounter{notedecoupage}

\begin{document}

\title{General relations for quantum gases in two and three
\\
dimensions:  Two-component fermions}

\author{F\'elix Werner}
\affiliation{Department of Physics, University of Massachusetts,
Amherst, MA 01003, USA}
\affiliation{Laboratoire Kastler Brossel, \'Ecole Normale
Sup\'erieure, UPMC and CNRS, 24 rue Lhomond, 75231 Paris Cedex 05, France}

\author{Yvan Castin}
\affiliation{Laboratoire Kastler Brossel, \'Ecole Normale
Sup\'erieure, UPMC and CNRS, 24 rue Lhomond, 75231 Paris Cedex 05, France}

\begin{abstract}
We derive exact general relations between various observables for $N$ spin-1/2 fermions 
with zero-range or short-range interactions, in continuous space or on a lattice, in two or three dimensions, in an arbitrary external potential.
Some of our results generalize known relations 
between
the large-momentum behavior of the momentum distribution,
the short-distance behaviors of the pair distribution function and of the one-body density matrix,
the norm of the regular part of the wavefunction,
the derivative of the energy with respect to the scattering length or to 
time, 
and 
the interaction energy (in the case of finite-range interactions).
The expression relating the energy to a functional of the momentum distribution
is also generalized.
Moreover, expressions are found (in terms of the regular part of the wavefunction) for the derivative of the energy 
 with respect to the effective range $r_e$ in $3D$, and to the effective range squared in $2D$. They express the fact
that the leading corrections to the eigenenergies due to a finite interaction-range are linear in
the effective range in $3D$ (and in its square in $2D$) with model-independent coefficients.
There are subtleties in the validity condition of this conclusion, for the $2D$ continuous space
(where it is saved by factors that are only logarithmically large in the zero-range limit)
and for the $3D$ lattice models (where it applies only for some magic dispersion relations on the lattice,
that sufficiently weakly break Galilean invariance and that do not have cusps at the border of the first
Brillouin zone;
an example of such relations is constructed).
Furthermore, the subleading short distance behavior
of the pair distribution function and the subleading $1/k^6$ tail of the momentum distribution are related
to $\partial E/\partial r_e$ (or to $\partial E/\partial (r_e^2)$ in $2D$).
The second order derivative of the energy with respect to the inverse (or the logarithm in the two-dimensional case) of the scattering length is found to be expressible, for any eigenstate, in terms of the eigenwavefunctions' regular parts; this implies that, at thermal equilibrium, this second order derivative, taken at fixed entropy, is 
negative.  Applications of the general relations are presented: 
We compute corrections to exactly solvable two-body and three-body problems
and find agreement with available numerics;  for the unitary gas in an isotropic harmonic trap,
we determine how the finite-$1/a$ and finite range energy corrections vary within each energy ladder (associated
to the $SO(2,1)$ dynamical symmetry) and we deduce the frequency shift and the collapse time of the breathing mode; for the bulk unitary gas, 
we compare to fixed-node Monte Carlo data,
and we estimate the deviation from the Bertsch parameter due to the finite interaction range in typical experiments.
\end{abstract}

\date{\today}

\maketitle

\section{General introduction}

The experimental breakthroughs of 1995 having led to the first realization
of a Bose-Einstein condensate in an atomic vapor
\cite{Cornell_bec,Hulet_bec,Ketterle_bec}
have opened the era of experimental studies of ultracold gases with 
non-negligible or even strong interactions, in dimension lower than or equal to three~\cite{RevueBlochDalibard,RevueTrentoFermions, HouchesBEC,HouchesLowD,Varenna}.
In these systems, the thermal de Broglie wavelength and the typical distance
between atoms are much larger than the range of the interaction potential. 
This so-called {\sl zero-range} regime has interesting
universal properties: Several quantities such as the thermodynamic 
functions of the gas depend on the interaction potential only through
the scattering length $a$, a length that can be defined in any dimension
and that characterizes the low-energy scattering amplitude of two atoms.

This universality property holds for the weakly repulsive
Bose gas in three dimensions \cite{LHY} up to the order of expansion
in $(n a^3)^{1/2}$ corresponding to Bogoliubov theory 
\cite{Wu1959,SeiringerLHY}, $n$ being the gas density. 
It also holds for the weakly repulsive 
Bose gas in two dimensions \cite{Schick,Popov,Lieb2D,MoraCastin}, even at the next order
beyond Bogoliubov theory \cite{MoraCastin2D}. For $a$ much larger than the range of the interaction potential, the ground state of $N$ 
bosons in two dimensions is 
a universal $N$-body bound state~\cite{BruchTjon3bosons2D,Fedorov3bosons2D,Platter4bosons2D,HammerSon,Lee2D}.
In one dimension, the universality holds for 
any scattering length, as exemplified by the fact that
the Bose gas with zero-range interaction is exactly solvable
by the Bethe ansatz both in the repulsive case \cite{LiebLiniger}
and in the attractive case \cite{McGuire,Herzog,Caux}.

For spin 1/2 fermions, the universality properties are expected to be even stronger.
The weakly interacting regimes in $3D$
\cite{LeeYang,HuangYang,Abrikosov,Galitski,
LiebSeiringerSolovej_FermionsT=0,
Seiringer_fermions}
and in $2D$ \cite{Bloom} are universal,
as well as for any scattering length in the Bethe-ansatz-solvable
$1D$ case
\cite{GaudinArticle,GaudinLivre}. 
Universality is also expected to hold for an arbitrary scattering length 
even in $3D$,
as was recently tested by experimental studies on the BEC-BCS crossover
using a Feshbach resonance, see~\cite{Varenna} and Refs. therein and e.~g.~\cite{HuletClosedChannel, HuletPolarized,ThomasVirielExp,thomas_entropie_PRL,thomas_entropie_JLTP,JinPhotoemission,KetterleGap,GrimmModesTfinie, ZwierleinPolaron,SylPolaron,SylEOS,MukaiyamaEOS,
NirEOS,ZwierleinEOS,JinPseudogap,AustraliensC,AustraliensT,JinC},
and in agreement with unbiased Quantum Monte Carlo calculations \cite{bulgacQMC,zhenyaPRL,Juillet,BulgacCrossover,zhenyas_crossover,ChangAFMC,VanHouckePrepa};
and in $2D$,
a similar universal crossover from BEC to BCS is expected  when the parameter $\ln (k_F a)$ 
[where $k_F$ is the Fermi momentum] varies from $-\infty$ to $+\infty$~\cite{Petrov2D,Miyake2D,Randeria2D,Zwerger2D,Leyronas4corps,Giorgini2D,Kohl2Dgap,Kohl2D}.
Mathematically, results on universality were obtained for the $N$-body problem  in $2D$ \cite{Teta}.
In $3D$, mathematical results were obtained for the $3$-body problem~(see, e.g.,~\cite{MinlosFaddeev1,MinlosFaddeev2,VugalterZhislin,Shermatov,Albeverio3corps}).
The universality for the fermionic equal-mass $N$-body problem in $3D$ remains mathematically unproven~\footnote{The proof given in~\cite{Teta} that, for a sufficiently large number of equal-mass fermions, 
the energy is unbounded from below, is actually incorrect, since the fermionic antisymmetry was not properly taken into account. 
A theorem was published without proof in \cite{MinlosProceeding} implying that
the spectrum of the Hamiltonian of $N_\up$ same-spin-state fermions of mass $m_\up$ interacting with a distinguishable particle of mass $m_\down$ 
is unbounded below,
not only
for $m_\up=m_\down$ and large enough $N_\up$,
but 
 also for $N_\up=3$ and $m_\up/m_\down$ larger than the critical mass ratio $5.29\dots$.
No proof was found yet for this theorem; it was only proven that no Efimov effect occurs for $N_\up=3$, $N_\down=1$ provided $m_\up/m_\down$ is sufficiently small~\cite{Minlos2011}.
It was recently shown that a four-body Efimov effect occurs in this $3+1$ body problem
(for an angular momentum $l=1$ and not for any other $l\le10$) and makes the spectrum unbounded below,
however for a widely different critical mass ratio $m_\up/m_\down\simeq 13.384$~\cite{CMP},
which sheds some doubts on \cite{MinlosProceeding}.
}.

Universality is also expected for mixtures
in $2D$~\cite{PetrovCristal,Leyronas4corps,LudoPedri}, and in $3D$ for Fermi-Fermi mixtures below a critical mass ratio~\cite{Efimov73,PetrovCristal,BaranovLoboShlyapMassesDiff,CMP}.
Above a critical mass ratio, the Efimov effect takes place, as it also takes place for bosons
\cite{Efimov,RevueBraaten}. In this case, the three-body problem depends 
on a single additional parameter, the three-body parameter. The Efimov physics is presently under active experimental
investigation \cite{GrimmEfimov,FlorenceEfimov,KhaykovichEfimov,NatureEfimov,HuletEfimov,KhaykovichEfimov2,JochimEfimovRF}. It is not the subject of this paper (see \cite{50pages}).

In the zero-range regime, it is intuitive that the short-range or high-momenta properties of the gas are dominated by two-body physics. 
For example the pair distribution function $g^{(2)}(\mathbf{r}_{12})$ 
of particles at distances $r_{12}$
much smaller than the de Broglie wavelength is expected 
to be proportional to the modulus squared of the zero-energy two-body
scattering-state wavefunction $\phi(\rr_{12})$, with a proportionality factor
$\Lambda_g$ depending on the many-body state of the gas.
Similarly the tail of the momentum distribution
$n(\mathbf{k})$, at 
wavevectors much larger than the inverse de Broglie wavelength, 
is expected to be proportional to the modulus squared of
the Fourier component $\tilde{\phi}(\mathbf{k})$ of the zero-energy scattering-state wavefunction,
with a proportionality factor $\Lambda_n$ depending on the many-body
state of the gas: 
Whereas two colliding atoms in the gas
have a center of mass wavevector of the order of the inverse de Broglie
wavelength, their relative wavevector can access much larger values,
up to the inverse of the interaction range,
simply because the interaction potential has a width in the space
of relative momenta of the order of the inverse of its range in real space.

For these intuitive reasons, and with the notable exception of one-dimensional
systems, one expects that the mean interaction energy $E_{\rm int}$
of the gas, being
sensitive to the shape of $g^{(2)}$ at distances of the order
of the interaction range, is not universal, but diverges
in the zero-range limit; one also expects
that, apart from the $1D$ case,
the mean kinetic energy, being dominated by the tail
of the momentum distribution, is not universal and diverges
in the zero-range limit, a well known fact
in the context of Bogoliubov theory for Bose gases
and of BCS theory for Fermi gases.
Since the total energy of the gas is universal, and $E_{\rm int}$
is proportional to $\Lambda_g$ while $E_{\rm kin}$ is proportional
to $\Lambda_n$, one expects that there exists a simple relation
between $\Lambda_g$ and $\Lambda_n$.

The precise link between the pair distribution function, the tail
of the momentum distribution and the energy of the gas
was first established for one-dimensional systems.
In \cite{LiebLiniger} the value of the
pair distribution function for $r_{12}=0$
was expressed in terms of the derivative of the gas energy with respect
to the one-dimensional scattering length, thanks to the Hellmann-Feynman
theorem. In \cite{Olshanii_nk} the tail
of $n(k)$ was also related to this derivative of the energy,
by using a simple and general property of the Fourier transform of a function
having discontinuous derivatives in isolated points.

In three dimensions, results in these directions were first obtained
for weakly interacting gases. For the weakly interacting Bose gas,
Bogoliubov theory contains the expected properties, in particular
on the short distance behavior of the pair distribution function
\cite{Huang_article,Holzmann,Glauber} 
and the fact that the momentum distribution
has a slowly decreasing tail.
For the weakly interacting spin-1/2 Fermi gas, it was shown that 
the BCS anomalous average (or pairing field) 
$\langle \hat{\psi}_\uparrow(\mathbf{r}_1) 
\hat{\psi}_\downarrow(\mathbf{r}_2)\rangle$ behaves at short
distances as the zero-energy two-body scattering wavefunction $\phi(\rr_{12})$
\cite{Bruun},
 resulting in a $g^{(2)}$ function indeed proportional
to $|\phi(\rr_{12})|^2$ at short distances. It was however understood
later that the corresponding proportionality factor $\Lambda_g$
predicted by BCS theory is incorrect \cite{CarusottoCastin},
e.g.\ at zero temperature the BCS prediction drops exponentially with $1/a$
in the non-interacting limit $a\to 0^-$, whereas the correct result
drops as a power law in $a$.

More recently, in a series of two articles \cite{TanEnergetics,TanLargeMomentum},
explicit expressions for the proportionality factors
$\Lambda_g$ and $\Lambda_n$ were obtained in terms of the derivative
of the gas energy with respect to the inverse scattering length, for a spin-1/2 interacting Fermi
gas in three dimensions, for an arbitrary value of the scattering length,
that is, not restricting to the weakly interacting limit. 
Later on, these results were rederived in \cite{Braaten,BraatenLong,ZhangLeggettUniv},
and also in \cite{WernerTarruellCastin} with very elementary methods
building on the aforementioned intuition that $g^{(2)}(\rr_{12})\propto |\phi(\rr_{12})|^2$
at short distances and $n(\kk)\propto |\tilde{\phi}(\kk)|^2$
at large momenta. These relations were tested by
numerical four-body calculations \cite{BlumeRelations}.
An explicit relation between $\Lambda_g$ and the interaction energy was derived in \cite{ZhangLeggettUniv}. Another fundamental relation discovered in \cite{TanEnergetics} and recently generalized in \cite{TanSimple,CombescotC}  to fermions in $2D$,
expresses the total energy as a functional of the momentum distribution and the spatial density. 

\section{Contents}

Here we derive generalizations of the relations
of \cite{LiebLiniger,Olshanii_nk,TanEnergetics,TanLargeMomentum,ZhangLeggettUniv,TanSimple,CombescotC} to two dimensional gases, 
and to the case of a small but non-zero interaction range (both on a lattice and in continuous space).
We also find entirely new results for
the first order derivative of the energy with respect to the effective range, 
as well as for the second order derivative with respect to the scattering length.
We shall also include rederivations of known relations using our elementary methods.
We treat in detail
the case of spin-1/2 fermions, with equal masses in the two spin states, both in three dimensions and
in two dimensions.
The discussion of spinless bosons and arbitrary mixtures is deferred to another article, as it may involve
the Efimov effect in three dimensions~\cite{CompanionBosons}.

This article is organized as follows.
Models, notations and some basic properties are introduced in Section~\ref{subsec:models}.
Relations for zero-range interactions are 
summarized in Table~\ref{tab:fermions} and derived
for pure states 
 in Section \ref{sec:ZR}. We then consider lattice models (Tab.~\ref{tab:latt} and Sec.~\ref{sec:latt}) and finite-range models in continuous space (Tab.~\ref{tab:V(r)} and Sec.~\ref{sec:V(r)}). In Section~\ref{sec:re} we derive a model-independent expression for the correction to the energy due to a finite range or a finite effective range of the interaction, and we relate this energy correction 
to the subleading short distance behavior of the pair distribution function and to the coefficient of
the $1/k^6$ subleading tail of the momentum distribution  (see Tab.~\ref{tab:re}).
The case of general statistical mixtures of pure states or of stationary states is discussed in Sec.~\ref{sec:stat_mix},
and the case of thermodynamic equilibrium states in Sec.~\ref{subsec:finiteT}.
Finally we present applications of the general relations: For two particles and three particles in harmonic traps we compute corrections to exactly solvable cases (Sec. \ref{subsec:appli_deux_corps} and Sec. \ref{subsec:appl_3body}).
For the unitary gas trapped in an isotropic harmonic potential, we determine how the equidistance between levels 
within a given energy ladder (resulting from the $SO(2,1)$ dynamical symmetry) is affected by finite $1/a$ and finite range
corrections, which leads to a frequency shift and a collapse of the breathing mode of the zero-temperature
gas~(Sec.~\ref{subsec:contactN}).
For the bulk unitary gas, we check that general relations are satisfied by existing fixed-node Monte Carlo 
data~\cite{Giorgini,Giorgini_nk,LoboGiorgini_g2} 
for correlation functions of the unitary gas~(Sec.~\ref{sec:FNMC}). 
We quantify the finite range corrections to the unitary gas energy in typical experiments,
which is required for precise measurements of its equation of state~(Sec.~\ref{subsec:app_manips}).
We conclude in Section \ref{sec:conclusion}.

\begin{table*}
\begin{tabular}{|cc|cc|}
\hline   
Three dimensions && Two dimensions& \\
\hline 
\vspace{-4mm}
& & &  \\
$\ds \psi(\rr_1,\ldots,\rr_N)\underset{r_{ij}\to0}{=}
\left( \frac{1}{r_{ij}}-\frac{1}{a} \right) \, A_{ij}\left(
\mathbf{R}_{ij}, (\mathbf{r}_k)_{k\neq i,j}
\right)
+O(r_{ij})\ $
&
(1a)
&
$\ds \psi(\rr_1,\ldots,\rr_N)\underset{r_{ij}\to0}{=}
\ln( r_{ij}/a ) \, A_{ij}\left(
\mathbf{R}_{ij}, (\mathbf{r}_k)_{k\neq i,j}
\right)
+O(r_{ij})\ $
&
(1b)
\vspace{-4mm}
\\
&&& \\
\hline 
\multicolumn{4}{|c|}{\vspace{-4mm}} \\
\multicolumn{3}{|c}{
$\ds
( A^{(1)},A^{(2)} )\equiv \sum_{i<j} \int \Big(\prod_{k\neq i,j} d^d\! r_k \Big) d^d\! R_{ij}
A^{(1)*}_{ij}(\mathbf{R}_{ij}, (\mathbf{r}_k)_{k\neq i,j})
A^{(2)}_{ij}(\mathbf{R}_{ij}, (\mathbf{r}_k)_{k\neq i,j})$ }
& (2) 
\\ 
\hline
\multicolumn{4}{|c|}{\vspace{-4mm}} \\
\multicolumn{3}{|c}{
$\ds
( A^{(1)},\mathcal{H} A^{(2)} )\equiv \sum_{i<j} \int \Big(\prod_{k\neq i,j} d^d\! r_k \Big) d^d\! R_{ij}
A^{(1)*}_{ij}(\mathbf{R}_{ij}, (\mathbf{r}_k)_{k\neq i,j}) \mathcal{H}_{ij}
A^{(2)}_{ij}(\mathbf{R}_{ij}, (\mathbf{r}_k)_{k\neq i,j})$ }
& (3) \vspace{-4mm}
\\
\multicolumn{4}{|c|}{} \\
\hline
\end{tabular}
\caption{Notation for the regular part $A$ of the $N$-body wavefunction appearing in the contact conditions (first line,
with $\RR_{ij}=(\rr_i+\rr_j)/2$ fixed), 
for the scalar product between such regular parts (second line) and for corresponding matrix elements of operators $\mathcal{H}_{ij}$
acting on $\RR_{ij}$ and on the $\rr_k$, $k\neq i,j$
(third line).
\label{tab:notations}}
\end{table*}

\section{Models, notations, and basic properties} \label{subsec:models}

We now introduce the three models used in this work to account for interparticle interactions and associated notations,
 together with some basic properties to be used in some of the derivations.

For a fixed number $N_\sigma$ of fermions in each spin state $\sigma=\uparrow,\downarrow$, one can consider that particles $1,\ldots,N_\uparrow$ have a spin~$\uparrow$ and particles $N_\uparrow+1,\ldots,N_\uparrow+N_\downarrow=N$ have a spin~$\downarrow$, so that the wavefunction $\psi(\rr_1,\ldots,\rr_N)$ 
(normalized to unity) changes sign when one exchanges the positions of two particles having the same spin~\footnote{ The corresponding state vector is $|\Psi\rangle=
[N!/(N_\uparrow!N_\downarrow!)]^{1/2} \hat{A} \left( |\uparrow,\ldots,\uparrow,\downarrow,\ldots,\downarrow\rangle \otimes |\psi\rangle \right)$ where there are $N_\uparrow$ spins $\uparrow$ and $N_\downarrow$ spins $\downarrow$,
and the operator $\hat{A}$ antisymmetrizes with respect to all particles. The wavefunction $\psi(\rr_1,\ldots,\rr_N)$ is then proportional to $\left( \langle \uparrow,\ldots,\uparrow,\downarrow,\ldots,\downarrow | \otimes \langle \rr_1,\ldots,\rr_N | \right) |\Psi\rangle$, with the proportionality factor
$(N!/N_\up!N_\down !)^{1/2}$.}.

\subsection{Zero-range model}

In this well-known model (see e.g. \cite{AlbeverioLivre,YvanHouchesBEC,PetrovJPhysB,Efimov, MaximLudo2D,RevueBraaten,YvanVarenna,WernerThese} and refs. therein)
the interaction potential is replaced by boundary conditions on the $N$-body wavefunction: For any pair of particles $i\neq j$, there exists a function $A_{ij}$, hereafter called regular part of $\psi$, such that 
[Tab. I, Eq.~(1a)] holds in the $3D$ case
and
[Tab. I, Eq.~(1b)] holds in the $2D$ case,
where the limit of vanishing distance $r_{ij}$ between particles $i$ and $j$ is taken for a fixed position 
of their center of mass $\mathbf{R}_{ij}=(\rr_i+\rr_j)/2$
and fixed positions  of the remaining particles $(\rr_k)_{k\neq i,j}$ different from
$\RR_{ij}$.
Fermionic symmetry of course imposes $A_{ij}=0$ if particles $i$ and $j$ have the same spin.
When none of the $\rr_i$'s coincide, there is no interaction potential and Schr\"odinger's equation reads 
$H\,\psi(\rr_1,\ldots,\rr_N)=E\,\psi(\rr_1,\ldots,\rr_N)$
with
$\ds H=-\frac{\hbar^2}{2 m}\sum_{i=1}^N 
\Delta_{\rr_i} + H_{\rm trap}$,
where $m$ is the atomic mass
and the trapping potential energy is
\be
H_{\rm trap}\equiv\sum_{i=1}^N U(\rr_i),
\label{eq:Htrap}
\ee
$U$ being an external trapping potential.
The crucial difference between the Hamiltonian $H$ and the non-interacting Hamiltonian is the boundary condition [Tab. I, Eqs. (1a,1b)].

\subsection{Lattice models} \label{sec:models:lattice}

These models are used for quantum Monte Carlo calculations \cite{bulgacQMC,zhenyaPRL,zhenyaNJP,Juillet,ChangAFMC,BulgacCrossover}. They can also be convenient for analytics, as used in \cite{MoraCastin,LudoYvanBoite, WernerTarruellCastin,MoraCastin2D} and in this work.
Particles live on a lattice, i. e. the coordinates are integer multiples of the lattice spacing $b$. The Hamiltonian is
\be
H=H_{\rm kin}+
H_{\rm int}+H_{\rm trap}
\label{eq:Hlatt}
\ee
with, in first quantization, the kinetic energy
\be
H_{\rm kin}=-\frac{\hbar^2}{2 m}\sum_{i=1}^N
\Delta_{\rr_i}, 
\label{eq:def_H0}
\ee
the interaction energy
\be
H_{\rm int}=g_0 \sum_{i<j} \delta_{\rr_i,\rr_j} b^{-d},
\label{eq:def_W}
\ee
and the trapping potential energy defined by (\ref{eq:Htrap});
i.e. in second quantization
\bea
H_{\rm kin}&=&\sum_\sigma\int_D \frac{d^dk}{(2\pi)^d}\,\epsk c_\sigma^\dagger(\kk)c_\sigma(\kk) 
\label{eq:Hkin_second_quant} 
\\
H_{\rm int}&=&g_0\sum_\rr b^d (\psi^\dagger_\up\psi^\dagger_\down\psi_\down\psi_\up)(\rr)
\label{eq:defW_2e_quant}
\\
H_{\rm trap}&=&\sum_{\rr,\sigma} b^d U(\rr) (\psi_\sigma^\dagger \psi_\sigma)(\rr).
\eea
Here $d$ is the space dimension, 
$\epsilon_\kk$ is the dispersion relation, $\hat{\psi}$ obeys discrete anticommutation relations
$\{\hat{\psi}_\sigma(\rr), \hat{\psi}^\dagger_{\sigma'}(\rr')\}=b^{-d} \delta_{\rr \rr'} \delta_{\sigma\sigma'}$.
The operator $ c_\sigma^\dagger(\kk)$ creates a particle in the plane wave state $|\kk\rangle$ defined by
$\langle \rr | \kk \rangle = e^{i \kk \cdot \rr}$ for any $\kk$ belonging to the first Brillouin zone
$D=\left(-\frac{\pi}{b},\frac{\pi}{b}\right]^d$. The corresponding anticommutation relations
are $\{c_\sigma(\kk),c^\dagger_{\sigma'}(\kk')\}= (2\pi)^d \delta_{\sigma\sigma'} \delta(\kk-\kk')$
if $\kk$ and $\kk'$ are both in the first Brillouin zone \footnote{Otherwise
$\delta(\kk-\kk')$ has to be replaced by the periodic version
$\sum_{\KK\in (2\pi/b)\mathbb{Z}^d} \delta(\kk-\kk'-\KK)$.}.
The operator $\Delta$ in (\ref{eq:def_H0}) is the lattice version of the Laplacian defined by
$-\frac{\hbar^2}{2m}\langle \rr | \Delta_\rr | \kk \rangle \equiv \epsilon_\kk \langle \rr | \kk \rangle$.
The simplest choice for the dispersion relation is $\ds\epsk=\frac{\hbar^2k^2}{2m}$~\cite{MoraCastin,Juillet,LudoYvanBoite, MoraCastin2D,ChangAFMC}. Another choice, used in~\cite{zhenyaPRL,zhenyaNJP}, is the dispersion relation of the Hubbard model: 
$\ds\epsk=\frac{\hbar^2}{m b^2}\sum_{i=1}^d\left[1-\cos(k_i b)\right]$. More generally, what follows applies to any $\epsk$ such that $\ds\epsk\underset{b\to0}{\rightarrow}\frac{\hbar^2k^2}{2m}$ sufficiently rapidly and $\epsilon_{-\kk}=\epsk$.

A key quantity is the zero-energy scattering state $\phi(\rr)$, defined by the two-body Schr\"odinger equation (with the center of mass at rest)
\be
\left(-\frac{\hbar^2}{m}\,\Delta_\rr +g_0 \frac{\delta_{\rr,\vn}}{b^d}\right) \phi(\rr) = 0
\ee
and by the normalization conditions
\bea
\phi(\rr)&\underset{r\gg b}{\simeq}& \frac{1}{r}-\frac{1}{a} \ \ \ {\rm in}\ 3D
\label{eq:normalisation_phi_tilde_3D}
\\
\phi(\rr)&\underset{r\gg b}{\simeq}& \ln(r/a) \ \ \ {\rm in}\ 2D.
\label{eq:normalisation_phi_tilde_2D}
\eea
A two-body analysis, detailed in Appendix~\ref{app:2body}, yields the relation between the scattering length and the bare coupling constant $g_0$,
in three and two dimensions:
\bea
\!\!\!\!\!\!\!\!\!\!\!\! &&\frac{1}{g_0} \stackrel{3D}{=}\frac{m}{4\pi\hbar^2 a}\!-\!\! \int_D \!\!\frac{d^3 k}{(2\pi)^3} \frac{1}{2\epsk}  
\label{eq:g0_3D}
\\
\!\!\!\!\!\!\!\!\!\!\!\! &&\frac{1}{g_0} \stackrel{2D}{=} \lim_{q\to 0} \left[ 
-\frac{m}{2\pi\hbar^2}\ln(\frac{a q e^\gamma}{2}) \!
+\!\!\int_D\!\! \frac{d^2 k}{(2\pi)^2} \mathcal{P} \frac{1}{2(\epsilon_{\mathbf{q}} - \epsk)} \right] 
\label{eq:g0_2D}
\eea
where $\gamma=0.577216\ldots$ is Euler's constant and $\mathcal{P}$ is the principal value.
This implies that (for constant $b$):
\bea
\frac{d(1/g_0)}{d(1/a)}=\frac{m}{4\pi\hbar^2 } & & \ \ \ {\rm in}\ 3D
\label{eq:dg0_da_3D}
\\
\frac{d(1/g_0)}{d(\ln a)}= 
-\frac{m}{2\pi\hbar^2}
& & \ \ \ {\rm in}\ 2D.
\label{eq:dg0_da_2D}
\eea
Another useful property derived in Appendix~\ref{app:2body} is
\bea
\phi(\vn)&=&-\frac{4\pi\hbar^2}{m g_0}\ \ \ {\rm in}\ 3D
\label{eq:phi0_vs_g0}
\\
\phi(\vn)&=&\frac{2\pi\hbar^2}{m g_0}\ \ \ {\rm in}\ 2D,
\label{eq:phi0_vs_g0_2D}
\eea
which, together with (\ref{eq:dg0_da_3D},\ref{eq:dg0_da_2D}), gives
\bea
|\phi(\vn)|^2&=&\frac{4\pi\hbar^2}{m}\frac{d(-1/a)}{dg_0}\ \ \ {\rm in}\ 3D
\label{eq:phi_tilde_3D}
\\
|\phi(\vn)|^2&=&\frac{2\pi\hbar^2}{m}\frac{d(\ln a)}{dg_0}\ \ \ {\rm in}\ 2D
\label{eq:phi_tilde_2D}.
\eea

In the zero-range limit ($b\to 0$ with $g_0$ adjusted in such a way that $a$ remains constant), it is expected
that the spectrum of the lattice model converges to the one of the zero-range model,
as explicitly checked for three particles in~\cite{LudoYvanBoite}, 
and that any eigenfunction $\psi(\rr_1,\dots,\rr_N)$ of the lattice model tends to the corresponding eigenfunction of the zero-range model 
provided all interparticle distances remain much larger than $b$.
For any stationary state, let us denote by $1/\ktyp$ the typical length-scale on which the zero-range model's wavefunction varies:
e.g. for the lowest eigenstates,
this
is on the order of the mean interparticle distance, or
on the order of $a$ in the regime where $a$ is small and positive and dimers are formed.
The zero-range limit is then reached if $\ktyp b\ll1$.
This
 notion of 
typical wavevector $\ktyp$
can also be applied to
the case of a thermal equilibrium state, since most significantly populated eigenstates then have a  $\ktyp$ on the same order;
it is then expected
that the thermodynamic potentials converge to the ones of the zero-range model when $b\to0$,
and that this limit is reached provided $\ktyp b \ll 1$.
For the homogeneous gas,
defining a thermal wavevector $k_T$ by $\hbar^2 k_T^2/(2m)=k_B\,T$,
 we have $\ktyp\sim\max(k_F,k_T)$ for $a<0$ and 
$\ktyp\sim\max(k_F,k_T,1/a)$
for $a>0$.

For lattice models, it will prove convenient to define the regular part  $A$ by
\begin{multline}
\psi(\rr_1,\ldots,\rr_i=\RR_{ij},\ldots,\rr_j=\RR_{ij},\ldots,\rr_N)=\phi(\vn) \\
\times A_{ij}(\RR_{ij},(\rr_k)_{k\neq i,j}).
\label{eq:def_A_reseau}
\end{multline}
In the zero-range regime $k_{\rm typ} b\ll1$, when the distance $r_{ij}$ between two particles of opposite spin is $\ll 1/\ktyp$ while all the other interparticle distances are much larger than $b$ and than $r_{ij}$, the many-body wavefunction is proportional to $\phi(\rr_j-\rr_i)$, with a proportionality constant given by~(\ref{eq:def_A_reseau}):
\be
\psi(\rr_1,\ldots,\rr_N)\simeq\phi(\rr_j-\rr_i)\,A_{ij}(\RR_{ij},(\rr_k)_{k\neq i,j}) 
\label{eq:psi_courte_dist}
\ee
where $\RR_{ij}=(\rr_i+\rr_j)/2$.
If moreover $r_{ij}\gg b$, $\phi$ can be replaced by its asymptotic form (\ref{eq:normalisation_phi_tilde_3D},\ref{eq:normalisation_phi_tilde_2D}); since
the contact conditions [Tab. I, Eqs. (1a,1b)] of the zero-range model must be recovered, we see that the lattice model's regular part tends to the zero-range model's regular part in the zero-range limit.

\subsection{Finite-range continuous-space models}

Such models are used in numerical few-body
correlated Gaussian  and many-body fixed-node Monte Carlo  calculations (see e.\ g.\ \cite{Pandha,Giorgini,BlumeUnivPRL,StecherLong,BlumeRelations, RevueTrentoFermions, Giorgini2D} and refs. therein).
They are also relevant to neutron matter \cite{GezerlisCarlson}.
The Hamiltonian reads
\be
H=H_0+\sum_{i=1}^{N_\up}\sum_{j=N_\up+1}^N\,V(r_{ij}),
\ee
$H_0$ being defined by (\ref{eq:def_H0}) where $\Delta_{\rr_i}$ now stands for the usual Laplacian,
and $V(r)$ is an
interaction potential between particles of opposite spin, which vanishes for $r>b$ or at least decays quickly enough for $r\gg b$.
The two-body zero-energy scattering state $\phi(r)$ is again defined by the Schr\"odinger equation $-(\hbar^2/m)\Delta_\rr\phi+V(r)\phi=0$ and the boundary condition (\ref{eq:normalisation_phi_tilde_3D}) or (\ref{eq:normalisation_phi_tilde_2D}).
The
zero-range regime
is again reached for $k_{\rm typ} b\ll1$ with $k_{\rm typ}$ the typical relative wavevector~\footnote{
For purely attractive interaction potentials such as the square-well potential, above a critical particle number, 
the ground state is a collapsed state and
the zero-range regime can only be reached for certain excited states (see e.g. \cite{LeChapitre} and refs. therein).}. Equation (\ref{eq:psi_courte_dist}) again holds in the zero-range regime, where $A$ now simply stands for the zero-range model's regular part.

\section{Relations in the zero-range limit}
\label{sec:ZR}

\begin{table*}
\begin{tabular}{|cc|cc|}
\hline  
  Three dimensions & & Two dimensions & \\
\hline
 \multicolumn{4}{|c|}{\vspace{-4mm}} 
\\
 \multicolumn{3}{|c}{$C\equiv {\displaystyle \lim_{k\to +\infty}} 
k^4 n_\sigma(\kk)$}
& (1) \vspace{-4mm}
\\
  \multicolumn{4}{|c|}{} \\
\hline 
\vspace{-4mm}
& & & \\
\vspace{-4mm}
$\ds C =
(4\pi)^2\ (A,A)
$
&
(2a)
&
$\ds C = 
(2\pi)^2\,(A,A)
$
&
(2b)
\\
& & & \\
\hline
& & &
\vspace{-4mm}
\\
$\ds
\int d^3R \,
 g_{\uparrow \downarrow}^{(2)} \left(\mathbf{R}+\frac{\mathbf{r}}{2},
\mathbf{R}-\frac{\mathbf{r}}{2}\right) 
\underset{r\to0}{\sim}
\frac{C}{(4\pi)^2}
\frac{1}{r^2}
$
&(3a)  \vspace{-4mm}
&
$\ds
\int d^2R \,
 g_{\uparrow \downarrow}^{(2)} \left(\mathbf{R}+\frac{\mathbf{r}}{2},
\mathbf{R}-\frac{\mathbf{r}}{2}\right)
\underset{r\to0}{\sim}
\frac{C}{(2\pi)^2}
\ln^2 r$
&
(3b)
\\
& & & 
\\
 \hline & & &
\vspace{-4mm}
\\
$\displaystyle\frac{dE}{d(-1/a)} = \frac{\hbar^2 C}{4\pi m} $ &
 (4a) &
$\ds\frac{dE}{d(\ln a)} = \frac{\hbar^2 C}{2\pi m} $ 
& (4b)  \vspace{-4mm}
 \\
&& & \\
\hline & & & \vspace{-4mm} \\
$\ds E - E_{\rm trap}  = \frac{\hbar^2 C}{4\pi m a}  $ 
&  &
$\ds E - E_{\rm trap}  = \lim_{\Lambda\to\infty}\left[-\frac{\hbar^2 C}{2\pi m} \ln \left(\frac{a \Lambda e^\gamma}{2}\right) \right.
$
&
\vspace{-3mm}
\\
& & & \\ 
$\ds +\sum_{\sigma} \int \frac{d^3\!k}{(2\pi)^3}  \frac{\hbar^2 k^2}{2m} 
\left[n_\sigma(\kk) - \frac{C}{k^4}\right]$
&(5a)
&
$\ds  +\sum_{\sigma} \left. \int_{k<\Lambda} \frac{d^2\!k}{(2\pi)^2}  \frac{\hbar^2 k^2}{2m} 
n_\sigma(\kk) \right]$
& (5b) \vspace{-4mm}
\\&
&&\\
\hline &&& \vspace{-4mm}  \\
$\ds \int d^3R \,
 g_{\sigma \sigma}^{(1)} \left(\mathbf{R}+\frac{\mathbf{r}}{2},
\mathbf{R}-\frac{\mathbf{r}}{2}\right)
\underset{r\to0}{=}
N_\sigma -\frac{C}{8\pi}\, r + O(r^2)\ \ \ $
&
(6a)  &
$\ds \int d^2R \,
 g_{\sigma \sigma}^{(1)} \left(\mathbf{R}+\frac{\mathbf{r}}{2},
\mathbf{R}-\frac{\mathbf{r}}{2}\right)
\underset{r\to0}{=}
N_\sigma +\frac{C}{4\pi}\, r^2\ln r + O(r^2)\ \ \ $
&
(6b)
\vspace{-4mm}
\\ &&& \\
\hline
&&
&
\vspace{-4mm}
\\ 
$\ds \frac{1}{3} \sum_{i=1}^3  \sum_\si \int d^3R \,
 g_{\sigma \sigma}^{(1)} \left(\mathbf{R}+\frac{r {\bf u_i}}{2},
\mathbf{R}-\frac{r {\bf u_i}}{2}\right)
\underset{r\to 0}{=}
N$
&
  &
$\ds \frac{1}{2} \sum_{i=1}^2 \sum_\si
 \int d^2R \,
 g_{\sigma \sigma}^{(1)} \left(\mathbf{R}+\frac{r {\bf u_i}}{2},
\mathbf{R}-\frac{r {\bf u_i}}{2}\right)
\underset{r\to 0}{=}
N
$&

\\
$\ds -\frac{C}{4\pi}r -\frac{m}{3\hbar^2}\left(
E-E_{\rm trap} - \frac{\hbar^2 C}{4\pi m a}
\right) r^2 + o(r^2)$
&(7a)&
$\ds +\frac{C}{4\pi}r^2
\left[\ln\left(\frac{r}{a}\right) -1\right]
-\frac{m}{2\hbar^2}\left(
E-E_{\rm trap}
\right) r^2 + o(r^2)$
&(7b) \vspace{-4mm}
\\&& & \\
\hline
&& & \vspace{-4mm}
 \\$\ds\frac{1}{2} \frac{d^2E_n}{d(-1/a)^2}
= \left(\frac{4\pi\hbar^2}{m}\right)^2 \sum_{n',E_{n'}\neq E_n} 
\frac{|(A^{(n')},A^{(n)})|^2}{E_n-E_{n'}}$
&(8a)  &
$\ds\frac{1}{2} \frac{d^2E_n}{d(\ln a)^2}
= \left(\frac{2\pi\hbar^2}{m}\right)^2 \sum_{n',E_{n'}\neq E_n} 
\frac{|(A^{(n')},A^{(n)})|^2}{E_n-E_{n'}} $
&
 (8b) \vspace{-4mm}
 \\&&
& \\
 \hline & & & \vspace{-4mm} \\
$\displaystyle\left(\frac{d\bar{E}}{d(-1/a)}\right)_{\!S} =\left(\frac{dF}{d(-1/a)}\right)_{T} = \frac{\hbar^2 C}{4\pi m} $ &
 (9a)  &
$\displaystyle\left(\frac{d\bar{E}}{d(\ln a)}\right)_{\!S} =\left(\frac{dF}{d(\ln a)}\right)_{T} = \frac{\hbar^2 C}{2\pi m}$ 
 &
(9b)\vspace{-4mm} \\
&& & 
\\ \hline
 && 
&\vspace{-4mm} \\
$\ds\left(\frac{d^2F}{d(-1/a)^2}\right)_T < 0 $ 
&(10a) 
&
$\ds\left(\frac{d^2F}{d(\ln a)^2}\right)_T < 0 $
&
(10b) \vspace{-4mm}
\\
&& &\\
\hline
&&& \vspace{-4mm} \\
$\ds\left(\frac{d^2\bar{E}}{d(-1/a)^2}\right)_{\!S} < 0 $
&(11a)&
$\ds\left(\frac{d^2\bar{E}}{d(\ln a)^2}\right)_{\!S} < 0 $
&(11b)\vspace{-4mm}\\
&&&\\
\hline
 & && \vspace{-4mm} \\
$\ds\frac{dE}{dt} = \frac{\hbar^2 C}{4\pi m} \frac{d(-1/a)}{dt}+
\Big<\frac{dH_{\rm trap}}{dt}\Big>$
&(12a) 
&
$\ds\frac{dE}{dt} = \frac{\hbar^2 C}{2\pi m} \frac{d(\ln a)}{dt}+
\Big<\frac{dH_{\rm trap}}{dt}\Big>$
&
(12b)\vspace{-4mm}\\
&& & \\
\hline
\end{tabular}
\caption{Relations for spin-1/2 fermions with zero-range interactions.
The definition (1) of $C$, as well as the relations in lines 3, 5, 6 and 7,
concern any (non-pathological) statistical mixture
of states which satisfy the contact conditions~[Tab.~I, line~1] (with real wavefunctions for line 7).
Line~2
 holds for any pure state; here $A$ is the regular part of the wavefunction appearing in the contact condition,
 and $(A,A)$ is its squared norm (defined in Tab.~I).
 Lines 4 and 8
hold for any stationary state.
Lines 9-11 hold at thermal equilibrium in the canonical ensemble.
Line 12 holds for any time-dependence of scattering length and trapping potential, and any corresponding time-dependent statistical mixture.
Many of the $3D$ relations were originally obtained in~\cite{TanEnergetics,TanLargeMomentum} (see text), while the $2D$ relation~(5b) 
was obtained in~\cite{CombescotC} for the homogeneous system and in \cite{TanSimple} (in a different form) for the general case.
\label{tab:fermions}}
\end{table*}

We now derive relations for the zero-range model. For some of the derivations we will use a lattice model and then take the zero-range limit.
We recall that we derive all relations for pure states in this section, the generalization to statistical mixtures and the discussion of thermal equilibrium being deferred to Sections~\ref{sec:stat_mix} and \ref{subsec:finiteT}.

\subsection{Tail of the momentum distribution}
\label{sec:C_nk}

In this subsection as well as in the following subsections~\ref{sec:g2}, \ref{sec:energy_thm}, 
\ref{sec:g1}, \ref{subsec:tdote},
we consider a many-body pure state whose wavefunction $\psi$ satisfies the contact condition~[Tab. I, Eqs. (1a,1b)].
We now show that
 the momentum distribution
$n_\sigma(\kk)$ has a $\sigma$-independent
tail proportional to $1/k^4$, with a coefficient denoted by $C$~[Tab. II, Eq. (1)].
$C$ is usually referred to as the ``contact''.
We shall also show that 
$C$ is related by [Tab.~II, Eqs.~(2a,2b)]
to the norm of the regular part $A$ of the wavefunction (defined in Tab.~I).
In $3D$ these results were obtained in~\cite{TanLargeMomentum} \footnote{The existence of the $1/k^4$ tail had already been observed within a self-consistent approximate theory~\cite{Haussmann_PRB}.}.
Here the momentum distribution is defined in second quantization by
$\ds n_\sigma(\kk) =\la\hat{n}_\sigma(\kk)\ra= \la {c}^\dagger_\si(\kk) {c}_\si(\kk) \ra$
where ${c}_\si(\kk)$ annihilates a particle of spin $\si$ in the plane-wave state $|\kk\ra$ defined by $\la\rr|\kk\ra=e^{i \kk\cdot\rr}$;
this corresponds to the normalization
\be
\int \frac{d^d k}{(2\pi)^d}\,n_\sigma(\kk) = N_\sigma.
\label{eq:def_nk_fermions}
\ee
In first quantization,
\be
n_\sigma(\kk)=\sum_{i:\sigma} \int \Big( \prod_{l\neq i} d^d r_l \Big)
\left| \int d^d r_i e^{-i \kk\cdot\rr_i} \psi(\rr_1,\ldots,\rr_N)\right|^2
\label{eq:nk_1e_quant}
\ee
where the sum is taken over all particles of spin $\sigma$: $i$ runs from $1$ to $N_\uparrow$ for $\sigma=\uparrow$,
and from $N_\uparrow+1$ to $N$ for $\sigma=\downarrow$.

\noindent{\underline{\it Three dimensions:}}
\\
The key point is that in the large-$k$ limit, the Fourier transform with respect to $\rr_i$ is dominated by the contribution of the short-distance divergence coming from the contact condition~[Tab. I, Eq. (1a)]:
\begin{multline}
\int d^3 r_i\, e^{-i \kk\cdot\rr_i} \psi(\rr_1,\ldots,\rr_N)
\underset{k\to\infty}{\simeq}\int d^3 r_i\, e^{-i \kk\cdot\rr_i} \\
\times \sum_{j,j\neq i} \frac{1}{r_{ij}} A_{ij}(\rr_j,(\rr_l)_{l\neq i,j}).
\label{eq:FT_sing_3D}
\end{multline}
A similar link between the short-distance singularity of the wavefunction and the tail of its Fourier transform was used to derive exact relations in $1D$ in~\cite{Olshanii_nk}.
From $\Delta(1/r)=-4\pi\delta(\rr)$, we have
$\ds \int d^3 r \,e^{-i \kk\cdot \rr}\frac{1}{r}=\frac{4\pi}{k^2}$, so that
\begin{multline}
\int d^3 r_i \,e^{-i \kk\cdot\rr_i} \psi(\rr_1,\ldots,\rr_N)
\underset{k\to\infty}{\simeq} \frac{4\pi}{k^2}
\sum_{j,j\neq i} e^{-i \kk\cdot\rr_j} \\
\times A_{ij}(\rr_j,(\rr_l)_{l\neq i,j}).
\end{multline}
One inserts this into (\ref{eq:nk_1e_quant}) and expands the modulus squared. 
After spatial integration over all the $\rr_l$, $l\neq i$,
the crossed terms rapidly vanish in the large-$k$ limit, as they are the product
of $e^{i\kk\cdot (\rr_j-\rr_{j'})}$ and of regular functions of $\rr_j$ and $\rr_{j'}$
\footnote{E.g.\ for $n_\downarrow(\kk)$ in the trapped three-body case, 
with particles $1$ and $2$ in state $\uparrow$ and particle $3$ in state $\downarrow$, one has $i=3$ and
$j,j'=1$ or $2$.  Then the crossed term $A_{31}(\rr_1,\rr_2)A_{32}(\rr_2,\rr_1)$ has to all orders finite derivatives with respect to $\rr_1$ and $\rr_2$,
except if $\rr_1=\rr_2$  where it vanishes as $|\rr_1-\rr_2|^{2s-2}$, $s>0$ not integer, see e.g.\ Eq.~(\ref{eq:separa_Aij}) and below that equation.
By a power counting argument, its Fourier transform with respect to $\rr_1-\rr_2$ contributes to the momentum distribution tail 
as $1/k^{2s+5}=o(1/k^4)$;  one recovers the ``three-close-particle" contribution mentioned
in a note of \cite{TanLargeMomentum}.}.
This yields
$n_\sigma(\kk)\underset{k\to\infty}{\sim}C/k^4$,
with the expression~[Tab. II, Eq. (2a)] of $C$ in terms of the norm $(A,A)$ defined in~[Tab. I, Eq. (2)].



\noindent{\underline{\it Two dimensions:}}
\\
The $2D$ contact condition~[Tab.~I, Eq.~(1b)] now gives
\begin{multline}
\int d^2 r_i \, e^{-i \kk\cdot\rr_i} \psi(\rr_1,\ldots,\rr_N)
\underset{k\to\infty}{\simeq}\int d^2 r_i \, e^{-i \kk\cdot\rr_i} \\
\times \sum_{j,j\neq i} \ln (r_{ij}) A_{ij}(\rr_j,(\rr_l)_{l\neq i,j}).
\label{eq:FT_sing_2D}
\end{multline}
From $\Delta(\ln r)=2\pi\delta(\rr)$, one has
$\!\!\ds\int\!\!  d^2 r \,e^{-i \kk\cdot \rr}\ln r=-\frac{2\pi}{k^2}$
and
\begin{multline}
\int d^2 r_i \,e^{-i \kk\cdot\rr_i} \psi(\rr_1,\ldots,\rr_N)
\underset{k\to\infty}{\simeq} -\frac{2\pi}{k^2}
\sum_{j,j\neq i} e^{-i \kk\cdot\rr_j} \\
\times A_{ij}(\rr_j,(\rr_l)_{l\neq i,j}).
\end{multline}
As in $3D$ this leads to
[Tab. II, Eq. (2b)].


\subsection{Pair distribution function at short distances}
\label{sec:g2}

The pair distribution function gives the probability density of finding a spin-$\uparrow$ particle at
$\rr_\uparrow$ and a spin-$\downarrow$ particle at $\rr_\downarrow$:
$\ds g_{\uparrow\downarrow}^{(2)}\left(\rr_\uparrow,\rr_\downarrow\right)=\langle
(\hat{\psi}^\dagger_\uparrow\hat{\psi}_\up)(\rr_\uparrow)
(\hat{\psi}^\dagger_\downarrow
\hat{\psi}_\downarrow)(\rr_\downarrow)\rangle=\int (\prod_{k=1}^{N} d^d r_k)
|\psi(\rr_1,\ldots,\rr_N)|^2
\sum_{i=1}^{N_\uparrow} \sum_{j=N_\uparrow+1}^{N}
\!\!\!\!\!\!\delta\left(\rr_\uparrow\!-\!\rr_i\right)\delta\left(\rr_\downarrow\!-\!\rr_j\right)$.
We set $\rr_{\uparrow,\downarrow}=\RR\pm \rr/2$ and we integrate over $\rr_i$ and
$\rr_j$:
\begin{multline}
g_{\uparrow\downarrow}^{(2)}\left(\RR+\frac{\rr}{2},\RR-\frac{\rr}{2}\right)
= \sum_{i=1}^{N_\uparrow}
\sum_{j=N_\uparrow+1}^{N} 
\int \Big(  \prod_{k\neq i,j} d^d r_k \Big) \\
\left| \psi\left(\rr_1,\ldots,\rr_i=\RR+\frac{\rr}{2},\ldots,\rr_j=\RR-\frac{\rr}{2},\ldots,\rr_N\right) \right|^2
\label{eq:def_g2_psi}
\end{multline}
Let us define the spatially integrated pair distribution function~\footnote{For simplicity, we refrain here from expressing $C$ as the integral of a ``contact density'' $\mathcal{C}(\RR)$ 
related to the small-$r$ behavior of the local pair distribution function
$g_{\uparrow \downarrow}^{(2)} \left(\mathbf{R}+\mathbf{r}/2,
\mathbf{R}-\mathbf{r}/2\right)$
as was done for the $3D$ case in~\cite{TanEnergetics,TanLargeMomentum,Braaten}; this $\mathcal{C}(\RR)$ is then also related to the large-$k$ tail of the Wigner distribution [i.e. the Fourier transform with respect to $\rr$ of the one-body density matrix $\la\psi^\dagger_\sigma(\RR-\rr/2)\psi_\sigma(\RR+\rr/2)\ra$], see Eq.~(30) of~\cite{TanEnergetics}.}
\be
G^{(2)}_{\uparrow\downarrow}(\rr)\equiv
\int d^dR \ 
 g_{\uparrow \downarrow}^{(2)} \left(\mathbf{R}+\frac{\mathbf{r}}{2},
\mathbf{R}-\frac{\mathbf{r}}{2}\right),
\ee
whose small-$r$ singular behavior we will show to be related to $C$ {\rm via}~[Tab.~II, Eqs.~(3a,3b)].

\noindent{\underline{\it Three dimensions:}}
\\
Replacing the wavefunction in~(\ref{eq:def_g2_psi}) by its asymptotic behavior given by the contact condition~[Tab. I, Eq. (1a)] immediately yields
\be
G^{(2)}_{\uparrow\downarrow}(\rr)
\underset{r\to 0}{\sim}
\frac{(A,A)}{r^2}.
\ee
Expressing $(A,A)$ in terms of $C$ through~[Tab. II, Eq. (2a)] finally gives
[Tab.~II, Eq.~(3a)].

In a measurement of all particle positions, the mean
total number of pairs of particles of opposite spin which are separated by a distance smaller than $s$ is
$N_{\rm pair}(s)=\int_{r<s} d^d r\, G^{(2)}_{\uparrow\downarrow}(\rr)$,
so that from [Tab.~II, Eq.~(3a)]
\be
N_{\rm pair}(s)\underset{s\to 0}{\sim} \frac{C}{4\pi}
s,
\label{eq:Npair_3D}
\ee
as obtained in~\cite{TanEnergetics,TanLargeMomentum}.

\noindent{\underline{\it Two dimensions:}}
\\
The contact condition~[Tab.~I, Eq.~(1b)] similarly leads to
[Tab.~II, Eq.~(3b)].
After integration over the region $r<s$ this gives
\be
N_{\rm pair}(s)\underset{s\to 0}{\sim} \frac{C}{4\pi}
s^2
\ln^2 s.
\label{eq:Npair_2D}
\ee

\subsection{First order derivative of the energy with respect to the scattering length}
\label{sec:dEda}

The relations [Tab. II, Eqs.~(4a,4b)] can be derived straightforwardly using the lattice model, see Sec.\ref{sec:dE_latt}.
Here we derive them by directly using the zero-range model, which is  more involved but also instructive.

\noindent{\underline{\it Three dimensions:}}
\\
Let us consider a wavefunction $\psi_1$ satisfying the contact condition~[Tab. I, Eq. (1a)] for  a scattering length $a_1$. We denote by $A^{(1)}_{ij}$ the regular part of $\psi_1$ appearing in the contact condition~[Tab. I, Eq. (1a)]. Similarly, $\psi_2$ satisfies the contact condition for a scattering length $a_2$ and a regular part $A^{(2)}_{ij}$. 
Then, as shown in Appendix~\ref{app:lemme} using the divergence theorem, the following lemma holds:
\be
\langle \psi_1, H \psi_2 \rangle - \langle H \psi_1, \psi_2 \rangle = 
\frac{4\pi\hbar^2}{m}\left(\frac{1}{a_1}-\frac{1}{a_2}\right) \ ( A^{(1)},A^{(2)} )
\label{eq:lemme_3D}
\ee
where the scalar product between regular parts is defined by~[Tab. I, Eq. (2)].
We then apply (\ref{eq:lemme_3D}) to the case where $\psi_1$ and $\psi_2$ are $N$-body stationary states of energy $E_1$ and $E_2$. The left hand side of (\ref{eq:lemme_3D}) then reduces to $(E_2-E_1) \langle \psi_1 | \psi_2 \rangle$. Taking the limit $a_2\to a_1$ gives 
\be
\frac{dE}{d(-1/a)} = \frac{4\pi\hbar^2}{m} (A,A)
\label{eq:thm_dE_3D}
\ee
for any stationary state.
Expressing $(A,A)$ in terms of $C$ thanks to~[Tab.~II, Eq.~(2a)] finally yields~[Tab.~II, Eq.~(4a)]. This result as well as~(\ref{eq:thm_dE_3D}) is contained in Ref.~\cite{TanEnergetics,TanLargeMomentum}\footnote{Our derivation is similar to the one given in the two-body case and sketched in the many-body case in Section~3 of~\cite{TanLargeMomentum}.}.
We recall that here and in what follows, the wavefunction is normalized: $\la\psi|\psi\ra=1$.

\noindent{\underline{\it Two dimensions:}}
\\
The $2D$ version of the lemma~(\ref{eq:lemme_3D}) is
\be
\langle \psi_1, H \psi_2 \rangle - \langle H \psi_1, \psi_2 \rangle = 
\frac{2\pi\hbar^2}{m}\ln\left(a_2/a_1\right) \ ( A^{(1)},A^{(2)} ),
\label{eq:lemme_2D}
\ee
as shown in Appendix~\ref{app:lemme}.
As in $3D$, we deduce that
\be
\frac{dE}{d(\ln a)} = \frac{2\pi\hbar^2}{m} (A,A),
\label{eq:thm_dE_2D}
\ee
which gives the desired~[Tab.~II, Eq.~(4b)] by using~[Tab.~II, Eq.~(2b)].


\subsection{Expression of the energy in terms of the momentum distribution}
\label{sec:energy_thm}
\noindent{\underline{\it Three dimensions:}}
\\
As shown in~\cite{TanEnergetics}, the mean total energy $E$
minus the mean trapping-potential energy
$E_{\rm trap}\equiv \left< H_{\rm trap}\right>$,
 has the simple expression in terms of the momentum distribution given in~[Tab.~II, Eq.~(5a)],
for any pure state $|\psi\ra$ satisfying the contact condition~[Tab. I, Eq. (1a)].
We give a simple rederivation of this result by using the lattice model (defined in Sec.~\ref{sec:models:lattice}).

We first treat the case where $|\psi\ra$ is an eigenstate of the zero-range model.
Let $|\psi_b\ra$ be the eigenstate of the lattice model that tends to $|\psi\ra$ for $b\to0$.
We first note that 
$C_b\equiv\la\psi_b|\hat{C}|\psi_b\ra$, where $\hat{C}$ is defined by [Tab.~III, Eqs.~(1a,1b)],
tends to 
the contact $C$ of the state $\psi$
[defined in~Tab.~II, Eq.~(1)]
when $b\to0$,
as shown in Appendix~\ref{app:C_b}.
Then, the key step is to use~[Tab.~III, Eq.~(3a)], which, after taking the expectation value in the state $|\psi_b\ra$,
yields the desired~[Tab.~II, Eq.~(5a)] in the zero-range limit since $\ds D\rightarrow\mathbb{R}^3$ and $\epsilon_\kk\to\hbar^2 k^2/(2m)$ for $b\to0$.

To generalize~[Tab.~II, Eq.~(5a)] to any pure state $|\psi\ra$
satisfying the contact condition~[Tab. I, Eq. (1a)], we use
the state $|\psi_b\ra$ defined in~Appendix~\ref{app:C_b_2}.
As shown in that appendix,
the expectation value of $\hat{C}$
taken in this state $|\psi_b\ra$
tends to the contact $C$ of $|\psi\ra$~[defined in Tab.~II, Eq.~(1)].
Moreover the expectation values of $H-H_{\rm trap}$ and of $\hat{n}_\sigma(\kk)$,
taken in this state $|\psi_b\ra$,
should tend to the corresponding expectation values taken in the state $|\psi\ra$.
This yields the desired relation.

Finally we mention the equivalent form of relation~[Tab.~II, Eq.~(5a)]:
\begin{multline}
E - E_{\rm trap}  = \lim_{\Lambda\to\infty}\Bigg[
 \frac{\hbar^2 C}{4\pi m}\left(\frac{1}{a}-\frac{2\Lambda}{\pi}\right) 
\\  +\sum_\sigma\int_{k<\Lambda} \frac{d^3k}{(2\pi)^3}  \frac{\hbar^2 k^2}{2m} 
n_\sigma(\kk) \Bigg].
\label{eq:energy_thm_3D_Lambda}
\end{multline}

\noindent{\underline{\it Two dimensions:}}
\\
The $2D$ version of (\ref{eq:energy_thm_3D_Lambda}) is
[Tab.~II, Eq.~(5b)].
This was shown for a homogeneous system in \cite{CombescotC} and in the general case in \cite{TanSimple}
\footnote{This relation was written in \cite{TanSimple} in a form containing a generalised function $\eta(\kk)$ (i.e. a distribution). We have
checked that this form is equivalent to our Eq.~(\ref{eq:energy_thm_2D_heaviside}),
using Eq.~(16b) of \cite{TanSimple},
$n_\sigma(\kk)-(C/k^4)\theta(k-q)=O(1/k^5)$ at large $k$, and
$\int d^2k\, \eta(\kk) f(\kk)=\int d^2k \,f(\kk)$
for any $f(\kk)=O(1/k^3)$.
This last property is implied in Eq.~(16a) in \cite{TanSimple}.
}. 
This can easily be rewritten in the following forms, which 
resemble~[Tab.~II, Eq.~(5a)]:
\begin{multline}
 E - E_{\rm trap}  = -\frac{\hbar^2 C}{2\pi m }  \ln\left(\frac{a q e^\gamma}{2}\right)
 +\sum_{\sigma} \int \frac{d^2\!k}{(2\pi)^2}  \frac{\hbar^2 k^2}{2m}  \\
\times 
\left[n_\sigma(\kk) - \frac{C}{k^4}\theta(k-q)\right]\ \ {\rm for\ any}\ q>0,
\label{eq:energy_thm_2D_heaviside}
\end{multline}
where the Heaviside function $\theta$ ensures that the integral converges at small $k$,
or equivalently
\begin{multline}
 E - E_{\rm trap}  = -\frac{\hbar^2 C}{2\pi m} \ln \left(\frac{a q e^\gamma}{2}\right) 
+\sum_{\sigma} \int \frac{d^2\!k}{(2\pi)^2}  \frac{\hbar^2 k^2}{2m}  \\
\times \left[n_\sigma(\kk) - \frac{C}{k^2(k^2+q^2)}\right]\ \ {\rm for\ any}\ q>0.
\label{eq:energy_thm_2D_yvan}
\end{multline}
To derive this we again use the lattice model. 
We note that, if the limit $q\to0$ is replaced by the limit $b\to0$ taken for fixed $a$, Eq.~(\ref{eq:g0_2D}) remains true (see Appendix~\ref{app:2body}); repeating the reasoning of Section~\ref{sec:energy_thm_latt} then shows that [Tab.~III, Eq.~(3b)] remains true; as in $3D$ we finally get in the limit $b\to0$
\begin{multline}
 E - E_{\rm trap}  = -\frac{\hbar^2 C}{2\pi m} \ln \left(\frac{a q e^\gamma}{2}\right) 
+\sum_{\sigma} \int \frac{d^2\!k}{(2\pi)^2}  \frac{\hbar^2 k^2}{2m}  \\
\times \left[n_\sigma(\kk) - \frac{C}{k^2}\mathcal{P}\frac{1}{k^2-q^2}\right]
\label{eq:energy_thm_2D_PP}
\end{multline}
for any $q>0$;
this is easily rewritten as [Tab.~II, Eq.~(5b)].

\subsection{One-body density matrix at short-distances}
\label{sec:g1}

The one-body density matrix is defined as
$\ds
g_{\sigma \sigma}^{(1)} \left(\rr,
\rr'\right)=\langle \hat{\psi}_\sigma^\dagger
\left(\rr\right)
\hat{\psi}_\sigma\left(\rr'\right) \rangle
$
where $\hat{\psi}_\sigma(\rr)$ annihilates a particle of spin $\sigma$ at point $\rr$.
Its spatially integrated version
\be
G^{(1)}_{\si\si}(\rr)\equiv\int d^dR \,
 g_{\sigma \sigma}^{(1)} \left(\mathbf{R}-\frac{\mathbf{r}}{2},
\mathbf{R}+\frac{\mathbf{r}}{2}\right)
\ee
is a Fourier transform of the momentum distribution:
\be  G^{(1)}_{\si\si}(\rr) = 
\int \frac{d^d k}{(2\pi)^d}\,e^{i\kk\cdot\rr} n_\si(\kk).
\label{eq:G1_vs_nk}
\ee
The expansion of $G^{(1)}_{\si\si}(\rr)$ up to first order in $r$ is given by
[Tab.~II, Eq.~(6a)] in $3D$, as first obtained in~\cite{TanEnergetics},
and by~[Tab.~II, Eq.~(6b)] in $2D$.
The expansion can be pushed to second order if one sums over spin and averages over 
$d$ orthogonal directions of $\rr$, see [Tab.~II, Eqs.~(7a,7b)]
where the ${\bf u_i}$'s are an orthonormal 
basis~\footnote{These last relations  also hold if one averages over all directions of $\rr$ uniformly on the unit sphere or unit circle.}.
Such a second order expansion was first obtained in $1D$ in~\cite{Olshanii_nk}; the following derivations however differ from the $1D$ case~\footnote{Our result does not follow from the well-known fact that, for a finite-range interaction potential in continuous space, $-\frac{\hbar^2}{2m}\sum_\si \Delta G^{(1)}_{\si\si}(\rr=\vn)$ equals the kinetic energy; indeed, the Laplacian does not commute with the zero-range limit in that case [cf.~also the comment below Eq.~(\ref{eq:g1_pour_MC})].}.

\noindent \underline{{\it Three dimensions:}}
\\
To derive [Tab.~II, Eqs.~(6a,7a)] we rewrite (\ref{eq:G1_vs_nk}) as
\begin{multline}
G^{(1)}_{\si\si}(\rr) = N_\si + 
\int \frac{d^3 k}{(2\pi)^3}\,\left(e^{i\kk\cdot\rr} -1\right)\frac{C}{k^4}
\\ + \int \frac{d^3 k}{(2\pi)^3}\, \left(e^{i\kk\cdot\rr} -1\right)
\left(n_\si(\kk)-\frac{C}{k^4}\right).
\end{multline}
The first integral equals $-(C/8\pi) r$. In the second integral, we use
\be
e^{i\kk\cdot\rr}-1\underset{r\to 0}{=}i\kk\cdot\rr-\frac{(\kk\cdot\rr)^2}{2}+o(r^2).
\label{eq:expand_exp}
\ee
The first term of this expansion gives a contribution to the integral  proportional to the total momentum of the gas, which vanishes since the eigenfunctions are real.
The second term is $O(r^2)$, which gives~[Tab.~II, Eq.~(6a)].
Equation~(7a) of Tab.~II follows from the fact that the contribution of the second term, after averaging over the directions of $\rr$, 
is given by the integral of $k^2 [n_\si(\kk)-C/k^4]$, which (after summation over spin)
is related to the total energy by~[Tab.~II, Eq.~(5a)].

\noindent \underline{{\it Two dimensions:}}
\\
To derive~[Tab.~II, Eqs.~(6b,7b)] we rewrite (\ref{eq:G1_vs_nk}) as
$G^{(1)}_{\si\si}(\rr) = N_\si + I(\rr)+J(\rr)$
with
\bea
I(\rr)=
\int \frac{d^2 k}{(2\pi)^2}\,\left(e^{i\kk\cdot\rr} -1\right)\frac{C}{k^4}\theta(k-q)  && \\
 J(\rr)= \int \frac{d^2 k}{(2\pi)^2}\,\left(e^{i\kk\cdot\rr} -1\right)\left(n_\si(\kk)-\frac{C}{k^4}\theta(k-q)\right) &&
\eea
where $q>0$ is arbitrary and the Heaviside function $\theta$ ensures that the integrals converge.

To evaluate $I(\rr)$ we use standard manipulations to 
rewrite it as $I(\rr)=Cr^2/(2\pi)\int_{qr}^{+\infty} dx [J_0(x)-1]/x^3$,
$J_0$ being a Bessel function. Expressing this integral with Mathematica in terms of an
hypergeometric function and a logarithm leads for $r\to 0$ to
$I(\rr)=C r^2/(8\pi)[\gamma-1-\ln 2 +\ln (qr)]+O(r^4)$.
To evaluate $J(\rr)$ we use the same procedure as in $3D$:  
expanding the exponential [see~(\ref{eq:expand_exp})] yields an integral which can be related to the total energy 
thanks to~(\ref{eq:energy_thm_2D_heaviside})
\footnote{As suggested by a referee, [Tab.~II, Eq.~(7b)] can be tested for the dimer wavefunction $\psi(\rr_1,\rr_2)=
\phi_{\rm dim}(r_{12})=-\kappa K_0(\kappa r)/\pi^{1/2}$ \cite{MaximLudo2D}, which has the energy $E=-\hbar^2 \kappa^2/m$ 
and the momentum distribution $n_{\sigma}(\kk)= 4\pi \kappa^2/(k^2+\kappa^2)^2$, 
where $\kappa=2/(ae^\gamma)$ and $K_0$ is a Bessel function. From Eq.~(\ref{eq:G1_vs_nk}) we
find $G_{\sigma\sigma}^{(1)}(\rr)=\kappa r K_1(\kappa r)$. From $C/(4\pi)=-mE/\hbar^2=\kappa^2$ and the known
expansion of $K_1$ around zero, we get the same low-$r$ expansion as in [Tab.~II, Eq.~(7b)]. To calculate $G_{\sigma\sigma}^{(1)}(\rr)$,
we used the fact that $K_0(\kappa r)$ is the $2D$ Fourier transform of $2\pi/(k^2+\kappa^2)$: it remains to take the derivative with respect
to $\kappa$ and to realize that $K_0'=-K_1$.}.

\subsection{Second order derivative of the energy with respect to the scattering length}
We denote by $|\psi_n\ra$ an orthonormal basis of $N$-body stationary states that vary smoothly with $1/a$, and  by $E_n$ the corresponding eigenenergies.
We will derive
[Tab. II, Eqs. (8a,8b)],
where the sum is taken on all values of $n'$ such that $E_{n'}\neq E_n$.
This implies that for the ground state energy $E_0$,
\bea
\frac{d^2E_0}{d(-1/a)^2} &<& 0\ \ \ {\rm in}\ 3D
\label{eq:d^2E0_3D}
\\
\frac{d^2E_0}{d(\ln a)^2} &<& 0\ \ \ {\rm in}\ 2D.
\label{eq:d^2E0_2D}
\eea
Eq.~(\ref{eq:d^2E0_3D}) was intuitively expected~\cite{LeticiaSoutenance}: Eq.~(\ref{eq:Npair_3D}) shows that $dE_0/d(-1/a)$ is proportional to the probability of finding two particles very close to each other, and it is natural that this probability decreases when one goes from the BEC limit ($-1/a\to-\infty$) to the BCS limit ($-1/a\to+\infty$), i.e. when the interactions become less attractive~\footnote{In the lattice model in $3D$, the coupling constant $g_0$ is always negative in the zero-range limit $|a|\gg b$, and is an increasing function of $-1/a$, as seen from (\ref{eq:g0_3D}).}.
Eq.~(\ref{eq:d^2E0_2D})
also agrees with intuition~\footnote{Eq.~(\ref{eq:Npair_2D}) shows that $dE_0/d(\ln a)$ is proportional to the probability of finding two particles very close to each other, and it is natural that this probability decreases when one goes from the BEC limit ($\ln a\to-\infty$) to the BCS limit ($\ln a\to+\infty$), i.e. when the interactions become less attractive [in the lattice model in $2D$, the coupling constant $g_0$ is always negative in the zero-range limit $a\gg b$, and is an increasing function of $\ln a$, as can be seen from (\ref{eq:g0_2D})].}.

For the derivation,  it is convenient to use the lattice model
(defined in Sec.~\ref{sec:models:lattice}): As shown in Sec.\ref{sec:d^2E_reseau} one easily obtains (\ref{eq:d^2E/dg0^2}) and [Tab.~III, Eq.~(6)], from which the result is deduced as follows.
$|\phi(\vn)|^2$ is eliminated using (\ref{eq:phi_tilde_3D},\ref{eq:phi_tilde_2D}). Then,
in $3D$, one uses
\be
\frac{d^2E_n}{d(-1/a)^2}=\frac{d^2E_n}{dg_0^{\phantom{0}2}} \left(\frac{dg_0}{d(-1/a)}\right)^2+\frac{dE_n}{dg_0}\frac{d^2 g_0}{d(-1/a)^2}
\label{eq:deriv_2_fois}
\ee
where the second term equals $2g_0\,dE_n/d(-1/a)\,m/(4\pi\hbar^2)$ and thus vanishes in the zero-range limit.
In $2D$, similarly, one uses the fact that
$d^2E_n/d(\ln a)^2$ is the zero-range limit of $(d^2E_n/dg_0^{\phantom{0}2}) \cdot (dg_0/d(\ln a))^2$.

\subsection{Time derivative of the energy}
\label{subsec:tdote}

We now consider the case where the scattering length $a(t)$ and the trapping potential $U(\rr,t)$ are varied with time. The time-dependent version of the zero-range model (see e.g.~\cite{CRAS}) is given by Schr\"odinger's equation
\be
i\hbar \frac{\partial}{\partial t} \psi(\rr_1,\ldots,\rr_N;t) = 
H(t)\,  \psi(\rr_1,\ldots,\rr_N;t)
\ee
when all particle positions are distinct, with
\be
H(t)=
\sum_{i=1}^N \left[
-\frac{\hbar^2}{2 m}\Delta_{\rr_i} + U(\rr_i,t)
\right],
\ee
and by the contact condition~[Tab. I, Eq. (1a)] in~$3D$ or~[Tab.~I, Eq.~(1b)] in~$2D$ for the scattering length $a=a(t)$.
One then has the relations
[Tab. II, Eqs.~(12a,12b)],
where $E(t)=\langle \psi(t) | H(t)|\psi(t)\rangle$ is the total energy and
$H_{\rm trap}(t)=\sum_{i=1}^N U(\rr_i,t)$ is the trapping potential part of the Hamiltonian.
In $3D$, this relation was first obtained in~\cite{TanLargeMomentum}.
 A very simple derivation of these relations using the lattice model is given in Section \ref{sec:dEdt_reseau}. Here we give a derivation within the zero-range model.

\noindent{\underline{\it Three dimensions:}}
\\
 We first note that the
generalization of the
 lemma (\ref{eq:lemme_3D})  to the case of two Hamiltonians $H_1$ and $H_2$ with corresponding trapping potentials $U_1(\rr)$ and $U_2(\rr)$ reads:
\begin{multline}
\langle \psi_1, H_2 \psi_2 \rangle - \langle H_1 \psi_1, \psi_2 \rangle = 
\frac{4\pi\hbar^2}{m}\left(\frac{1}{a_1}-\frac{1}{a_2}\right) \ ( A^{(1)},A^{(2)}) \\+ 
\langle \psi_1 | \sum_{i=1}^N \left[U_2(\rr_i,t)-U_1(\rr_i,t)\right] |\psi_2 \rangle.
\label{eq:lemme_modif_3D}
\end{multline}
Applying this relation for $|\psi_1\rangle=|\psi(t)\rangle$ and $|\psi_2\rangle=|\psi(t+\delta t)\rangle$ [and correspondingly $a_1=a(t)$,
$a_2=a(t+\delta t)$ and $H_1=H(t)$, $H_2=H(t+\delta t)$] gives:
\begin{multline}
\la \psi(t), H(t+\delta t) \psi(t+\delta t)\ra -
\la H(t)\psi(t),\psi(t+\delta t)\ra = \\ \frac{4\pi\hbar^2}{m}
\left(\frac{1}{a(t)}-\frac{1}{a(t+\delta t)}\right) (A(t),A(t+\delta t)) \\
+\la\psi(t)| \sum_{i=1}^N \left[U(\rr_i,t+\delta t)-U(\rr_i,t)\right]
|\psi(t+\delta t)\ra.
\label{eq:interm_dt}
\end{multline}
 Dividing by $\delta t$, taking the limit $\delta t\to0$,
and using the expression~[Tab. II, Eq. (1a)] of $(A,A)$ in terms of $C$,
the right-hand-side of (\ref{eq:interm_dt})  reduces to the right-hand-side of [Tab. II, Eq.~(12a)].
Using twice Schr\"odinger's equation, one rewrites the left-hand-side of (\ref{eq:interm_dt}) 
as $i\hbar \frac{d}{dt} \la\psi(t)|\psi(t+\delta t)\ra$ and
one Taylor expands this last expression to obtain [Tab. II, Eq.~(12a)].

\noindent{\underline{\it Two dimensions:}}
\\
~[Tab.~II, Eq.~(12b)] is derived similarly from the lemma
\begin{multline}
\langle \psi_1, H_2 \psi_2 \rangle - \langle H_1 \psi_1, \psi_2 \rangle 
=
\frac{2\pi\hbar^2}{m}\ln(a_2/a_1) ( A^{(1)},A^{(2)}) \\ + 
\langle \psi_1 | \sum_{i=1}^N \left[U_2(\rr_i,t)-U_1(\rr_i,t)\right] |\psi_2 \rangle.
\label{eq:lemme_modif_2D}
\end{multline}
\section{Relations for lattice models}\label{sec:latt}
In this Section, 
it will prove convenient to introduce an {\it operator} $\hat{C}$ by
[Tab. III, Eqs.~(1a,1b)] and to {\it define} $C$ by its expectation value 
in the state of the system, 
\be
C = \langle \hat{C}\rangle
\label{eq:defCreseau}
\ee
In the zero-range limit, this new definition of $C$ coincides with the definition [Tab. II, Eq. (1)]
of Section~\ref{sec:ZR}, as shown in Appendix~\ref{app:C_b}.

\begin{table*}[t!]
\begin{tabular}{|cc|cc|}
\hline   
Three dimensions & & Two dimensions &  \\
\hline
\vspace{-4mm}
& && \\
$\ds \hat{C}\equiv\frac{4\pi m}{\hbar^2}\frac{dH}{d(-1/a)}$
&(1a)& $\ds \hat{C}\equiv\frac{2\pi m}{\hbar^2}\frac{dH}{d(\ln a)}$
&(1b)
\vspace{-4mm}
\\
& & & \\
\hline
  \multicolumn{4}{|c|}{} \vspace{-4mm}\\
 \multicolumn{3}{|c}{\vspace{-4mm} $\ds H_{\rm int}=\frac{\hbar^4}{m^2} \frac{\hat{C}}{g_0}$} 
 &(2)
\\
  \multicolumn{4}{|c|}{} \\
\hline
& & & \vspace{-4mm}\\
$\ds H - H_{\rm trap}  = \frac{\hbar^2 \hat{C}}{4\pi m a}$
&
&
$\ds H - H_{\rm trap}  = \lim_{q\to0}\Bigg\{-\frac{\hbar^2 \hat{C}}{2\pi m} \ln \left(\frac{a q e^\gamma}{2}\right)$
&
\\
$\ds +\sum_{\sigma} \int_D \frac{d^3\!k}{(2\pi)^3}  \epsk
\left[\hat{n}_\sigma(\kk) - \hat{C}\left(\frac{\hbar^2}{2m\epsk}\right)^2\right]$ &
(3a)&
$\ds +\sum_{\sigma} \int_D \frac{d^2\!k}{(2\pi)^2}  \epsk
\left[\hat{n}_\sigma(\kk) - \hat{C}\frac{\hbar^2}{2m\epsk}\mathcal{P}\frac{\hbar^2}{2m(\epsk-\epsq)}\right]\Bigg\}$
&
(3b) \vspace{-4mm}
 \\
& & &\\
\hline
& & & \vspace{-4mm}\\
$\ds C=(4\pi)^2\ (A,A)$
&(4a)&
 $\ds C=(2\pi)^2\ (A,A)$
&(4b) \vspace{-4mm}
\\
& && \\
\hline
& && \vspace{-4mm} \\
$\ds \frac{dE}{d(-1/a)}=\frac{\hbar^2 C}{4\pi m}$
&(5a)&
$\ds \frac{dE}{d(\ln a)}=\frac{\hbar^2 C}{2\pi m}$
&(5b) \vspace{-4mm}
\\
& & &\\
\hline
 \multicolumn{4}{|c|}{\vspace{-4mm}} \\
 \multicolumn{3}{|c}{$\ds\frac{1}{2} \frac{d^2E_n}{dg_0^2}
= |\phi(\vn)|^4 \sum_{n',E_{n'}\neq E_n} 
\frac{|(A^{(n')},A^{(n)})|^2}{E_n-E_{n'}}$}
&(6) \vspace{-4mm}
\\
  \multicolumn{4}{|c|}{} \\
\hline
 \multicolumn{4}{|c|}{\vspace{-4mm}} \\
 \multicolumn{3}{|c}{$\ds \left(\frac{d^2F}{dg_0^2}\right)_T <0$,\ \ \ 
$\ds \left(\frac{d^2E}{dg_0^2}\right)_S <0$}
&(7) \vspace{-4mm}
\\
  \multicolumn{4}{|c|}{} \\
\hline
&& & \vspace{-4mm}\\
$\ds\sum_\RR b^3 (\psi^\dagger_\up\psi^\dagger_\down\psi_\down\psi_\up)(\RR)=\frac{\hat{C}}{(4\pi)^2}|\phi(\vn)|^2$
&(8a)&
 $\ds\sum_\RR b^2 (\psi^\dagger_\up\psi^\dagger_\down\psi_\down\psi_\up)(\RR)=\frac{\hat{C}}{(2\pi)^2}|\phi(\vn)|^2$
&(8b) \vspace{-4mm}
\\
& & &\\
\hline
  \multicolumn{4}{|c|}{In the zero-range regime $\ktyp b\ll1$}
\\ \hline
& & & \vspace{-4mm}\\
  $\ds\sum_{\RR} b^3
 g_{\uparrow \downarrow}^{(2)} \left(\mathbf{R}+\frac{\mathbf{r}}{2},
\mathbf{R}-\frac{\mathbf{r}}{2}\right)
\simeq
\frac{C}{(4\pi)^2}|\phi(\rr)|^2$,\ for $r\ll \ktyp^{-1}\ \ $
&(9a)&
 $\ds\sum_{\RR} b^2
 g_{\uparrow \downarrow}^{(2)} \left(\mathbf{R}+\frac{\mathbf{r}}{2},
\mathbf{R}-\frac{\mathbf{r}}{2}\right)
\simeq
\frac{C}{(2\pi)^2}|\phi(\rr)|^2$,\ for $r\ll \ktyp^{-1}\ \ $
&(9b) \vspace{-4mm}
\\
& &&
\\ \hline
 \multicolumn{4}{|c|}{\vspace{-4mm}} \\
\multicolumn{3}{|c}{$\ds n_\sigma(\kk)\simeq C \left(\frac{\hbar^2}{2m\epsk}\right)^2,\ $ for $k\gg\ktyp$}
&(10) \vspace{-4mm}
\\
\multicolumn{4}{|c|}{}
\\ 
\hline
\end{tabular}
\caption{Relations for spin-1/2 fermions for lattice models. $\hat{C}$ is defined in line 1 and
$C=\langle\hat{C}\rangle$.
Lines 2, 3 and 8 are relations between operators.
Line 4 holds for any pure state [the regular part $A$ being defined in Eq.~(\ref{eq:def_A_reseau}) in the text].
Lines 5-6 hold for any stationary state.
Line 7 holds at thermal equilibrium in the canonical ensemble. 
Lines 9-10 are expected to hold in the zero-range regime $\ktyp b \ll 1$, where $\ktyp$ is the typical wavevector, for any stationary state or at thermal equilibrium.
\label{tab:latt}}
\end{table*}

\subsection{Interaction energy and $\hat{C}$}
The interaction part $H_{\rm int}$ of the lattice model's Hamiltonian is obviously equal to $\ds g_0\frac{dH}{dg_0}$ 
[see Eqs. (\ref{eq:Hlatt},\ref{eq:def_H0},\ref{eq:def_W})].
Rewriting this as $\ds \frac{1}{g_0}\,\frac{dH}{d(-1/g_0)}$,
and using the simple expressions (\ref{eq:dg0_da_3D},\ref{eq:dg0_da_2D}) for $d(1/g_0)$, we get the relation [Tab. III, Eq. (2)] between $H_{\rm int}$ and $\hat{C}$, both in $3D$ and in $2D$.

\subsection{Total energy minus trapping potential energy in terms of momentum distribution and $\hat{C}$}\label{sec:energy_thm_latt}
Here we derive [Tab. III, Eqs. (3a,3b)].
We start from the expression [Tab. III, Eq. (2)] of the interaction energy and eliminate $1/g_0$ thanks to (\ref{eq:g0_3D},\ref{eq:g0_2D}).
The desired expression of $H-H_{\rm trap}=H_{\rm int}+H_{\rm kin}$ then simply follows from the expression (\ref{eq:Hkin_second_quant}) of the kinetic energy.

\subsection{Interaction energy and regular part}
In the forthcoming subsections \ref{C_vs_AA_latt}, \ref{sec:dE_latt} and \ref{sec:d^2E_reseau},
we will use the following lemma: 
For any wavefunctions $\psi$ and $\psi'$,
\be
\la\psi'|H_{\rm int}|\psi\ra =g_0 |\phi({\bf 0})|^2\ ( A',A)
\label{eq:lemme_W}
\ee
where $A$ and $A'$ are the regular parts related to $\psi$ and $\psi'$ through (\ref{eq:def_A_reseau}), and the scalar product between regular parts is naturally defined as the discrete version of~[Tab. I, Eq. (2)]:
\begin{multline}
( A',A )\equiv \sum_{i<j} 
\sum_{(\rr_k)_{k\neq i,j}} \sum_{\RR_{ij}} b^{(N-1)d}
A'^*_{ij}(\mathbf{R}_{ij}, (\mathbf{r}_k)_{k\neq i,j}) \\
\times A_{ij}(\mathbf{R}_{ij}, (\mathbf{r}_k)_{k\neq i,j}).
\end{multline}
The lemma simply follows from
\begin{multline}
\la\psi'|H_{\rm int}|\psi\ra=g_0\sum_{i<j} 
\sum_{(\rr_k)_{k\neq i,j}} b^{(N-2)d} \sum_{\rr_j} b^d\\
\times 
(\psi'^*\psi)(\rr_1,\ldots,\rr_i=\rr_j,\ldots,\rr_j,\ldots,\rr_N).
\label{eq:note_W}
\end{multline}

\subsection{Relation between $\hat{C}$ and $(A,A)$}\label{C_vs_AA_latt}
Lemma (\ref{eq:lemme_W}) with $\psi'=\psi$ writes
\be
\langle \psi | H_{\rm int} | \psi \rangle
= g_0 |\phi({\bf 0})|^2\ ( A,A).
\label{eq:W_AA}
\ee
Expressing $\langle \psi | H_{\rm int} | \psi \rangle$ in terms of $C= \la\psi| \hat{C}|\psi \ra$ thanks to [Tab. III, Eq. (2)],
and using
the expressions (\ref{eq:phi_tilde_3D},\ref{eq:phi_tilde_2D}) of $|\phi(\vn)|^2$, we get [Tab. III, Eqs. (4a,4b)].

\subsection{First order derivative of an eigenenergy with respect to the coupling constant}
\label{sec:dE_latt}
For any stationary state,
the Hellmann-Feynman theorem, together with
the definition [Tab. III, Eqs. (1a,1b)] of $\hat{C}$
and the relation [Tab. III, Eqs. (4a,4b)] between $C$ and $(A,A)$, immediately yields [Tab.~III, Eqs.~(5a,5b)].

\subsection{Second order derivative of an eigenenergy with respect to the coupling constant}\label{sec:d^2E_reseau}
We denote by $|\psi_n\ra$ an orthonormal basis of $N$-body stationary states which vary smoothly with $g_0$, and  by $E_n$ the corresponding eigenenergies.
We apply second order perturbation theory to determine how an eigenenergy varies for an infinitesimal change of $g_0$. This gives:
\be
\frac{1}{2}\frac{d^2 E_n}{dg_0^{\phantom{0}2}}=\sum_{n', E_{n'}\neq E_n} \frac{\left|\la\psi_{n'}|H_{\rm int}/g_0|\psi_n\ra\right|^2}{E_n-E_{n'}},
\label{eq:d^2E/dg0^2}
\ee
where the sum is taken over all values of $n'$ such that $E_{n'}\neq E_n$.
Lemma (\ref{eq:lemme_W}) then yields~[Tab.~III, Eq.~(6)].

\subsection{Time derivative of the energy}\label{sec:dEdt_reseau}

The relations [Tab. II, Eqs. (12a,12b)] remain exact for the lattice model. Indeed, $dE/dt$ equals $\la dH/dt\ra$ from
the Hellmann-Feynman theorem.
In $3D$, we can rewrite this quantity as $\la dH_{\rm trap}/dt\ra + d(-1/a)/dt\,\la dH/d(-1/a)\ra$, and the desired result follows from the definition [Tab. III, Eq. (1a)] of $\hat{C}$. The derivation of the $2D$ relation [Tab. II, Eq. (12b)] is analogous.

\subsection{On-site pair distribution operator}
Let us define a spatially integrated pair distribution operator
\be
\hat{G}^{(2)}_{\uparrow\downarrow}(\rr)\equiv
\sum_{\RR} b^d
(\psi^\dagger_\up{\psi_\up})\left(\mathbf{R}+\frac{\mathbf{r}}{2}\right)
(\psi_\down^{\dagger}{\psi_\down})\left(\mathbf{R}-\frac{\mathbf{r}}{2}\right).
\ee
Using the relation [Tab.~III, Eq.~(2)] between $\hat{C}$ and $H_{\rm int}$, expressing $H_{\rm int}$ in terms of $\hat{G}^{(2)}_{\uparrow\downarrow}(\mathbf{0})$ thanks to the second-quantized form (\ref{eq:defW_2e_quant}), and expressing $g_0$ in terms of $\phi(\vn)$ thanks to~(\ref{eq:phi0_vs_g0},\ref{eq:phi0_vs_g0_2D}), we immediately get:
\bea
\hat{G}^{(2)}_{\uparrow\downarrow}(\vn)
&=&\frac{\hat{C}}{(4\pi)^2}|\phi(\vn)|^2
\label{eq:g2_latt}
\ \ \ {\rm in}\ 3D
\\
\hat{G}^{(2)}_{\uparrow\downarrow}(\vn)
&=&\frac{\hat{C}}{(2\pi)^2}|\phi(\vn)|^2
\ \ {\rm in}\ 2D.
\eea
[Here, $|\phi(\vn)|^2$ may of course be eliminated using~(\ref{eq:phi0_vs_g0},\ref{eq:phi0_vs_g0_2D}).]
These relations are analogous to the one obtained previously within a different field-theoretical model, see Eq.~(12) in~\cite{Braaten}.

\subsection{Pair distribution function at short distances}
\label{subsec:G2_short_dist_latt}
The last result can be generalized to finite but small $r$,
see [Tab.~III, Eqs.~(9a,9b)]
where the zero-range regime $\ktyp b\ll1$ was introduced at the end of Sec.~\ref{sec:models:lattice}.
Here we justify this for the case where the expectation values
$g^{(2)}_{\uparrow\downarrow}\left(\mathbf{R}+\frac{\mathbf{r}}{2},\mathbf{R}-\frac{\mathbf{r}}{2}\right)=\la(\psi^\dagger_\up\psi_\up)\left(\mathbf{R}+\frac{\mathbf{r}}{2}\right)
(\psi^{\dagger}_\down\psi_{\down})\left(\mathbf{R}-\frac{\mathbf{r}}{2}\right)\ra$ and $C=\la\hat{C}\ra$ are taken in an arbitrary stationary state $\psi$ in the zero-range regime;
this implies that the same result holds for a thermal equilibrium state in the zero-range regime, see Section~\ref{subsec:finiteT}.
We first note that the expression (\ref{eq:def_g2_psi}) of $g^{(2)}_{\uparrow\downarrow}$ 
in terms of the wavefunction is valid for the lattice model with the obvious replacement of the integrals by sums, so that
\begin{multline}
G^{(2)}_{\uparrow\downarrow}(\rr)\equiv\left<\hat{G}^{(2)}_{\uparrow\downarrow}(\rr)\right>=\sum_\RR b^d \sum_{i=1}^{N_\up}\sum_{j=N_\up+1}^N
\sum_{(\rr_k)_{k\neq i,j}} \!\!\! b^{(N-2)d} \\
\times \left| \psi\left(\rr_1,\ldots,\rr_i=\RR+\frac{\rr}{2},\ldots,\rr_j=\RR-\frac{\rr}{2},\ldots,\rr_N\right) \right|^2.
\end{multline}
For $r\ll1/\ktyp$, we can replace $\psi$ by the short-distance expression (\ref{eq:psi_courte_dist}),
assuming that the multiple sum is dominated by the configurations where all the distances $|\rr_k-\RR|$ and $r_{k k'}$ 
are much larger than $b$ and $r$:
\be
G^{(2)}_{\uparrow\downarrow}(\rr)\simeq (A,A)\ |\phi(\rr)|^2.
\label{eq:G2_AA}
\ee
Expressing $(A,A)$ in terms of $C$ thanks to [Tab.~III, Eqs.~(4a,4b)] gives the desired~[Tab.~III, Eqs.~(9a,9b)].

\subsection{Momentum distribution at large momenta}\label{subsec:nk_latt}

Assuming again that we are in the zero-range regime $\ktyp b\ll1$, we will justify
[Tab.~III, Eq.~(10)] both in $3D$ and in $2D$. We start from
\be
n_\sigma(\kk)=\sum_{i:\sigma} \sum_{(\rr_l)_{l\neq i}} b^{d(N-1)}
\left|\sum_{\rr_i} b^d e^{-i\kk\cdot\rr_i}\psi(\rr_1,\dots,\rr_N)
\right|^2.
\label{eq:nk_psi_latt}
\ee
We are interested in the limit $k\gg\ktyp$.
Since $\psi(\rr_1,\ldots,\rr_N)$ is a function of $\rr_i$ which varies on the scale of $1/\ktyp$, except when $\rr_i$ is close to another particle $\rr_j$ where it varies on the scale of $b$,
we can replace $\psi$ by its short-distance form (\ref{eq:psi_courte_dist}):
\begin{multline}
\sum_{\rr_i} b^d e^{-i\kk\cdot\rr_i}\psi(\rr_1,\ldots,\rr_N)\simeq
\tilde{\phi}(\kk) \\ \times \sum_{j,j\neq i} e^{-i\kk\cdot\rr_j}A_{ij}(\rr_j,(\rr_l)_{l\neq i,j}),
\label{eq:TF_approx}
\end{multline}
where $\tilde{\phi}(\kk)=\la\kk|\phi\ra=\sum_\rr b^d e^{-i\kk\cdot\rr}\phi(\rr)$.
Here we excluded the configurations where more than two particles are at distances $\lesssim b$, which are expected to have a negligible contribution to (\ref{eq:nk_psi_latt}).
Inserting (\ref{eq:TF_approx}) into (\ref{eq:nk_psi_latt}),
expanding the modulus squared, and neglecting the cross-product terms in the limit $k\gg\ktyp$, we obtain
\be
n_\sigma(\kk)\simeq|\tilde{\phi}(\kk)|^2 (A,A).
\label{eq:nk_AA_latt}
\ee
Finally, $\tilde{\phi}(\kk)$ is easily computed for the lattice model: for $k\neq0$, the two-body Schr\"odinger equation (\ref{eq:schro_2corps}) directly gives
$\tilde{\phi}(\kk)=-g_0\phi(\vn)/(2\epsk)$, and $\phi(\vn)$ is given by (\ref{eq:phi0_vs_g0},\ref{eq:phi0_vs_g0_2D}), which yields [Tab.~III, Eq.~(10)].

\subsection{Minorization of $C$ by the order parameter}

\noindent{\sl (This subsection is supplementary to the published paper)}

None of the previous relations involve the macroscopic quantum properties of the spin-1/2 Fermi 
gas, such as superfluidity and off-diagonal long range order. For an arbitrary state $|\psi\rangle$ in which the gas is pair-condensed, with a nonzero {\sl order parameter} $\Delta(\rr)$ of arbitrary position dependence, 
one obtains an additional relation, in the form of the following minorization:
\[
C\geq \frac{m^2}{\hbar^4} \sum_{\rr} b^d |\Delta(\rr)|^2
\]
where $d=2$ or $3$ is the dimension of space. 

This inequality is straightforwardly obtained in a $U(1)$ symmetry breaking point of view, where the order parameter is related to 
the pairing field in the lattice model in 3D \cite{YvanVarenna} and in 2D [G. Tonini, F. Werner, Y. Castin, Eur. Phys. J. D {\bf 39}, 283 (2006)] by
\[
\Delta(\rr) \equiv g_0\langle \psi_\downarrow(\rr) \psi_\uparrow(\rr)\rangle.
\]
We then split the operator $\hat{O}=(\psi_\downarrow\psi_\uparrow)(\rr)$ as the sum of its expectation value $\langle \hat{O}\rangle$
and of fluctuations $\delta\hat{O}$. From the identity $\langle \hat{O}^\dagger \hat{O}\rangle = |\langle \hat{O}\rangle|^2+
\langle (\delta\hat{O})^\dagger (\delta \hat{O})\rangle$ and the nonnegativeness of the last term in that identity, we obtain
\[
\langle (\psi_\uparrow^\dagger\psi_\downarrow^\dagger\psi_\downarrow\psi_\uparrow)(\rr)\rangle \geq |\langle (\psi_\downarrow\psi_\uparrow)(\rr)\rangle|^2 = \frac{|\Delta(\rr)|^2}{g_0^2}.
\]
It remains to sum this inequality over $\rr$ and to use [Tab. III, Eq.~(2)] and the expression (\ref{eq:defW_2e_quant}) of $H_{\rm int}$ to obtain
the announced minorization.

The generalisation to the $U(1)$-symmetry preserving case is straightforward. When the gas is pair-condensed, the two-body density operator $\hat{\rho}_2$, defined by
\[
\langle \rr_1,\rr_2 | \hat{\rho}_2 |\rr_1',\rr_2'\rangle \equiv \langle \psi_\uparrow^\dagger(\rr_1') \psi_\downarrow^\dagger(\rr_2')
\psi_\downarrow(\rr_2)\psi_\uparrow(\rr_1)\rangle,
\]
has a normalised eigenvector $|\varphi_0\rangle$ with an eigenvalue $\bar{N}_0$ of order $N/2$, and this is the only macroscopically
populated two-particle mode. $\bar{N}_0$ is the mean number of condensed pairs and $\langle \rr_1,\rr_2|\varphi_0\rangle=\varphi_0(\rr_1,\rr_2)$ is the corresponding
pair condensate wavefunction. 
In this framework, the pairing field $\langle \psi_\downarrow(\rr_2)\psi_\uparrow(\rr_1)\rangle$ is replaced by the pair-condensed
field $\bar{N}_0^{1/2} \varphi_0(\rr_1,\rr_2)$ so that the order parameter is replaced by
\[
\Delta(\rr)=g_0 \bar{N}_0^{1/2} \varphi_0(\rr,\rr)
\]
We then introduce the splitting 
\[
\hat{\rho}_2= \bar{N}_0 |\varphi_0\rangle \langle\varphi_0| + \delta \hat{\rho}_2
\]
where both $\hat{\rho}_2$ and $\delta \hat{\rho}_2$ are hermitian nonnegative, hence the chain 
\[
\langle (\psi_\uparrow^\dagger\psi_\downarrow^\dagger\psi_\downarrow\psi_\uparrow)(\rr)\rangle = \langle \rr,\rr|\hat{\rho}_2|\rr,\rr\rangle
\geq \bar{N}_0 |\varphi_0(\rr,\rr)|^2= \frac{|\Delta(\rr)|^2}{g_0^2}
\]
The summation over $\rr$ as in the symmetry-breaking case leads to the announced minorization.

Our minorization extends to the continuous-space limit $b\to 0$ where, in particular,
our definition of the order parameter in 3D reconnects to the one (28) of reference 
\cite{RevueTrentoFermions}, as can be shown from the normalisation condition 
(\ref{eq:normalisation_phi_tilde_3D}) of $\phi(\rr)$ and from its value (\ref{eq:phi0_vs_g0}) 
at $\rr=\mathbf{0}$.

\section{Relations for a finite-range interaction in continuous space}
\label{sec:V(r)}

In this Section~\ref{sec:V(r)}, we restrict for simplicity to the case of a stationary state. It is then convenient to define $C$ by~[Tab. IV, Eqs.~(1a,1b)].

\begin{table*}
\begin{tabular}{|cc|cc|}
\hline
Three dimensions & & Two dimensions & \\
\hline
& & &  \vspace{-4mm} \\
$\ds C\equiv\frac{4\pi m}{\hbar^2}\frac{dE}{d(-1/a)}$
& (1a)
& $\ds C\equiv\frac{2\pi m}{\hbar^2}\frac{dE}{d(\ln a)}$
& (1b) \vspace{-4mm}
\\
& & & \\
\hline
& & & \vspace{-4mm} \\
$\ds E_{\rm int}=\frac{C}{(4\pi)^2}\int d^3r\,V(r) |\phi(r)|^2$
& (2a)
& $\ds E_{\rm int}=\frac{C}{(2\pi)^2}\int d^2r\,V(r) |\phi(r)|^2$
& (2b) \vspace{-4mm}
\\
& & & \\
\hline
& & & \vspace{-4mm} \\
 $\ds E-E_{\rm trap}=\frac{\hbar^2 C}{4\pi m a}$
 & 
& $\ds E-E_{\rm trap}=\lim_{R\to\infty}\Bigg\{\frac{\hbar^2 C}{2\pi m}\ln\left(\frac{R}{a}\right)$
& 
\\
 $\ds +\sum_{\si}\int \frac{d^3 k}{(2\pi)^3}\,\frac{\hbar^2 k^2}{2m}\left[n_\si(\kk)-\frac{C}{(4\pi)^2}|\tilde{\phi}'(k)|^2\right]$
 & (3a)
& $\ds  +\sum_{\si}\int \frac{d^2 k}{(2\pi)^2}\,\frac{\hbar^2 k^2}{2m}\left[n_\si(\kk)-\frac{C}{(2\pi)^2}|\tilde{\phi}'_R(k)|^2\right]\Bigg\}$
& (3b) \vspace{-4mm}
\\
& & & \\
\hline
  \multicolumn{4}{|c|}{In the zero-range regime $\ktyp b\ll1$}
\\ \hline
&&& \vspace{-4mm} \\
  $\ds\int d^3R\,
 g_{\uparrow \downarrow}^{(2)} \left(\mathbf{R}+\frac{\mathbf{r}}{2},
\mathbf{R}-\frac{\mathbf{r}}{2}\right)
\simeq
\frac{C}{(4\pi)^2}|\phi(\rr)|^2$\ \ \ for $r\ll \ktyp^{-1}$
& (4a)
& $\ds\int d^2R\,
 g_{\uparrow \downarrow}^{(2)} \left(\mathbf{R}+\frac{\mathbf{r}}{2},
\mathbf{R}-\frac{\mathbf{r}}{2}\right)
\simeq
\frac{C}{(2\pi)^2}|\phi(\rr)|^2$\ \ \ for $r\ll \ktyp^{-1}$
& (4b) \vspace{-4mm}
\\
& &&
\\ \hline
&  && \vspace{-4mm}\\
$\ds n_\si(\kk)\simeq\frac{C}{(4\pi)^2}|\tilde{\phi}(\kk)|^2$\ \ \ for $k\gg\ktyp$
&
(5a) &
$\ds n_\si(\kk)\simeq\frac{C}{(2\pi)^2}|\tilde{\phi}(\kk)|^2$\ \ \ for $k\gg\ktyp$
& (5b) \vspace{-4mm}
\\
& &&
\\ 
\hline
\end{tabular}
\caption{Relations for spin-1/2 fermions with a finite-range interaction potential $V(r)$ in continuous space, for any stationary state. $C$ is defined in line 1. All relations 
remain valid at thermal equilibrium in the canonical ensemble; the derivatives of the energy in line~1 then have to be taken at constant entropy.
Equations~(1a,2a,4a) are contained in~\cite{ZhangLeggettUniv} (for $\ktyp b\ll1$).
The functions $\phi'(r)$ and $\phi'_R(r)$ are given by Eqs.~(\ref{eq:defphip3d},\ref{eq:defphip2d}) 
and $\tilde{\phi}'(k)$, $\tilde{\phi}'_R(k)$ are their Fourier transforms.
\label{tab:V(r)}}
\end{table*}

\subsection{Interaction energy}
As for the lattice model, we find that the interaction energy is proportional to $C$,
see [Tab.~IV, Eqs.~(2a,2b)].
It was shown in~\cite{ZhangLeggettUniv} that the $3D$ relation is asymptotically 
valid in the zero-range limit.
Here we show that it remains exact for any finite value of the range and we generalize it to $2D$.

For the derivation, we set
\be
V(r)=g_0 W(r)
\ee
where $g_0$ is a dimensionless coupling constant which allows to tune $a$.
The Hellmann-Feynman theorem then gives $E_{\rm int}=g_0 dE/dg_0$. The result then follows by writing $dE/dg_0=dE/d(-1/a)\cdot d(-1/a)/dg_0$ in $3D$ and
$dE/dg_0=dE/d(\ln a)\cdot d(\ln a)/dg_0$ in $2D$,
and by using the definition~[Tab. IV, Eqs.~(1a,1b)] of $C$
as well as the following lemmas:
\bea
g_0\frac{d(-1/a)}{d g_0}&=& \frac{m}{4\pi\hbar^2}\int d^3 r\, V(r) |\phi(r)|^2
\ \ \ {\rm in}\ 3D
\label{eq:lemme_g0_vs_a_3D}
\\
g_0\frac{d(\ln a)}{d g_0}&=& \frac{m}{2\pi\hbar^2}\int d^2 r\, V(r) |\phi(r)|^2
\ \ \ {\rm in}\ 2D.
\label{eq:lemme_g0_vs_a_2D}
\eea
To derive these lemmas, we
consider two values of the scattering length $a_i,\ i=1,2$, and the corresponding scattering states $\phi_i$ and coupling constants $g_{0,i}$. The corresponding two-particle relative-motion Hamiltonians are $H_i=-(\hbar^2/m)\,\Delta_\rr + g_{0,i} W(r)$. Since $H_i \phi_i=0$, we have
\be
\lim_{R\to\infty} \int_{r<R} d^d r \left( \phi_1 H_2 \phi_2
- \phi_2 H_1 \phi_1 \right) = 0.
\ee
The contribution of the kinetic energies can be computed from the divergence theorem and the large-distance form of $\phi$~\footnote{We assume, to facilitate the derivation, that $V(r)=0$ for $r>b$, but the result is expected to hold for any $V(r)$ which vanishes quickly enough at infinity.}.
\setcounter{fnnumberter}{\thefootnote}
The contribution of the potential energies is proportional to $g_{0,2}-g_{0,1}$. Taking the limit $a_2\to a_1$ gives the results (\ref{eq:lemme_g0_vs_a_3D},\ref{eq:lemme_g0_vs_a_2D}).
Lemma (\ref{eq:lemme_g0_vs_a_3D}) was also used in~\cite{ZhangLeggettUniv} and the above derivation is essentially identical to the one of~\cite{ZhangLeggettUniv}.
For this $3D$ lemma, there also exists an alternative derivation based on  the two-body problem in a large box~\footnote{We consider two particles of opposite spin in a cubic box of side $L$ with periodic boundary conditions, and we work in the limit where $L$ is much larger than $|a|$ and $b$. In this limit, there exists a ``weakly interacting'' stationary state $\psi$ whose energy is given by the ``mean-field'' shift 
$E=g/L^3$ with $g=4\pi\hbar^2 a/m$. The Hellmann-Feynman theorem gives $g_0\,dE/dg_0=E_{\rm int}[\psi]$.
But the wavefunction $\psi(\rr_1,\rr_2)\simeq\Phi(r_{12})/L^3$ where $\Phi$ is the zero-energy scattering state normalized by $\Phi\to1$ at infinity. Thus $E_{\rm int}=\int d^3 r\,V(r) |\Phi(r)|^2/L^3$. The desired Eq.~(\ref{eq:lemme_g0_vs_a_3D}) then follows, since $\Phi=-a \phi$.}.

\subsection{Relation between energy and momentum distribution}
\noindent
\underline{\it Three dimensions:}
The natural counterpart, for a finite-range interaction potential, of the zero-range-model expression of the energy as a functional of the momentum distribution [Tab.~II, Eqs.~(5a)] is given
by [Tab.~IV, Eq.~(3a)],
where $\tilde{\phi}'(k)$  is the zero-energy scattering state in momentum space with 
the incident wave contribution $\propto \delta(\kk)$ subtracted out: 
$\tilde{\phi}'(k)=\tilde{\phi}(k)+a^{-1}(2\pi)^3\delta(\kk)=\int d^3r\,e^{-i\kk\cdot\rr}\phi'(r)$ with
\be
\phi'(r)=\phi(r)+\frac{1}{a}.
\label{eq:defphip3d}
\ee
This is simply obtained by adding the kinetic energy to [Tab.~IV, Eq.~(2a)]
and by using the lemma:
\be
\int d^3 r\, V(r) |\phi(r)|^2 = \frac{4\pi\hbar^2}{m a}-\int \frac{d^3 k}{(2\pi)^3}\frac{\hbar^2k^2}{m}|\tilde{\phi}'(k)|^2.
\label{eq:lemme_phi'(k)_3D}
\ee
To derive this lemma, we start from Schr\"odinger's equation $-(\hbar^2/m)\Delta\phi+V(r)\phi=0$, which implies
\be
\int d^3 r\, V(r) |\phi(r)|^2=\frac{\hbar^2}{m}\int d^3 r \, \phi\Delta\phi.
\label{eq:phiDeltaphi}
\ee
Applying the divergence theorem 
 over the sphere of radius $R$, using the asymptotic expression (\ref{eq:normalisation_phi_tilde_3D}) of $\phi$
 and taking the limit $R\to\infty$ then yields
\be
\int d^3 r \, \phi\Delta\phi = \frac{4\pi}{a}-\int d^3 r\, (\mathbf{\nabla} \phi)^2.
\ee
We then replace $\nabla \phi$ by $\nabla\phi'$. Applying the Parseval-Plancherel relation to $\partial_i \phi'$,
and using the fact that $\phi'(r)$ vanishes at infinity, we get:
\be
\int d^3 r\,(\nabla\phi')^2 = \int \frac{d^3 k}{(2\pi)^3}\,k^2 |\tilde{\phi}'(k)|^2
\ee
The desired result (\ref{eq:lemme_phi'(k)_3D}) follows.

\noindent
\underline{\it Two dimensions:}
An additional regularisation procedure for small momenta is required in $2D$, as was the case for the zero-range
model~[Tab. II, Eq.~(5b)] and for the lattice model~[Tab. III, Eq.~(3b)]. One obtains
[Tab.~IV, Eq.~(3b)],
where $\tilde{\phi}_R'(k)=\int d^{2}r\,e^{-i\kk\cdot\rr}\phi'_R(r)$ with
\be
\phi_R'(r)=\left[\phi(r)-\ln(R/a)\right]\,\theta(R-r).
\label{eq:defphip2d}
\ee
This follows from [Tab.~IV, Eq.~(2b)] and from the lemma:
\begin{multline}
\int d^2r\,V(r)|\phi(r)|^2=\lim_{R\to\infty}\left\{\frac{2\pi\hbar^2}{m}\ln\left(\frac{R}{a}\right)
\right. \\ \left. -\int\frac{d^2k}{(2\pi)^2}\,\frac{\hbar^2k^2}{m}|\tilde{\phi}'_R(k)|^2\right\}.
\label{eq:lemme_phi'(k)_2D}
\end{multline}
The derivation of this lemma again starts with the 2D version of (\ref{eq:phiDeltaphi}). The divergence theorem then 
gives~[\thefnnumberter]
\be
\int d^2r\,\phi\Delta\phi=\lim_{R\to\infty}\left\{2\pi\ln\left(\frac{R}{a}\right)-\int_{r<R} d^2r\,(\mathbf{\nabla}\phi)^2\right\}.
\ee
We can then replace $\int_{r<R}d^2r\,(\nabla \phi)^2$ by $\int d^2r\,(\nabla\phi'_R)^2$, since $\phi'_R(r)$ is continuous at $r=R$~[\thefnnumberter] so that $\nabla\phi'_R$ does not contain any delta distribution. The Parseval-Plancherel relation can be applied to $\partial_i \phi'_R$, since this function is square-integrable. Then, using the fact that $\phi'_R(r)$ vanishes at infinity, we get
\be
\int d^2r\,(\nabla\phi'_R)^2 = \int \frac{d^2k}{(2\pi)^2}\,k^2|\tilde{\phi}'_R(k)|^2,
\ee
and the lemma (\ref{eq:lemme_phi'(k)_2D}) follows.

\subsection{Pair distribution function at short distances}
\label{subsec:g2_V(r)}
In the zero-range regime $k_{\rm typ} b \ll 1$, the short-distance behavior of the pair distribution function is given by the same expressions [Tab.~III, Eqs.~(9a,9b)] as for the lattice model.
Indeed, Eq.~(\ref{eq:G2_AA}) is derived in the same way as for the lattice model;
one can then use the
zero-range model's expressions~[Tab.~II, Eqs.~(2a,2b)]
of $(A,A)$ in terms of $C$, since the finite range model's quantities $C$ and $A$ tend to the zero-range model's ones in the zero-range limit.
In $3D$, the result [Tab.~III, Eq.~(9a)] is contained in~\cite{ZhangLeggettUniv}.

\subsection{Momentum distribution at large momenta}\label{subsec:nk_V(r)}
In the zero-range regime $\ktyp b\ll1$ the momentum distribution at large momenta $k\gg\ktyp$ is given by
\bea
n_\sigma(\kk)&\simeq&\frac{C}{(4\pi)^2}|\tilde{\phi}(\kk)|^2
\ \ \ {\rm in}\ 3D
\label{eq:nk_V(r)_3D}
\\
n_\sigma(\kk)&\simeq&\frac{C}{(2\pi)^2}|\tilde{\phi}(\kk)|^2
\ \ \ {\rm in}\ 2D.
\eea
Indeed, Eq.~(\ref{eq:nk_AA_latt}) is derived as for the lattice model,
and $(A,A)$ can be expressed in terms of $C$ as in the previous subsection \ref{subsec:g2_V(r)}.

\section{Derivative of the energy with respect to the effective range}
\label{sec:re}

Assuming that the zero-range model is solved, we first show 
that the first correction to the energy due to a finite range 
of the interaction potential $V(r)$ 
can be explicitly obtained and only depends
on the $s$-wave effective range of the interaction.
We then enrich the discussion using the many-body
diagrammatic point of view,
where the central object is the full two-body $T$-matrix, to recall 
in particular that the situation is more subtle for lattice models 
\cite{zhenyaNJP}.
Finally, we relate $\partial E/\partial r_e$ to a subleading term of the short distance
behavior of the pair distribution function in Sec.\ref{subsec:re_dans_g2} and  to
the coefficient of the $1/k^6$ subleading tail of $n_{\sigma}(\kk)$ in Sec.~\ref{subsec:unsurk6}.

\subsection{Derivation of the explicit formulas}
\label{subsec:dotef}
\begin{table*}[t!]
\begin{tabular}{|cc|cc|}
\hline
Three dimensions & & Two dimensions &  \\
\hline
&&& \vspace{-4mm} \\
\ \ \ \ $\ds \left(\frac{\partial E}{\partial r_e}\right)_{\!a}= 2\pi (A, (E-\mathcal{H}) A)$ \ \ \ \ & (1a) & 
\ \ \ \ $\ds \left(\frac{\partial E}{\partial(r_e^2)}\right)_a = \pi(A, (E-\mathcal{H}) A)$ \ \ \ \ & (1b) \vspace{-4mm} \\
&&& \\ 
\hline
\multicolumn{4}{|c|}{\vspace{-4mm}} 
\\
\multicolumn{3}{|c}{
$\ds \mathcal{H}_{ij} \equiv -\frac{\hbar^2}{4m}\Delta_{\RR_{ij}} -\frac{\hbar^2}{2m}\sum_{k\neq i,j}\Delta_{\rr_k}
+2 U(\RR_{ij}) + \sum_{k\neq i,j} U(\rr_k)$} & (2) \vspace{-4mm} \\
\multicolumn{4}{|c|}{} \\
\hline
&&& \vspace{-4mm} \\
$\ds\bar{G}^{(2)}_{\uparrow\downarrow}(\rr)\underset{r\to 0}{=}\frac{C}{(4\pi)^2} \left(\frac{1}{r}-\frac{1}{a}\right)^2-\frac{m}{2\pi\hbar^2} \frac{\partial E}{\partial r_e}
+O(r)$ & (3a) & 
$\ds\bar{G}^{(2)}_{\uparrow\downarrow}(\rr)\underset{r\to 0}{=}\frac{C}{(2\pi)^2} \ln^2(r/a)-\frac{m}{2\pi\hbar^2} \frac{\partial E}{\partial (r_e^2)}
r^2\ln^2 r +O(r^2\ln r)$
& (3b) \vspace{-4mm} \\
&&& \\ 
\hline
&&& \vspace{-4mm} \\
$\ds \bar{n}_\sigma(k) - \frac{C}{k^4} \underset{k\to\infty}{\sim} \frac{1}{k^6} \left[\frac{16\pi m}{\hbar^2}
\frac{\partial E}{\partial r_e} -8\pi^2 (A,\Delta_{\RR} A) \right]$ &  (4a)  & 
$\ds \bar{n}_\sigma(k) - \frac{C}{k^4} \underset{k\to\infty}{\sim} \frac{1}{k^6} \left[\frac{8\pi m}{\hbar^2}
\frac{\partial E}{\partial (r_e^2)} -4\pi^2 (A,\Delta_{\RR} A) \right]$ 
&  (4b) \vspace{-4mm} \\
&&& \\ 
\hline
\end{tabular}

\caption{For spin-1/2 fermions, derivative of the energy with respect to the effective range $r_e$, or to its square in $2D$, 
taken at $r_e=0$ for a fixed value of the scattering length.
The functions $A$ (assumed to be real) are the ones of the zero-range regime. The compact notations for the scalar products and the matrix elements
are defined in Tab.~I. $\bar{n}_\sigma(k)$ is the average of $n_\sigma(\kk)$ over the direction of $\kk$. $\bar{G}^{(2)}_{\uparrow\downarrow}(\rr)$
is the pair distribution function integrated over the center of mass of the pair and averaged 
over the direction of $\rr$.
\label{tab:re}}
\end{table*}

\noindent{\underline{\it Three dimensions:}}
\\
In $3D$, the leading order finite-range correction to the zero-range model's spectrum
depends on the interaction potential $V(r)$ only {\it via} its effective range $r_e$,
and is given 
by the expression [Tab.~V, Eq.~(1a)],
where the derivative is taken in $r_e=0$ for a fixed value of the scattering length, the function $A$ is assumed to be real without loss
of generality.
As a first way
to obtain this result we use a modified version of the zero-range model, 
where the boundary condition [Tab. I, Eq. (1a)] is replaced by
\begin{multline}
 \psi(\rr_1,\ldots,\rr_N)\underset{r_{ij}\to0}{=}
\left( \frac{1}{r_{ij}}-\frac{1}{a}+\frac{m}{2\hbar^2}\mathcal{E} r_e
 \right) \\ \times \, A_{ij}\left(
\mathbf{R}_{ij}, (\mathbf{r}_k)_{k\neq i,j}
\right)
+O(r_{ij}),
\label{eq:CL_re}
\end{multline}
where
\begin{multline}
\mathcal{E}=E-2 U(\RR_{ij})-\left(\sum_{k\neq i,j} U(\rr_k)\right)
+\frac{1}{A_{ij}\left(
\mathbf{R}_{ij}, (\mathbf{r}_k)_{k\neq i,j}
\right)} \\ \times
\left[
\frac{\hbar^2}{4m}\Delta_{\RR_{ij}}+\frac{\hbar^2}{2m}\sum_{k\neq i,j}\Delta_{\rr_k}\right]
A_{ij}\left(
\mathbf{R}_{ij}, (\mathbf{r}_k)_{k\neq i,j}
\right).
\label{eq:E=}
\end{multline}
Equations (\ref{eq:CL_re},\ref{eq:E=}) generalize the ones already used for 3 bosons in free space in \cite{Efimov93,PetrovBosons}
(the predictions of \cite{Efimov93} and \cite{PetrovBosons} have been confirmed using different approaches, see \cite{PlatterRangeCorrections} and Refs. therein, and \cite{MoraBosons,LudoMattia} respectively; moreover, a derivation of these equations was given in 
\cite{Efimov93}).
Such a model was also used in the two-body case, see e.g. \cite{Greene,Tiesinga,Naidon}, and the modified scalar product that makes it hermitian
was constructed in \cite{LudovicOndeL}.

For the derivation of [Tab.~V, Eq.~(1a)], we consider a stationary state $\psi_1$ of the zero-range model, satisfying the boundary condition [Tab. I, Eq. (1a)] with a scattering length $a$ and a regular part $A^{(1)}$, and the corresponding finite-range stationary state $\psi_2$ satisfying (\ref{eq:CL_re},\ref{eq:E=}) with the same scattering length $a$ and a regular part $A^{(2)}$.
As in Appendix~\ref{app:lemme} we get (\ref{eq:ostro}), as well as 
(\ref{eq:int_surface_3D}) with $1/a_1-1/a_2$ replaced by $m\mathcal{E}r_e/(2\hbar^2)$.
This yields [Tab.~V, Eq.~(1a)].

A deeper physical understanding and a more self-contained derivation may be achieved 
by going back to the actual
 finite range model $V(r;b)$ for the interaction potential, such that the scattering
length remains fixed when the range $b$ tends to zero. The Hellmann-Feynman theorem gives
\be
\frac{dE}{db}\! =\!\!  \sum_{i=1}^{N_\uparrow} \sum_{j=N_\uparrow+1}^{N} \!\int\! d^3r_1\ldots d^3r_N |\psi(\rr_1,\ldots,\rr_N)|^2 \partial_b V(r_{ij};b).
\label{eq:HellFeyn}
\ee
We need to evaluate $|\psi|^2$ for a typical configuration with two atoms $i$ and $j$ within the potential range $b$; in the limit $b\to 0$ one may then
assume that the other atoms are separated by much more than $b$ and are at distances from $\RR_{ij}=(\rr_i+\rr_j)/2$ much larger than $b$. This motivates
the factorized ansatz
\be
\psi(\rr_1,\ldots,\rr_N) \simeq \chi(r_{ij}) A_{ij}(\RR_{ij},(\rr_k)_{k\neq i,j}).
\label{eq:fact_ansatz}
\ee
We take a rotationally invariant $\chi$, because we assume the absence
of scattering resonance in the partial waves other than $s$-wave
\footnote{More precisely, one first takes a general, non-rotationally invariant function $\chi(\rr)$, that
one then expands in partial waves of angular momentum $l$, that is in spherical harmonics. 
Performing the reasoning to come for each $l$, one finds at the end that the $l=0$ channel finite range correction
dominates for small $b$, in the absence of $l$-wave resonance for $l\neq 0$.}: The $p$-wave scattering amplitude, that vanishes quadratically with the relative wavenumber $k$,
is then $O(b^3 k^2)$, resulting
in an energy contribution $O(b^3)$ negligible at the present order.

Inserting the ansatz (\ref{eq:fact_ansatz}) into Schr\"odinger's equation $H\psi = E \psi$, and
neglecting the trapping potential within the interaction range
$r_{ij}\leq b$, as justified in the Appendix~\ref{app:piege_dans_interaction},
gives \footnote{Since $\mathcal{E}$ depends on $\RR_{ij}$ and
the $(\rr_k)_{k\neq i,j}$, $\chi$ actually depends on these variables and not only on $r_{ij}$. This dependence however rapidly vanishes in the limit $b\to 0$,
if one restricts to the distances $r_{ij} \lesssim b$, for the normalization (\ref{eq:norma_chi}): $\partial_{\mathcal{E}}\chi/\chi = O(mb^2/\hbar^2)$.}
\be
\mathcal{E} \chi(r_{ij}) \simeq [-\frac{\hbar^2}{m} \Delta_{\rr_{ij}}+ V(r_{ij};b)] \chi(r_{ij}),
\label{eq:espc}
\ee
where $\mathcal{E}$ is given by (\ref{eq:E=}).
For $\mathcal{E}>0$, we set $\mathcal{E}=\hbar^2 k^2/m$ with $k>0$, and
$\chi$ is a finite energy scattering state; to match the normalization of the zero energy scattering state
$\phi$ used in this article, see (\ref{eq:normalisation_phi_tilde_3D}), we take for $r$ out of the interaction potential
\be
\chi(r) \underset{r\to\infty}{=} \frac{1}{f_k} \frac{\sin(kr)}{kr} + \frac{e^{ikr}}{r},
\label{eq:norma_chi}
\ee
where $f_k$ is the scattering amplitude. The optical theorem,  implying that 
\be
f_k=-\frac{1}{ik+u(k)},
\label{eq:introu}
\ee
where $u(k)\in \mathbb{R}$, ensures that $\chi$ is real
\footnote{$u(k)$ is related to the $s$-wave collisional phase shift $\delta_0(k)$
by $u(k)=-k/\tan \delta_0(k)$.}.

Inserting the ansatz (\ref{eq:fact_ansatz}) into the Hellmann-Feynman expression 
(\ref{eq:HellFeyn})  gives
\begin{multline}
\frac{dE}{db} \simeq \sum_{i<j} \int d^3R_{ij} \int (\prod_{k\neq i,j} d^3r_k) \ A^2_{ij}(\RR_{ij},(\rr_k)_{k\neq i,j})
\\
 \times \int d^3r_{ij} \ \chi^2(r_{ij}) \partial_b V(r_{ij};b)
\end{multline}
To evaluate the integral of $\chi^2 \partial_b V$, we use 
the following lemma (whose derivation is given in the next paragraph):
\be
\frac{4\pi\hbar^2}{m} [u_2(k)-u_1(k)]= \int_{\mathbb{R}^3}\!\!\! d^3r\, \chi_1(r) \chi_2(r) [V(r;b_1)-V(r;b_2)]
\label{eq:lemme_utile}
\ee
where $\chi_1$ and $\chi_2$ are the same energy $\mathcal{E}$ scattering states for two different values $b_1$ and $b_2$ of the potential range.
Then dividing this expression by $b_1-b_2$, taking the limit $b_1\to b_2$,  and afterwards the limit $b_2\to 0$ for which the low-$k$ expansion holds:
\be
u(k) = \frac{1}{a} - \frac{1}{2} r_e k^2 +O(b^3 k^4)
\label{eq:devuk_3D}
\ee
$r_e$ being the effective range of the interaction potential of range $b$, 
we obtain [Tab.~V, Eq.~(1a)] \footnote{In general, when $N_\uparrow\geq 2$ and $N_\downarrow\geq 2$,
the functions $A_{ij}$ have $1/r_{kl}$ divergences when $r_{kl}\to 0$.
This is apparent in the dimer-dimer scattering problem \cite{PetrovPRL}. As a consequence, in the integral
of [Tab.~V, Eq.~(1a)], one has to exclude the manifold where at least two particles are at the same
location. The same exclusion has to be performed in $2D$}.

As a side result of this physical approach, the modified contact conditions (\ref{eq:CL_re}) may be rederived.
One performs an analytical continuation of the out-of-potential wavefunction (\ref{eq:norma_chi}) 
to the interval $r\leq b$ \cite{CombescotC} and one takes the zero-$r$ limit of that continuation
\footnote{The wavefunction is not an analytic function of $r$ for a compact support interaction potential, since
a non-zero compact support function is not analytic.}. In simple words, this amounts
to expanding (\ref{eq:norma_chi}) in powers of $r$:
\be
\chi(r) = \frac{1}{r}-\frac{1}{a} + \frac{1}{2} k^2 r_e + O(r).
\ee
Inserting this expansion in (\ref{eq:fact_ansatz}) and using $k^2=m\mathcal{E}/\hbar^2$ gives (\ref{eq:CL_re}).

The lemma (\ref{eq:lemme_utile}) is obtained by multiplying Schr\"odinger's equations for 
$\chi_1$ (respectively for $\chi_2$) by $\chi_2$ (respectively by $\chi_1$),
taking the difference of the two resulting equations, integrating this difference over the sphere $r<R$ and using the divergence theorem to convert
the volume integral of $\chi_2\Delta_\rr \chi_1-\chi_1\Delta_\rr \chi_2$ into a surface integral, where the asymptotic forms
(\ref{eq:norma_chi}) for $r=R\to +\infty$ may be used.
When $\mathcal{E}<0$, we set $\mathcal{E}=-\hbar^2\kappa^2/m$ with $\kappa>0$ and we perform analytic continuation of the $\mathcal{E}>0$ case
by replacing $k$ with $i\kappa$. From (\ref{eq:norma_chi}) it appears that $\chi(r)$ now diverges exponentially at large distances, as $e^{\kappa r}/r$,
if $1/f(i\kappa)\ne 0$.
If the interaction potential is a compact support potential, or simply tends to zero more rapidly than $\exp(-2\kappa r)$, the lemma
and the final conclusion 
[Tab.~V, Eq.~(1a)] still hold; the functions $u_1(i\kappa)$ and $u_2(i\kappa)$ remain real, since the series expansion of $u(k)$ has only
even powers of $k$.
\\
\noindent{\underline{\it Two dimensions:}}
\\
The above physical reasoning may be directly generalized to $2D$ \footnote{We consider here a truly $2D$ gas. In experiments,
quasi-$2D$ gases are produced by freezing the $z$ motion in a harmonic oscillator ground state of size $a_z=[\hbar/(m\omega_z)]^{1/2}$:
At zero temperature, a $2D$ character appears for $\hbar^2 k_F^2/(2m)\ll \hbar\omega_z$. From the quasi-$2D$ scattering amplitude given
in \cite{LudoScatteringLowD} (see also \cite{PetrovShlyapCollisions2D}) we find the effective range squared,
$r_e^2 = -(\ln 2)\, a_z^2$. Anticipating on subsection \ref{subsec:wwlftdpov}
we also find $\rho_e=R_1=0$. It would be interesting to 
see if the finite range energy corrections dominate over the corrections due to the $3D$ nature of the gas,
both effects being controlled by the same small parameter $(k_F r_e)^2$.},
giving [Tab.~V, Eq.~(1b)],
where the derivative is taken for a fixed scattering length in $r_e=0$. 
The main difference with the $3D$ case [Tab.~V, Eq.~(1a)] 
is that the energy $E$ now varies quadratically with the effective range $r_e$,
as already observed numerically for three-boson-bound states in \cite{HelfrichHammer}.
In the derivation, the first significant difference with the $3D$ case occurs in the normalization of the two-body scattering state: (\ref{eq:norma_chi})
is replaced with
\be
\chi(r) \underset{r\to\infty}{=} \frac{\pi}{2i} \left[\frac{1}{f_k} J_0(kr) + H_0^{(1)}(kr)\right]
\label{eq:asympt2d}
\ee
where $H_0^{(1)}=J_0+i N_0$ is a Hankel function, 
$J_0$ and $N_0$ are Bessel functions of the first and second kinds. The optical theorem implies $|f_k|^2+\mbox{Re}\, f_k=0$
so that
\be
f_k = \frac{-1}{1+i u(k)} \ \ \ \ \ \mbox{with}\ \ \ \ \ u(k)\in \mathbb{R},
\label{eq:def_fk_a_2D}
\ee
and $\chi$ is real.  The low-$k$ expansion for a potential of range $b$
takes the form \cite{Verhaar2D,Khuri}
\be
u(k) = \frac{2}{\pi} \left[\ln\left(e^{\gamma} ka/2\right)+ 
\frac{1}{2} (k r_e)^2+\ldots\right],
\label{eq:devuk_2D}
\ee
where $\gamma=0.577216\ldots$ is Euler's constant,
the logarithmic term being obtained in the zero-range Bethe-Peierls model and the $k^2$ term corresponding to finite effective range
corrections (with the sign convention of \cite{Verhaar2D} such that $r_e^2>0$ for a hard disk potential).
The subsequent calculations are similar to the $3D$ case, also for the negative energy case where analytic continuation
gives rises to the special functions $I_0(\kappa r)$ and $K_0(\kappa r)$.
For example, at positive energy, the lemma (\ref{eq:lemme_utile}) 
takes in $2D$ the form 
\be
\frac{\pi^2\hbar^2}{m} [u_1(k)-u_2(k)]= \int_{\mathbb{R}^2}\!\!\! 
d^2r\, \chi_1(r) \chi_2(r) [V(r;b_1)-V(r;b_2)]
\label{eq:lemme_utile_2D}
\ee
The fact that one can neglect the trapping potential within
the interaction range is again 
justified in Appendix~\ref{app:piege_dans_interaction}.
Finally, we note that the expansion of the asymptotic form (\ref{eq:asympt2d}) for $r\to 0$, and for $k\to 0$,
\be
\chi(r) =\ln(r/a) - \frac{1}{2} (k r_e)^2 + O(r^2\ln r)
\ee
allows to determine the $2D$ version of the modified zero-range model (\ref{eq:CL_re}),
\begin{multline}
\psi(\rr_1,\ldots,\rr_N)\underset{r_{ij}\to0}{=}
\left( \ln(r_{ij}/a)-\frac{m}{2\hbar^2}\mathcal{E} r_e^2
\right) \\ \times \, A_{ij}\left(
\mathbf{R}_{ij}, (\mathbf{r}_k)_{k\neq i,j}
\right)
+O(r_{ij})
\end{multline}
where $\mathcal{E}$ is defined as in $3D$ by (\ref{eq:E=}).
To complete this $2D$ derivation, one has to check that the $p$-wave interaction
brings a negligible contribution to the energy. The $p$-wave scattering amplitude at low relative wavenumber $k$
vanishes as $k^2 R_1^2$ where $R_1^2$ is the $p$-wave scattering surface \cite{AdhikariGibson}. One could believe
that $r_e\approx R_1 \approx b$, one would then  conclude that the $p$-wave
contribution to the energy, scaling as $R_1^2$, cannot be neglected as compared to the $s$-wave finite range
correction, scaling as $r_e^2$. Fortunately, as shown in subsection \ref{subsec:wwlftdpov}, this expectation is too naive, and
[Tab.~V, Eq.~(1b)] is saved by a logarithm, $r_e$ being larger than $R_1$ by a factor $\ln(a/b)\gg 1$ in the zero range limit
\footnote{As in $3D$ one may also be worried by the dependence of $\chi$ with $\RR_{ij}$ and
the $(\rr_k)_{k\neq i,j}$ {\sl via} its dependence with the energy $\mathcal{E}$. We reach the estimate $\partial_{\mathcal{E}}\chi(b)/\chi(b)
\approx m r_e^2/[\hbar^2\ln(a/b)]$ that vanishes more rapidly than $r_e^2$ in the zero-range limit.}.

\subsection{What we learn from diagrammatic formalism}
\label{subsec:wwlftdpov}

In the many-body diagrammatic formalism \cite{PitaevskiiLifchitz,FetterWalecka}, 
the equation of state of the homogeneous gas (in the thermodynamic limit) is accessed from the single particle Green's function,
which can be expanded in powers of the interaction potential, each term of the expansion
being represented by a Feynman diagram. The internal momenta of the diagrams can however be as large as 
$\hbar/b$, where $b$ is the interaction range. A standard approach to
improve the convergence of the perturbative series for strong interaction potentials is to perform the so-called
{\sl ladder} resummation. The resulting Feynman diagrams then involve the two-body $T$-matrix of the interaction, rather than the bare
interaction potential $V$. For the spin-$1/2$ Fermi gas, where there is {\sl a priori} no Efimov effect, one then expects that 
the internal momenta of the Feynman diagrams are on the order of $\hbar k_{\rm typ}$ only, where the typical wavenumber $k_{\rm typ}$
was defined in subsection \ref{sec:models:lattice}. As put forward in \cite{zhenyaNJP}, the interaction parameters
controlling the first deviation of the gas energy from its zero-range limit are then the ones appearing in the first deviations
of the two-body $T$-matrix element $\langle \kk_1,\kk_2|T(E+i0^+)|\kk_3,\kk_4\rangle$ from its zero-range limit, where all the $\kk_i$
are on the order of $k_{\rm typ}$ and $E$ is on the order of $\hbar^2 k_{\rm typ}^2/m$. 
The single particle Green's function
is indeed a sum of integrals of products of $T$-matrix elements and of ideal-gas Green's functions.

We explore this idea in this subsection.
For an interaction potential $V(r)$, we confirm the results of subsection \ref{subsec:dotef}. 
In addition to the effective range $r_e$ characterizing the on-shell $T$-matrix elements (that is the scattering amplitude),
the diagrammatic point of view introduces a length $\rho_e$ characterizing the $s$-wave low-energy {\sl off-shell} $T$-matrix elements, 
and a length $R_1$ characterizing the $p$-wave on-shell scattering; we will show that the contributions
of $\rho_e$ and $R_1$ are negligible as compared to the one of the effective range $r_e$.
Moreover, in the case of lattice models, a length $R_e$ characterizing the breaking 
of the Galilean invariance appears \cite{zhenyaNJP}. Its contribution is in general of the same order
as the one of $r_e$. Both contributions can be zeroed for appropriately tuned matterwave dispersion relations
on the lattice.
Finally, in the case of a continuous space model with a delta interaction potential 
plus a spherical cut-off
in momentum space, and in the case of a lattice model with a spherical momentum cut-off,
we show that the breaking of Galilean invariance does not disappear
in the infinite cut-off limit.

\subsubsection{For the continuous space interaction $V(r)$}

When each pair of particles $i$ and $j$ interact in continuous space {\sl via} the potential $V(r_{ij})$, one can use
Galilean invariance to restrict the $T$-matrix to the center of mass frame, where $\kk'\equiv \kk_1=-\kk_2$ and $\kk\equiv \kk_3= -\kk_4$.
Further using rotational invariance, one can restrict this internal $T$-matrix to  fixed total angular momentum $l$, with
matrix elements characterized by the function $t_l(k',k;E)$ whose low-energy behavior was extensively studied \cite{AdhikariGibson,Gibson3D}.
This function is said to be {\sl on-shell} iff $k=k'=(mE)^{1/2}/\hbar$, in which case it is simply noted as $t_l(E)$, otherwise it is said to
be {\sl off-shell}.

\noindent{\underline{\it Three dimensions:}}
\\
We assume that the interaction potential, of compact support of range $b$, is everywhere 
non-positive (or infinite). We recall that we are here in the {\sl resonant}
regime, with a $s$ wave scattering length $a$ such that $|a|\gg b$.
The potential is assumed to have the {\sl minimal} depth leading to the desired value of $a$, so as to exclude deeply bound dimers.
In particular, at resonance ($1/a=0$), there is no two-body bound state.
To invalidate the usual variational argument \cite{PandhaPRA,BlattWeisskopf,BaymCollapse,WernerThese} (that shows,  for a non-positive interaction potential, 
that the spin-$1/2$ fermions have deep $N$-body bound states in the large $N$ limit), 
we allow that $V(r)$ has a hard core of range $b_{\rm hard}< b$.  We directly restrict to the $s$-wave case ($l=0$), since the non-resonant $p$-wave interaction bring
a negligible $O(b^3)$ contribution, as already discussed in subsection \ref{subsec:dotef}.

The first deviation of the on-shell $s$-wave $T$-matrix from its zero-range limit is characterized by the effective range
$r_e$, already introduced in Eq.~(\ref{eq:devuk_3D}).  The effective range is given by the well-known Smorodinski formula \cite{Khuri}:
\be
\label{eq:smorodinski}
\frac{1}{2} r_e  = \int_0^{+\infty} dr \left[(1-r/a)^2 - u_0^2(r)\right]
\ee
in terms of the zero-energy scattering state $\phi(r)$, with $u_0(r)=r\phi(r)$ and $\phi$ is normalized as in Eq.~(\ref{eq:normalisation_phi_tilde_3D}).
Note that $u_0(r)$ is zero for $r\leq b_{\rm hard}$.
As $r_e$ deviates from its resonant ($|a|\to \infty$) value by terms $O(b^2/a)$, the discussion of its $1/a=0$ value is sufficient here.
The function $u_0$ then solves
\be
0 = -\frac{\hbar^2}{m}  u_0''(r) + V(r) u_0(r)
\label{eq:esaen}
\ee
with the boundary conditions $u_0(b_{\rm hard})=0$ and $u_0(r)=1$ for $r>b$.
Due to the absence of two-body bound states, $u_0$ is the ground two-body state 
and it has a constant sign, $u_0(r)\geq 0$ for all $r$. Since $V\leq 0$, Eq.~(\ref{eq:esaen})
implies that $u_0''\leq 0$, the function $u_0$ is concave. Combined with the boundary conditions, this leads to
$0 \leq u_0(r) \leq 1,$ for all $r$. Then from Eq.~(\ref{eq:smorodinski}):
\be
2 b_{\rm hard} \le r_e \le 2 b
\ee
For the considered model, this proves that $k_{\rm typ} r_e\to 0$ in the zero-range limit $b\to 0$, which is a key property
for the present work. Note that the absence of two-body bound states at resonance is the crucial hypothesis
ensuring that $r_e\geq 0$; it was not explicitly stated in the solution of problem 1 in Sec.~131 of \cite{LandauLifschitzMecaQ}.
Without this hypothesis, $r_e$ at resonance can be arbitrarily large and negative even for $V(r)\leq 0$ for all $r$,
see an explicit example in \cite{LeChapitre}.

In the $s$-wave channel, the first deviations of the {\sl off-shell} $T$-matrix from its zero-range value introduces, in addition to $r_e$,
another length that we call $\rho_e$, such that \cite{Gibson3D} \footnote{We have checked that the hypothesis of a non-resonant
interaction in \cite{Gibson3D} is actually not necessary to obtain (C16) and (C18) of that reference, that lead to (\ref{eq:rhoe_3D}).}
\begin{multline}
\frac{t_0(k,k';E)}{t_0(E)} -1 \underset{k,k',E\to 0}{\sim}  \left(\frac{2mE}{\hbar^2} -k^2 -k'^{2}\right) \frac{1}{2} \rho_e^2 \\
\mbox{with} \ \ \ \frac{1}{2} \rho_e^2 = \int_0^{+\infty} dr\, r [(1-r/a)-u_0(r)].
\label{eq:rhoe_3D}
\end{multline}
For our minimal-depth model at resonance, we conclude that $0\leq \rho_e^2 \leq b^2$, so it appears, in the finite-range correction
to the energy, at a higher order than $r_e$, and it cannot contribute to [Tab.~V, Eq.~(1a)].

\noindent{\underline{\it Two dimensions:}}
\\
The specific feature of the $2D$ case is that the minimal-depth attractive potential ensuring the desired scattering length
$a$ only weakly dephases the matter-wave over its range, when $\ln(a/b)\gg 1$. This is apparent e.g.\ if $V(r)$ is a square-well potential of range $b$,
$V(r)=-\frac{\hbar^2 k_0^2}{m} \theta(b-r)$: One has $-k_0 b J_0'(k_0b)/J_0(k_0b)=1/\ln(a/b)$, where $J_0$ is a Bessel function,
which shows that, for the minimal-depth solution, the matter-wave phase shift $k_0 b$ vanishes as $[2/\ln(a/b)]^{1/2}$ in the zero-range limit.
This property allows to treat the potential perturbatively.

There are three relevant parameters describing the low-energy behavior of the $T$-matrix beyond the zero-range limit. 
The first one is the effective range $r_e$ for the $s$-wave on-shell $T$-matrix, see Eq.~(\ref{eq:devuk_2D}).
It is given by the bidimensional Smorodinski formula \cite{Verhaar2D,Khuri}:
\be
\frac{1}{2} r_e^2 = \int_0^{+\infty} dr\, r [\ln^2(r/a)-\phi^2(r)]
\label{eq:smorodinski2d}
\ee
where the zero-energy scattering state $\phi(r)$ is normalized as in Eq.~(\ref{eq:normalisation_phi_tilde_2D}).
The second parameter is the length $\rho_e$ associated to the $s$-wave off-shell $T$-matrix: The $2D$ equivalent of
Eq.~(\ref{eq:rhoe_3D}) is \cite{AdhikariGibson}:
\begin{multline}
\frac{t_0(k,k';E)}{t_0(E)} -1 \underset{k,k',E\to 0}{\sim}  \left(\frac{2mE}{\hbar^2} -k^2 -k'^{2}\right) \frac{1}{2} \rho_e^2 \\
\mbox{with} \ \ \ \frac{1}{2} \rho_e^2 = \int_0^{+\infty} dr\, r [\phi(r)-\ln(r/a)].
\label{eq:rhoe_2D}
\end{multline}
The third parameter is the length $R_1$ characterizing the low-energy $p$-wave scattering. For the $l$-wave scattering state
of energy $E=\hbar^2 k^2/m$, $k>0$, we generalize Eq.~(\ref{eq:asympt2d}) as
\be
\chi^{(l)}(r) \underset{r\to\infty}{=} \frac{\pi}{2i} k^l\left[\frac{1}{f_k^{(l)}} J_l(kr) + H_l^{(1)}(kr)\right].
\ee
The $l$-wave scattering amplitude then vanishes as
\be
f_k^{(l)} \underset{k\to 0}{\sim} i\frac{\pi}{2} k^{2l} R_l^{2l}
\ee
and the leading behavior of the off-shell $l$-wave $T$-matrix is characterized by the same length $R_l$ as the on-shell one
\cite{AdhikariGibson}.

The situation thus looks critical in $2D$: Three lengths squared characterize the low-energy $T$-matrix, one may naively
expect that they are of the same order $\approx b^2$ and that they all three contribute to the finite-range correction to the gas
energy at the same level, whereas [Tab.~V, Eq.~(1b)] singles out the effective range $r_e$. By a perturbative
treatment of the minimal-depth finite-range potential $V(r)$ of fixed scattering length $a$, we however obtain in the zero-range
limit the following hierarchy, see Appendix~\ref{app:param2d}:
\bea
\label{eq:hier1}
r_e^2 &\underset{b\to 0}{\sim}&  2\rho_e^2 \ln(a/b) \\
\label{eq:hier2}
\rho_e^2 &\underset{b\to 0}{=}& \frac{1}{2} \frac{\int_{\mathbb{R}^2} d^2r \, r^2 V(r)}{\int_{\mathbb{R}^2} d^2r \, V(r)} 
\left[1+O\left(\frac{1}{\ln(a/b)}\right)\right] \\
\label{eq:hier3}
R_1^2 &\underset{b\to 0}{\sim}&  \frac{\rho_e^2}{2\ln(a/b)}
\eea
This validates [Tab.~V, Eq.~(1b)] when $\ln(a/b)\gg 1$.

\subsubsection{Lattice models}

We restrict here for simplicity to the $3D$ case.
To obtain a non-zero $T$-matrix element $\langle \kk_1,\kk_2|T(E+i0^+)|\kk_3,\kk_4\rangle$, due to the conservation of the total quasi-momentum,
we have to restrict to $\kk_1+\kk_2=\kk_3+\kk_4\equiv \KK$ (modulo a vector of the reciprocal lattice).
As the interactions in the lattice model are purely on-site,
the matrix element only depends on the total quasi-momentum $\KK$
and the energy $E$, and is noted as $t(\KK,E)$ in what follows. We recall that the bare coupling constant $g_0$ is adjusted to have 
a fixed scattering length $a$ on the lattice, see Eq.~(\ref{eq:g0_3D}), which leads to
\be
g_0 = \frac{4\pi\hbar^2a/m}{1- K_3\, a/b} 
\ee
where the numerical constant $K_3$ depends on the lattice dispersion relation $\epsilon_\kk$.
One then gets \cite{zhenyaNJP}
\begin{multline}
\frac{1}{t(\KK,E)} = \frac{m}{4\pi\hbar^2 a}  -\int_D \frac{d^3q}{(2\pi)^3} 
\Big( \frac{1}{2\epsilon_\qq} \\+  
\frac{1}{E+i0^+-\epsilon_{\frac{1}{2}\KK+\qq}-\epsilon_{\frac{1}{2}\KK-\qq}} \Big)
\label{eq:mattmsr}
\end{multline}
where $a$ is the $s$-wave scattering length and the dispersion relation $\epsilon_\qq$ is extended by periodicity from the first Brillouin zone
$D$ to the whole space. The low-$\KK$ and low-energy limit of that expression was worked out
in \cite{zhenyaNJP}, it involves the effective range $r_e$ and an extra length $R_e$ quantifying the breaking of Galilean invariance:
\be
\frac{1}{t(\KK,E)}= \frac{m}{4\pi\hbar^2} \left(\frac{1}{a}+ i k -\frac{1}{2} r_e  k^2   -\frac{1}{2} R_e K^2\right) + \ldots
\ee
where the relative wavenumber $k$ such that $E-\frac{\hbar^2 K^2}{4m}=\frac{\hbar^2 k^2}{m}$ is either real non-negative 
or purely imaginary with a positive imaginary part. The two lengths are given by
\bea
\label{eq:re_reseau}
r_e &=&\!\! \int_{\mathbb{R}^3\setminus D} \frac{d^3q}{\pi^2 q^4} + \!\! \int_{D} \frac{d^3q}{\pi^2} \left[\frac{1}{q^4}-
\left(\frac{\hbar^2}{2m\epsilon_\qq}\right)^2\right]\\
R_e &=& - \int_{\stackrel{\circ}{D}} \frac{d^3q}{4\pi^2} \left(\frac{\hbar^2}{2m\epsilon_\qq}\right)^2
\left[1-\frac{m}{\hbar^2} \frac{\partial^2\epsilon_{\qq}}{\partial q_x^2}\right] \nonumber \\ 
&-& 
\int_{-\frac{\pi}{b}}^{\frac{\pi}{b}} \int_{-\frac{\pi}{b}}^{\frac{\pi}{b}}\frac{dq_y dq_z}{8\pi^2} \frac{\hbar^2}{m\epsilon^2_{(\frac{\pi}{b},q_y,q_z)}}
\frac{\partial \epsilon_{(\frac{\pi}{b},q_y,q_z)}}{\partial q_x}
\label{eq:Re}
\eea
where the dispersion relation $\epsilon_\kk$ was supposed to be twice differentiable on the interior $\stackrel{\circ}{D}$ of the first Brillouin zone
and to be invariant under permutation of the coordinate axes.
As compared to \cite{zhenyaNJP} we have added the second term (a surface term) in Eq.~(\ref{eq:Re}) to include the case where the dispersion
relation has cusps at the border of the first Brillouin zone \footnote{This term is obtained by distinguishing three integration
zones before taking the limit $K_x\to 0$, so as to fold back the vectors $\qq\pm \frac{1}{2} \KK$ inside the first Brillouin zone:
 the left zone $-\frac{\pi}{b}< q_x < -\frac{\pi}{b} +\frac{1}{2}K_x$ where
$\epsilon_{\qq-\frac{1}{2}\KK}$ is written as $\epsilon_{\qq+\frac{2\pi}{b}\mathbf{e}_x-\frac{1}{2}\KK}$, the right
zone $\frac{\pi}{b}-\frac{1}{2}K_x < q_x < \frac{\pi}{b}$ where $\epsilon_{\qq+\frac{1}{2}\KK}$ 
is written as $\epsilon_{\qq-\frac{2\pi}{b}\mathbf{e}_x+\frac{1}{2}\KK}$,
and the central zone. The surface term can also be obtained by interpreting
$\partial_{q_x}^2$ in the sense of distributions, after having shifted the integration domain $D$ by $\frac{\pi}{b} \mathbf{e}_x$
for mathematical convenience. The second order derivative in the first term of Eq.~(\ref{eq:Re}) is of course
taken in the sense of functions.}.
\setcounter{notedecoupage}{\thefootnote}
As mentioned in the introduction of the present section, we then expect that, in the lattice model, the first deviation of any many-body eigenenergy $E$ from the zero-range limit
is a linear function of the {\sl two} parameters $r_e$ and $R_e$ with model-independent coefficients:
\be
E(b) \underset{b\to 0}{=} E(0) + \frac{\partial E}{\partial r_e} r_e + \frac{\partial E}{\partial R_e} R_e + o(b)
\label{eq:dEdRe}
\ee
This feature was overlooked in the early version \cite{50pages} of this work.
It invalidates the discussion of $\partial T_c/\partial r_e$ given in \cite{50pages}.

We illustrate this discussion with a few relevant examples.
For a parabolic dispersion relation $\epsilon_\kk=\frac{\hbar^2 k^2}{2m}$, 
the constant $K_3=2.442\, 749\, 607\, 806\, 335\ldots$ \cite{MoraCastin,boite}
and the effective range \cite{YvanVarenna,LeChapitre} were already calculated, first numerically then analytically;
in the quantity $R_e$, the first term vanishes but there is still breaking of Galilean invariance due to the non-zero surface term
that can be deduced from Eq.~(\ref{eq:une_integrale}):
\be
r_e=b\,\frac{12\sqrt{2}}{\pi^3} \arcsin\frac{1}{\sqrt{3}}\simeq 0.337 b \  \ \mbox{and} \ \ R_e=-\frac{1}{12} r_e
\ee
A popular model for Quantum Monte Carlo simulations is the Hubbard model, that leads to the dispersion
relation $\epsilon_\kk^{\rm Hub}=\frac{\hbar^2}{mb^2}[3-\cos(k_xb)-\cos(k_yb)-\cos(k_zb)]$ (as already
mentioned in subsection \ref{sec:models:lattice}). This leads to $K_3\simeq 3.175\, 911\, 6$. 
Again, both $r_e$ and $R_e$ differ from zero:
\be
r_e \simeq -0.305\, 718 b \ \ \mbox{and}\ \ R_e \simeq -0.264\, 659 b
\ee
In an attempt to reduce the dependence of the Monte Carlo results on the grid spacing $b$, a zero-effective-range dispersion relation was
constructed \cite{LeChapitre,CarlsonAFQMC},
\be
\epsilon_\kk =\frac{\hbar^2 k^2}{2m} [1-C(k b/\pi)^2],
\label{eq:disperrez}
\ee
with $C\simeq 0.257\, 022$, and used
in real simulations \cite{CarlsonAFQMC}. The corresponding $K_3 \simeq 2.899\, 952$. Unfortunately this leads to a sizeable $R_e$:
\be
R_e \simeq-0.168 b.
\ee
As envisioned in \cite{zhenyaNJP} one may look for dispersion relations with $r_e=R_e=0$. We have found an example of
such a {\sl magic} dispersion relation:
\be
\epsilon_\kk = \epsilon_\kk^{\rm Hub} [1+\alpha X + \beta X^2] \ \ \mbox{with} \ \ X=\frac{\epsilon_\kk^{\rm Hub}}{6\hbar^2/mb^2}.
\label{eq:magic}
\ee
Two sets of parameters are possible. The first choice is
\be
\alpha \simeq 1.470\, 885 \ \mbox{and}\ \beta\simeq -2.450\, 725,
\label{eq:magic1}
\ee
which leads to $K_3\simeq 3.137 \, 788$. The second choice is
\be
\alpha \simeq -1.728\, 219 \ \mbox{and}\  \beta\simeq 12.838\, 540,
\label{eq:magic2}
\ee
which leads to $K_3\simeq 1.949\, 671$.
Other examples of magic dispersion relation can be found \cite{Juillet_juillet}.

\subsubsection{The single-particle momentum cut-off model}
\label{subsubsec:tspmcom}

A continuous space model used in particular in \cite{zhenyas_crossover} takes a Dirac delta interaction potential
$g_0 \delta(\rr_i-\rr_j)$ between particles $i$ and $j$, and regularizes the theory by introducing a 
cut-off $\Lambda$ on all the single-particle wavevectors. Due to the conservation of momentum one needs to evaluate
the $T$-matrix only between states with the same total momentum $\hbar\KK$. Due to the contact interaction
the resulting matrix element depends only on $\KK$ and on $E$, and is noted
as $t(\KK,E)$. Expressing $g_0$ in terms of the 
$s$-wave scattering length as in \cite{zhenyas_crossover} one gets
\begin{multline}
\frac{1}{t(\KK,E)} = \frac{m}{4\pi\hbar^2 a}  -\int_{\mathbb{R}^3}  \frac{d^3q}{(2\pi)^3}
\Bigg[ \frac{\theta(\Lambda-q)}{2\epsilon_\qq} \\+
\frac{\theta(\Lambda-|\frac{1}{2}\KK+\qq|)\, \theta(\Lambda-|\frac{1}{2}\KK-\qq|)}{E+i0^+-\epsilon_{\frac{1}{2}\KK+\qq}-\epsilon_{\frac{1}{2}\KK-\qq}} \Bigg]
\end{multline}
where $\epsilon_\qq=\hbar^2 q^2/(2m)$ for all $\qq$.
Introducing the relative wavenumber $k$ such that $E-\frac{\hbar^2 K^2}{4m}=\frac{\hbar^2 k^2}{m}$, $k\in \mathbb{R}^+$ or $k\in i\mathbb{R}^+$,
we obtain the low wavenumbers expansion
\be
\frac{1}{t(\KK,E)} = \frac{m}{4\pi\hbar^2} \left(\frac{1}{a} + ik- \frac{K}{2\pi}  -\frac{1}{2} r_e k^2 -\frac{1}{2} R_e K^2 \right)
+\ldots
\label{eq:tamhbe}
\ee
The effective range is given by $r_e=4/(\pi \Lambda)$ and the length $R_e=r_e/12$
\footnote{The integration can be performed in spherical coordinates of polar axis the direction of $\KK$.}. 
The unfortunate feature of this model is the occurrence of a term linear in $K$, that does not disappear even if  $\Lambda\to +\infty$: 
The model thus does {\sl not} reproduce the universal zero-range model in the large cut-off limit, as soon as pairs of
particles have a non-zero total momentum.
Note that here one cannot exchange the order of the integration over $\qq$ and the $\Lambda\to\infty$ limit.
As a concrete illustration of the breaking of the Galilean invariance, for $a>0$ and in the limit $\Lambda\to +\infty$,
it is found (e.g.\ by calculating the pole of the $T$-matrix) that the total energy of a free-space dimer of total momentum $\hbar\KK$ is
\be
E_{\rm dim}^{\rm model} (\KK)= \frac{\hbar^2 K^2}{4m} -\frac{\hbar^2}{m} \left(\frac{1}{a}-\frac{K}{2\pi}\right)^2
\ee
and that this dimer state exists only for $K<2\pi/a$ \footnote{This problem does not show up in recent studies of the fermionic
polaron problem \cite{CGL2009,ZwergerPunk} since the momentum cut-off is introduced only for the majority atoms and not for the impurity, see 
\cite{TheseGiraud}.}.

\subsubsection{The single-particle momentum cut-off lattice model}
\label{subsubsec:tspmcolm}

A spherical momentum cut-off was also introduced for a lattice model in \cite{bulgacQMC,BulgacCrossover,BulgacPG,BulgacPG2}.
Our understanding is that this amounts to taking the following dispersion relation inside the first Brillouin
zone: $\epsilon_{\kk}=\hbar^2 k^2/(2m)$ for $k < \pi/b$, $\epsilon_\kk=+\infty$ otherwise.
The $T$-matrix is then given by Eq.~(\ref{eq:mattmsr}), where for $\KK\neq\mathbf{0}$ one extends $\epsilon_\kk$ by
periodicity out of the first Brillouin zone. By distinguishing three zones within the integration domain
for $\qq$, similarly to the note [\thenotedecoupage], 
and restricting for simplicity to $E=\hbar^2 K^2/(4m)$, we find the same undesired term $-K/(2\pi)$ as in Eq.~(\ref{eq:tamhbe}),
implying that the model does not reproduce the unitary gas even for $b\to 0$.

\subsection{The Juillet effect for lattice models}
\label{subsec:juillet}

\begin{figure*}[t]
\includegraphics[width=0.4\linewidth,clip=]{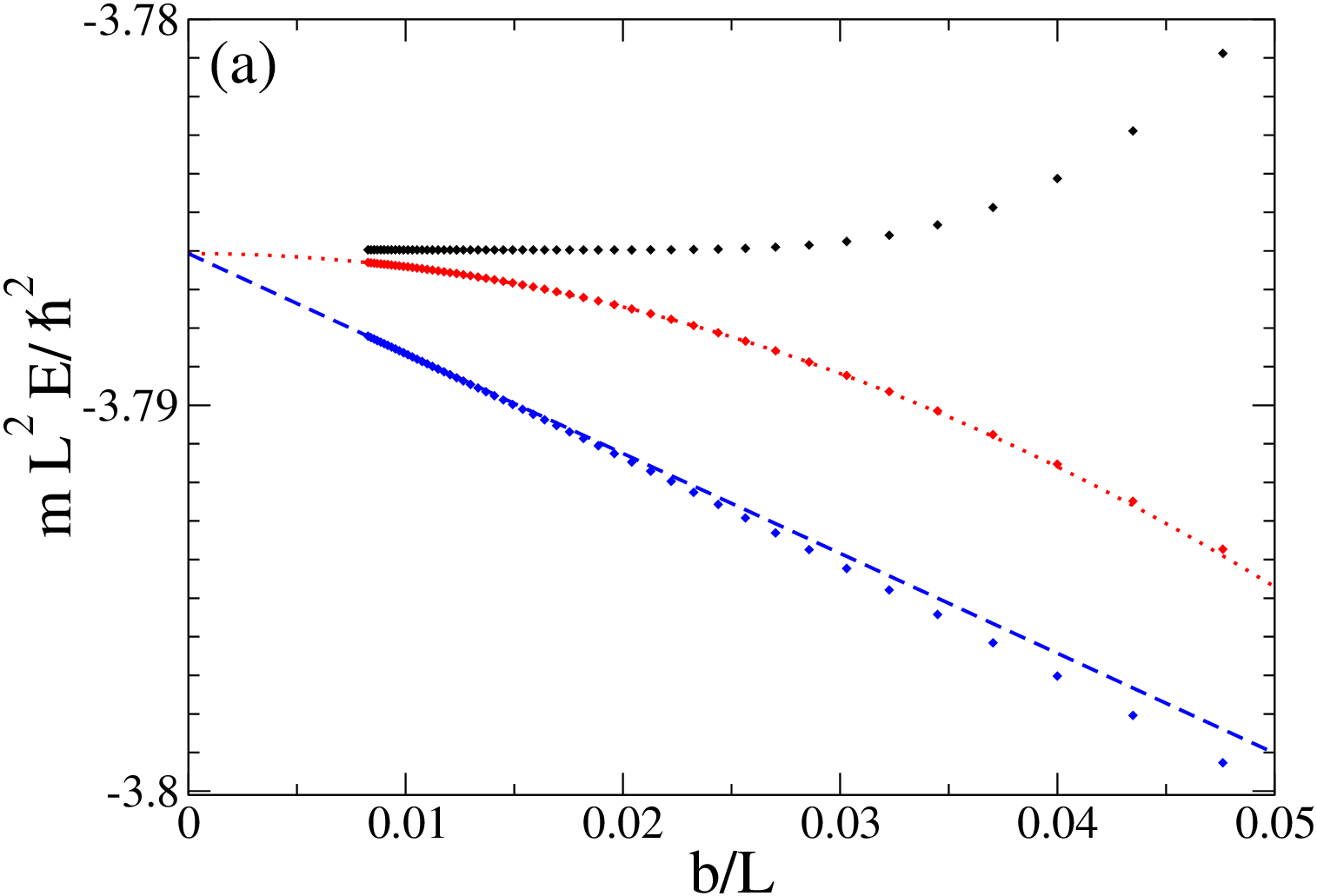}
\includegraphics[width=0.4\linewidth,clip=]{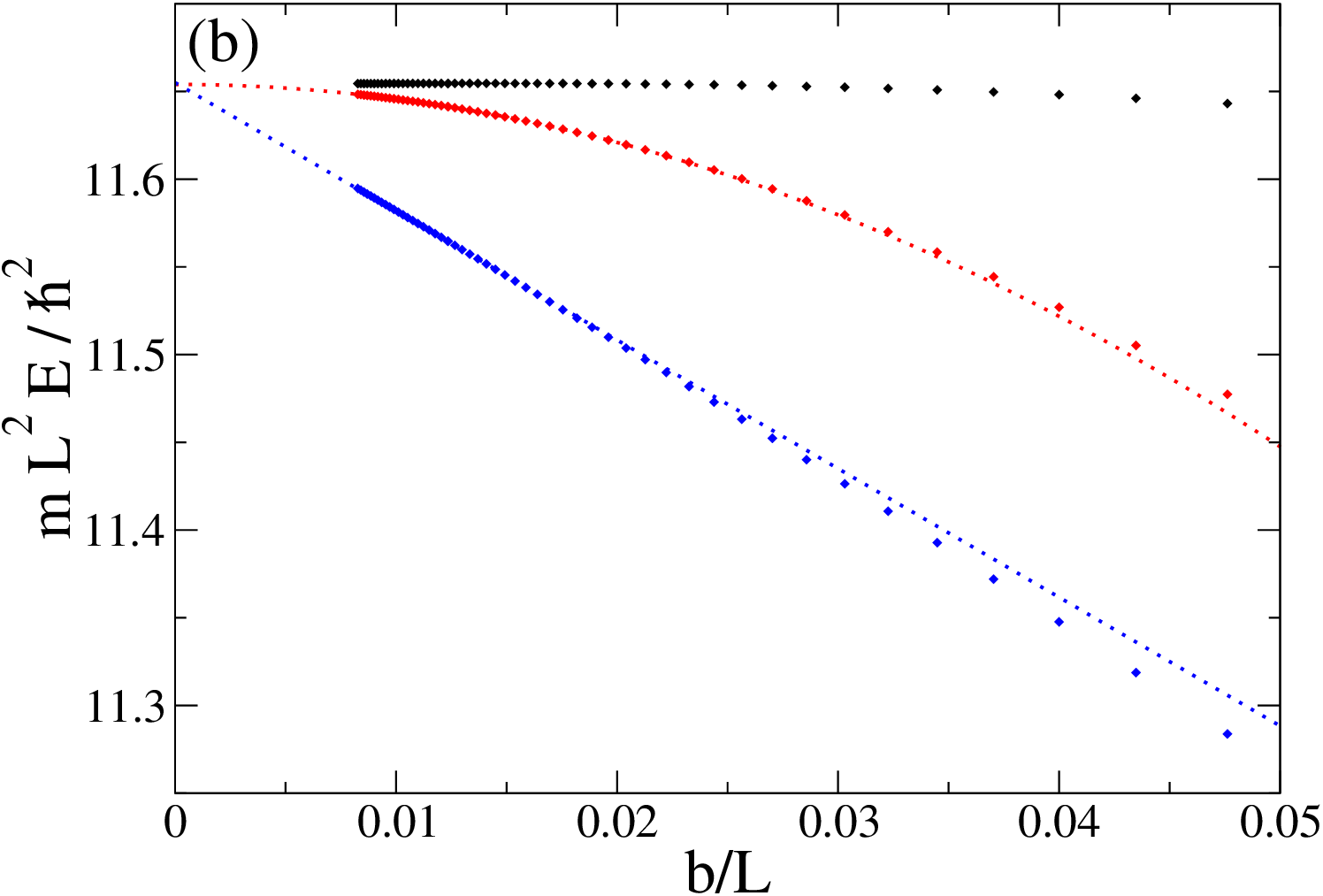}
\caption{(Color online) Illustration of the Juillet effect for the lattice model: In the cubic box $[0,L]^3$ with periodic
boundary conditions, ground state energy of two opposite spin fermions as a function of the grid spacing $b$,
for an infinite scattering length ($1/a=0$), for a total momentum equal to $\mathbf{0}$ in (a) and  equal to
$\frac{2\pi\hbar}{L}\mathbf{e}_z$ in (b). Three dispersion relations $\epsilon_\kk$ are considered,
the quartic one of Eq.~(\ref{eq:disperrez}) with zero effective range $r_e=0$ (in blue, lower set), and the magic one 
(\ref{eq:magic}) with $r_e=R_e=0$  
with the parameters of Eq.~(\ref{eq:magic1}) (in black, upper set) and of Eq.~(\ref{eq:magic2}) (in red, middle set).
The fact that the energy varies linearly in $b$ for the $r_e=0$ quartic dispersion relation at zero total
momentum is the Juillet effect explained in Sec.~\ref{subsec:juillet}, and the corresponding dashed
line is the analytical result (\ref{eq:analyE1}). At non-zero total momentum
the quartic dispersion relation leads to an energy variation linear in $b$ 
as expected e.g.\ from the fact that its has a non-zero $R_e$ [the dotted line is a linear fit for $b/L\leq 0.01$].
The magic dispersion relations
lead to a $O(b^2)$ variation of the energy both at zero and non-zero total momentum [the dotted lines are purely quadratic fits
performed for $b/L\leq 0.02$].
\label{fig:juillet}
}
\end{figure*}

With the lattice dispersion relation $\epsilon_\kk$ of
(\ref{eq:disperrez}), adjusted to have a
zero effective range $r_e=0$, Olivier Juillet numerically
observed, for two particles in the cubic box $[0,L]^3$ with 
periodic boundary conditions and zero total momentum, 
that the first energy correction to
the zero-range limit $b\to 0$ is linear in $b$
\cite{Juillet_juillet},
which seems to contradict [Tab.~V, Eq.~(1a)].  This is illustrated in Fig.~\ref{fig:juillet}.
This cannot be explained by a non-zero $R_e$ [defined in Eq.~(\ref{eq:Re})]
because the two opposite-spin fermions have here a zero total momentum.

This Juillet effect, as we shall see, is due to the fact that
the {\sl integral} of $1/\epsilon_\kk$ over $\kk$ in the first Brillouin 
zone and the corresponding {\sl discrete sum} for the finite size quantization
box differ for $b/L\to 0$ not only by a constant term but also by a term linear in $b$, when the dispersion relation
has a cusp at the surface of the first Brillouin zone, such
as Eq.~(\ref{eq:disperrez}).
The Juillet effect thus disappears in the thermodynamic limit.
This explains why it does not show up in the diagrammatic
point of view of Sec.~\ref{subsec:wwlftdpov}, which was considered
in the thermodynamic limit, so that only momentum integrals appeared.
This also shows that the Juillet effect does not invalidate [Tab.~V, Eq.~(1a)] 
since it was derived for an interaction that is smooth in momentum space.

In \cite{boite} it was shown that the lattice model spectrally
reproduces the zero-range model when the grid spacing $b\to 0$.
We now simply extend the reasoning of \cite{boite} for two particles to first order in $b$
included. For an eigenenergy $E$ which does not belong to the non-interacting
spectrum, the exact implicit equation is
\be
\frac{1}{g_0} + \frac{1}{L^3} \sum_{\kk\in D} \frac{1}{2\epsilon_{\kk}-E}=0
\ee
where the notation with a discrete sum over $\kk$ implicitly
restricts $\kk$ to $\frac{2\pi}{L} \mathbb{Z}^3$.
By adding and subtracting terms, and using the expressions (\ref{eq:g0_3D}) and (\ref{eq:re_reseau}) for the
bare coupling constant $g_0$ and the effective range $r_e$, one obtains
the useful form:
\begin{multline}
\frac{1}{g} -\frac{m^2 E r_e}{8\pi \hbar^4} 
+\frac{1}{L^3} \Bigg[-\frac{1}{E} +
\!\!\sum_{\kk\in D^*} F(\epsilon_\kk) 
+\!\! \sum_{\kk\in{\mathbb{R}^{3}}^*} \frac{E}{(\hbar^2 k^2/m)^2} 
\Bigg] \\
=
R_1 + E R_2 - E R_3
\label{eq:juillet_utile}
\end{multline}
with $g=4\pi\hbar^2 a/m$ and $F(\epsilon)=(2\epsilon-E)^{-1} -(2\epsilon)^{-1}
-E/(2\epsilon)^2$. We have defined
\be
R_1 \equiv \int_D \frac{d^3k}{(2\pi)^3} \frac{1}{2\epsilon_\kk}
-\frac{1}{L^3} \sum_{\kk\in D^*} \frac{1}{2\epsilon_\kk},
\label{eq:R1}
\ee
proportional to the function $C(b)$ introduced in \cite{boite}.
The quantities $R_2$ and $R_3$ have the same structure:
$R_2$ is obtained by replacing in $R_1$ the function
$1/(2\epsilon_\kk)$ by $1/(2\epsilon_\kk)^2-
1/(\hbar^2 k^2/m)^2$, in the integral and in the sum;
$R_3$ is obtained by replacing in $R_1$ the function
$1/(2\epsilon_\kk)$ by $1/(\hbar^2 k^2/m)^2$ 
and the set $D$ by $\mathbb{R}^3\setminus D$, both for the integration
and for the summation.

We now take $b\to 0$ in Eq.~(\ref{eq:juillet_utile}), keeping terms
up to $O(b)$ included. Since $F(\epsilon)=O(1/\epsilon^3)$ at large
$\epsilon$, we can replace $F(\epsilon_{\kk})$ by its $b\to 0$ limit
$F(\hbar^2 k^2/2m)$, and the summation set $D^*$ by its $b\to 0$ limit
\footnote{One has $\epsilon_\kk= \frac{\hbar^2 k^2}{2m} [1+O(k^2 b^2)]$.
For the finite number low energy terms, we directly use this fact.
For the other terms, such that $\epsilon_\kk \gg |E|$ and 
$\gg (2\pi\hbar)^2/(mL^2)$, we use 
$F(\epsilon_\kk)-F(\frac{\hbar^2 k^2}{2m})\simeq
(\epsilon_\kk-\frac{\hbar^2 k^2}{2m}) 
F'(\frac{\hbar^2 k^2}{2m})=O(b^2/k^4)$ which is integrable at 
large $k$ in $3D$ and leads to a total error $O(b^2)$.}:
\be
\sum_{\kk\in D^*} F(\epsilon_\kk) \underset{b\to 0}{=}
\sum_{\kk\in {\mathbb{R}^3}^*} F\left(\frac{\hbar^2 k^2}{2m}\right) 
+O(b^2)
\ee
In the quantities $R_i$, we perform the change of variables 
$\kk=2\pi \qq/b$, and we write the dispersion relation as 
\be
\epsilon_{\kk} = \frac{(2\pi \hbar)^2}{m b^2} \eta_{\kk b/(2\pi)}
\label{eq:gdr}
\ee
where the dimensionless $\eta_{\qq}$ does not depend on the lattice spacing $b$.
We then find that $b R_1$, $R_2/b$ and $R_3/b$ are 
differences between a converging integral
and a three-dimensional Riemann sum with a vanishing cell volume $(b/L)^3$.
As these differences vanish as $O(b)$, we conclude that $R_2=O(b^2)$ 
and $R_3=O(b^2)$ can be neglected in Eq.~(\ref{eq:juillet_utile}).
This however leads only to $R_1=O(1)$, so that more mathematical work
needs to be done, as detailed in the Appendix~\ref{app:R1}, to obtain
\be
\frac{\hbar^2}{m} L R_1\underset{b\to 0}{=} \frac{\mathcal{C}}{4\pi^2}
+\frac{\pi R_e^{\rm surf}}{2 L}
+O(b/L)^2.
\label{eq:devR1}
\ee
The numerical constant $\mathcal{C}\simeq 8.91363$
was calculated and called $C(0)$ in \cite{boite}.
$R_e^{\rm surf}$ remarkably is the surface contribution to the quantity
$R_e$ in Eq.~(\ref{eq:Re}), it scales as $b$. 
It is non-zero only when the dispersion
relation has a cusp at the surface of the first Brillouin zone.
In this case, $R_1$ varies to first order in $b$, which comes in addition
to the expected linear contribution of the $E r_e$ term in 
Eq.~(\ref{eq:juillet_utile}): This leads to the Juillet effect.
More quantitatively, the first deviation of the eigenenergy from its zero-range limit $E^0$,
shown as a dashed line in Fig.~\ref{fig:juillet}a,
is \footnote{The contribution proportional to $r_e$ in 
Eq.~(\ref{eq:analyE1}) can also be obtained from [Tab.~V, Eq.~(1a)] 
and from the fact that $\sum_{\kk\neq\mathbf{0}} e^{i\kk\cdot\rr}/k^2 \sim L^3/(4\pi r)$
for $r\to 0$.}:
\be
E-E^0 \underset{b\to 0}{\sim} \frac{\frac{m^2 E^0 r_e}{8\pi\hbar^4}+\frac{m\pi R_e^{\rm surf}}{2\hbar^2 L^2}}
{\frac{1}{L^3}\sum_{\kk\in \mathbb{R}^3}\frac{1}{\left(\frac{\hbar^2 k^2}{m}-E^0\right)^{2}}}
\label{eq:analyE1}
\ee

\subsection{Link between $\partial E/\partial r_e$ and the subleading short distance behavior
of the pair distribution function}
\label{subsec:re_dans_g2}

As shown by [Tab.~II, Eqs.~(3a,3b)] the short distance behavior of the pair distribution function
(averaged over the center of mass position of the pair) diverges as $1/r^2$ in $3D$ and as $\ln^2 r$ in $2D$,
with a coefficient proportional to $C$, that is related to the derivative of the energy
with respect to the scattering length $a$. Here we show that a subleading term in this short distance
behavior is related to the derivative of the energy with respect to the effective range $r_e$.
To this end, we explicitly write the next order term in the contact conditions [Tab.~I, Eqs.~(1a,1b)].

\noindent \underline{\sl Three dimensions:}
Including the next order term in [Tab.~I, Eq.~(1a)] gives
\begin{multline}
\label{eq:psios}
\psi(\rr_1,\ldots,\rr_N)\underset{r_{ij}\to0}{=}
\left( \frac{1}{r_{ij}}-\frac{1}{a} \right) \, A_{ij}\left(
\mathbf{R}_{ij}, (\mathbf{r}_k)_{k\neq i,j}
\right) \\
+ r_{ij} \ B_{ij} \left(\mathbf{R}_{ij}, (\mathbf{r}_k)_{k\neq i,j}\right) 
+ \sum_{\alpha=1}^{3} r_{ij,\alpha} L_{ij}^{(\alpha)}\left(\mathbf{R}_{ij}, (\mathbf{r}_k)_{k\neq i,j}\right)  \\
+O(r_{ij}^2)
\end{multline}
where we have distinguished between a {\sl singular} part linear with the interparticle distance
$r_{ij}$ and a {\sl regular} part linear in the relative coordinates of $i$ and $j$
($r_{ij,\alpha}$ is the component along axis $\alpha$ of the vector $\rr_{ij}$).
Injecting this form into Schr\"odinger's equation, keeping the resulting $\propto 1/r_{ij}$ terms and
using notation [Tab.~V, Eq.~(2)] gives
\be
B_{ij}(\RR_{ij},(\rr_k)_{k\neq i,j})= -\frac{m}{2\hbar^2} (E-\mathcal{H}_{ij}) A_{ij}\left(
\mathbf{R}_{ij}, (\mathbf{r}_k)_{k\neq i,j}\right)
\ee
[Tab.~V, Eq.~(1a)] thus becomes 
\be
\frac{\partial E}{\partial r_e} = -\frac{4\pi\hbar^2}{m} (A,B)
\ee
We square (\ref{eq:psios}) and as in Sec.~\ref{sec:g2} we integrate over $\RR_{ij}$,
the $\rr_k$'s and we sum over $i<j$. We further average $G^{(2)}_{\uparrow\downarrow}(\rr)$
over the direction of $\rr$ to eliminate the contribution of the regular term $L_{ij}$,
defining $\bar{G}^{(2)}_{\uparrow\downarrow}(\rr)=[G^{(2)}_{\uparrow\downarrow}(\rr)+G^{(2)}_{\uparrow\downarrow}(-\rr)]/2$.
We obtain [Tab.~V, Eq.~(3a)].

\noindent \underline{\sl Two dimensions:} Including next order terms in [Tab.~I, Eq.~(1b)] gives
\footnote{ From Schr\"odinger's equation, $\Delta_{\rr_{ij}}\psi$ diverges at most as $\psi$ itself,
that is as $\ln r_{ij}$, for $r_{ij}\to 0$.
The particular solution $f(r)=\frac{1}{4}r^2 (\ln r-1)$ of
$\Delta_{\rr} f(r)=\ln r$ fixes the form of the subleading term in $\psi$.}:
\begin{multline}
\label{eq:psios2d}
\psi(\rr_1,\ldots,\rr_N)\underset{r_{ij}\to0}{=}
\ln(r_{ij}/a) \, A_{ij}\left(
\mathbf{R}_{ij}, (\mathbf{r}_k)_{k\neq i,j}
\right) \\
+ r^2_{ij} \ln r_{ij} \ B_{ij} \left(\mathbf{R}_{ij}, (\mathbf{r}_k)_{k\neq i,j}\right)
+ \sum_{\alpha=1}^{2} r_{ij,\alpha} L_{ij}^{(\alpha)}\left(\mathbf{R}_{ij}, (\mathbf{r}_k)_{k\neq i,j}\right) \\
+O(r_{ij}^2)
\end{multline}
Proceeding as in $3D$ we obtain 
\be
B_{ij}(\RR_{ij},(\rr_k)_{k\neq i,j})= -\frac{m}{4\hbar^2} (E-\mathcal{H}_{ij}) A_{ij}\left(
\mathbf{R}_{ij}, (\mathbf{r}_k)_{k\neq i,j}\right)
\ee
[Tab.~V, Eq.~(1b)] thus becomes
\be
\frac{\partial E}{\partial (r_e^2)} = -\frac{4\pi\hbar^2}{m} (A,B)
\ee
These equations  finally leads to [Tab.~V, Eq.~(3b)].

\subsection{Link between $\partial E/\partial r_e$ and the $1/k^6$ subleading tail of
the momentum distribution}
\label{subsec:unsurk6}

A general idea given in \cite{Olshanii_nk} is that singular terms in the dependence of $\psi$ on 
the interparticle distance $r_{ij}$ (at short distances) reflect into power-law tails in the momentum 
distribution $n_{\sigma}(\kk)$ given by Eq.~(\ref{eq:nk_1e_quant}). In Sec.~\ref{sec:C_nk}, we restricted to the leading order.
Here we include the subleading term and we perform the same reasoning as in Sec.~\ref{sec:C_nk} to obtain
\footnote{In $3D$ we used the identity $\int d^3 r e^{i\kk\cdot\rr}/r = 4\pi/k^2$ and its derivatives with respect
to $k_\alpha$; e.g.\ taking the Laplacian with respect to $\kk$ gives $\int d^3 r\, e^{i\kk\cdot\rr} r = -8\pi/k^4$.
Equivalently, one can use the relation $\int d^3r e^{i\kk\cdot\rr} \frac{u(r)}{r} = \frac{4\pi}{k^2} u(0)
-\frac{4\pi}{k^4} u^{(2)}(0)+ O(1/k^6)$ and its derivatives with respect to $k_\alpha$;
this relations holds for any $u(r)$ which has a series expansion in $r=0$ and 
rapidly decreases at $\infty$. In $2D$ for $k>0$ we used the identity 
$\int d^2 r e^{i\kk\cdot\rr} \ln r = -2\pi/k^2$ and its derivatives with respect
to $k_\alpha$. The regular terms involving $L_{ij}^{(\alpha)}$
have (as expected) a negligible contribution to the tail of $n_\sigma(\kk)$.}
\footnote{The configurations with three close particles contribute to the tail of $n_{\sigma}(\kk)$ 
as $1/k^{5+2s}$, see a note of \cite{TanLargeMomentum}, with $s$ defined in Sec.~\ref{subsec:appl_3body}, which is negligible
for $s>1/2$.}
\be
\bar{n}_\sigma(k) \underset{k\to\infty}{=}  \frac{C}{k^4} + \frac{D}{k^6}+ \ldots
\ee
where $\bar{n}_\sigma(k)=\frac{1}{d}\sum_{i=1}^{d}  n_\sigma(k\uu_i)$ 
and $D$ is the linear combination of $\partial E/\partial r_e$ and $(A, \Delta_\RR A)$ given in
[Tab.~V, Eqs.~(4a,4b)].
Physically, the extra term $(A, \Delta_\RR A)$ results from the fact that the wavevector $\kk_1$ of a particle in an $\uparrow\downarrow$  colliding pair is a linear
combination of the relative wavevector $\kk_{\rm rel}$  and of the center-of-mass wavevector $\KK$ of the pair, so that, even
if the probability distribution of $\kk_{\rm rel}$ was exactly scaling as $1/k_{\rm rel}^4$, a non-zero $\KK$ would generate
a subleading $1/k_1^6$ contribution in the single particle momentum distribution.

This is apparent for the simple case of a free space dimer:
When the dimer is at rest, $\psi(\rr_1,\rr_2)=\phi_{\rm dim}(r_{12})$, $A_{12}(\RR_{12})$ is uniform and the extra term vanishes. 
When it has a momentum $\KK$, $\psi(\rr_1,\rr_2)=e^{i\KK\cdot \RR_{12}} \phi_{\rm dim}(r_{12})$, 
which shifts the single particle momentum distribution, $n_\uparrow^{\rm mov}(\kk)=n_\uparrow^{\rm rest}(\kk-\KK/2)$. Applying this shift 
to the momentum tail $C/k^4$ gives, after continuous average over the direction of $\kk$, a subleading $\delta D^{\rm mov}/k^6$ contribution,
with $\delta D^{\rm mov}=C K^2/2$ in $3D$ and $\delta D^{\rm mov}=C K^2$ in $2D$. Remarkably, the ratio of the extra term to
$C$ is proportional to the pair-center-of-mass kinetic energy.

In the $N$-body case, one can generalize this property by defining the mean center-of-mass kinetic
energy of a $\uparrow\downarrow$ pair at {\sl vanishing} pair diameter, which is allowed in quantum mechanics
since the center-of-mass operators and the relative-particle operators commute
\footnote{Similarly, a ``contact current'' was recently introduced in \cite{Tan2011}, 
whose spatial integral is proportional to $(A, \mathbf{\nabla}_\RR A)$.}. 
By a direct generalisation of the pair distribution function
of Sec.~\ref{sec:g2}, one has for the opposite-spin pair density operator
$\langle \rr_\uparrow,\rr_\downarrow|\hat{\rho}_{\uparrow\downarrow}^{(2)}|\rr_\uparrow',\rr_\downarrow'\rangle=
\langle\hat{\psi}_\uparrow^\dagger(\rr_\uparrow')\hat{\psi}^\dagger_\downarrow(\rr_\downarrow')\hat{\psi}_\downarrow(\rr_\downarrow)\hat{\psi}_\uparrow
(\rr_\uparrow)\rangle$. Whereas the usual pair-center-of-mass density operator is obtained by taking the trace over the
relative coordinates $\rr=\rr_\uparrow-\rr_\downarrow$, we rather define it here by taking the limit of vanishing relative coordinates,
\be
\langle \RR|\hat{\rho}_{\rm CoM}^{(2)}|\RR'\rangle = \mathcal{N} \lim_{r\to 0} 
\frac{\langle \RR+\frac{\rr}{2}, \RR-\frac{\rr}{2} | \hat{\rho}_{\uparrow\downarrow}^{(2)} 
|\RR'+\frac{\rr}{2},\RR'-\frac{\rr}{2}\rangle}{\phi^2(\rr)}
\ee
where the factor $\mathcal{N}$ is such that $\hat{\rho}_{\rm CoM}^{(2)}$ has a unit trace and $\phi(\rr)$ is the zero-energy
scattering state of Eqs.(\ref{eq:normalisation_phi_tilde_3D},\ref{eq:normalisation_phi_tilde_2D}).
Proceeding as in Sec.~\ref{sec:g2} we obtain
\begin{multline}
\langle \RR|\hat{\rho}_{\rm CoM}^{(2)}|\RR'\rangle = \mathcal{N} \sum_{i<j} \int (\prod_{k\neq i,j}d^dr_k) A_{ij}^*(\RR',
(\rr_k)_{k\neq i,j}) \\ \times A_{ij}(\RR,(\rr_k)_{k\neq i,j})
\end{multline}
By taking the expectation value of $-(\hbar^2/4m)\Delta_{\RR}$ within $\hat{\rho}_{\rm CoM}^{(2)}$, we finally obtain for the mean pair-center-of-mass kinetic energy 
at vanishing diameter:
\be
E_{\rm kin\ pair-CoM}^{r_{\uparrow\downarrow}\to 0} = -\frac{\hbar^2}{4m} \frac{(A,\Delta_{\RR} A)}{(A,A)}
\ee
where the denominator is  $\propto C$, see [Tab.~II, Eqs.~(2a,2b)].

\section{Generalization to arbitrary statistical mixtures} \label{sec:stat_mix}

In this section, we generalize some of the relations derived in the previous sections for pure states to the case of arbitrary statistical mixtures.
Let us first discuss zero-range interactions.
We consider a statistical mixture of pure states $\psi_n$ with occupation probabilities $p_n$, which is arbitrary, but non-pathological in the following sense~\cite{TanEnergetics}:
Each $\psi_n$ satisfies the contact condition
[Tab. I, Eqs. (1a,1b)];
moreover, $p_n$ decays sufficiently quickly at large $n$ so that
we have $\ds C=\sum_n p_n C_n$,
where 
$C_n$ (resp. $C$) is defined by [Tab. II, Eq.~(1)] with $n_\sigma(\kk)=\la c^\dagger_\sigma(\kk)c_\sigma(\kk)\ra$ and $\la\,.\,\ra=\la\psi_n|\,.\,|\psi_n\ra$ (resp. $\la\,.\,\ra=\sum_n p_n \la\psi_n|\,.\,|\psi_n\ra$).
Then, the relations in lines 3, 5, 6 and 7 of Table II, which were derived in Sec.~\ref{sec:ZR} for any pure state satisfying the contact conditions, obviously generalize to such a statistical mixture.
The relations for the time derivative of $E$ (Tab. \ref{tab:fermions} line 12) hold for any time-evolving pure state satisfying the contact conditions for a time-dependent $a(t)$, and thus also for any statistical mixture of such time-evolving pure states.

For lattice models, one can obviously take an average of 
the definition of $\hat{C}$~[Tab.~III, Eqs.~(1a,1b)] to define $C=\la\hat{C}\ra$ for in any statistical mixture;
taking averages of the relations between operators~[Tab.III, lines 2,3,8] then gives relations valid for any statistical mixture.

\section{Thermodynamic equilibrium in the canonical ensemble} \label{subsec:finiteT}

We turn to the case of thermal
equilibrium in the canonical ensemble. We shall use the notation
\be
\lambda\equiv\left\{
\begin{array}{lr}
-1/a & {\rm in}\ 3D
\\
\frac{1}{2}\ln a & {\rm in}\ 2D.
\end{array}
\right.
\label{eq:def_lambda}
\ee

\subsection{First order derivative of $E$}

The thermal average in the canonical ensemble $\overline{dE/d\lambda}$ 
can be rewritten in the following  more familiar way,
as detailed in Appendix \ref{app:adiab}:
\be
\overline{\left(\frac{dE}{d\lambda}\right)}
=
\left(
\frac{dF}{d\lambda}
\right)_{\!T}
=
\left(
\frac{d\bar{E}}{d\lambda}
\right)_{\!S}
\label{eq:relation_T}
\ee
where $\overline{(\dots)}$ is the canonical thermal average,
$F$ is the free energy
and $S$ is the entropy.
Taking the thermal average of [Tab.~\ref{tab:fermions}, Eqs.~(4a,4b)]
(which was shown above for any stationary state)
 thus gives~[Tab.~\ref{tab:fermions}, Eqs.~(9a,9b)].

\subsection{Second order derivative of $E$}

Taking a thermal average of the line 8 in Tab.~\ref{tab:fermions} we get after a simple manipulation:
\begin{multline}
\label{eq:manipulation}
\frac{1}{2} \overline{\left( \frac{d^2 E}{d\lambda^2}\right)} = \left( \frac{4\pi\hbar^2}{m}\right)^2
 \frac{1}{2 Z}\sum_{n,n';E_n\neq E_{n'}}\frac{e^{-\beta E_n}-e^{-\beta E_{n'}}}{E_n-E_{n'}} \\
\times |(A^{(n')},A^{(n)})|^2
\end{multline}
where $Z=\sum_n \exp(-\beta E_n)$.
This implies
\be
 \overline{\left( \frac{d^2 E}{d\lambda^2}\right)}<0.
 \label{eq:d2Ebar<0}
 \ee
Moreover one can check that
\be
\left(\frac{d^2F}{d\lambda^2}\right)_T - \overline{\left( \frac{d^2 E}{d\lambda^2}\right)} = -\beta\left[\,
\overline{\left( \frac{dE}{d\lambda}\right)^{\phantom{,}2}}
- \overline{\left( \frac{dE}{d\lambda}\right)}^{\phantom{,}2} \right]<0,
\ee
which implies [Tab.~II, Eqs.~(10a,10b)].
In usual cold atom experiments, however, there is no thermal reservoir imposing
a fixed temperature to the gas, one rather can achieve
adiabatic transformations by a slow variation of the scattering
length of the gas 
\cite{SalomonMolecules,HuletMolecules,GrimmCrossover} where the entropy is fixed
\cite{LandauLifschitzPhysStat,CarrCastin,WernerAF}. 
One also more directly accesses
the mean energy $\bar{E}$ of the gas rather than its
free energy, even if the entropy is also measurable \cite{thomas_entropie_PRL,thomas_entropie_JLTP}. The second order derivative of $\bar{E}$ with
respect to $\lambda$ for a fixed entropy is thus the relevant
quantity to consider.
As shown in Appendix \ref{app:adiab} 
one has in the canonical ensemble:
\be
\label{eq:d2us}
\left(\frac{d^2\bar{E}}{d\lambda^2}\right)_S
=
\overline{\left(\frac{d^2E}{d\lambda^2}\right)}
+\frac
{
\left[\mbox{Cov}\!\left(E,\frac{dE}{d\lambda}\right)\right]^2
-\mbox{Var}(E)\mbox{Var}\!\left(\frac{dE}{d\lambda}\right)
}
{k_B T\,\mbox{Var}(E)}
\ee
where $\mbox{Var}(X)$ and $\mbox{Cov}(X,Y)$ stand for the variance
of the quantity $X$ and the covariance of the quantities $X$ and $Y$
in the canonical ensemble, respectively.
From the Cauchy-Schwarz inequality 
$[\mbox{Cov}(X,Y)]^2\leq \mbox{Var}(X)\mbox{Var}(Y)$,
and from the inequality (\ref{eq:d2Ebar<0}), we thus obtain~[Tab.~II, Eqs.~(11a,11b)].

For lattice models, the inequalities~[Tab.~III, Eq.~(7)] are derived in the same way, by taking $\lambda$ now equal to $g_0$, and starting from the expression [Tab.~III, Eq.~(6)] of $d^2E_n/dg_0^2$.
For the case of a finite-range interaction potential $V(r)$ in continuous space, the relations~[Tab. IV, lines~1-3] which were derived for an arbitrary stationary state are generalized to the thermal equilibrium case in the same way.
Finally, the relations which asymptotically hold in the zero-range regime, [Tab.~III lines 9-10] for lattice models and [Tab.~IV lines 4-5] for finite-range interaction potential models, which were justified for any eigenstate in the zero-range regime $\ktyp b<<1$ where the typical relative wavevector $\ktyp$ is defined in terms of the considered eigenstate as described in Section~\ref{subsec:models},
remain true at thermal equilibrium with $\ktyp$ now defined as the typical density- and temperature-dependent wavevector described in Section~\ref{subsec:models},
since all the eigenstates which are thermally populated with a non-negligible weight are expected to have a typical wavevector smaller or on the order of the thermal-equilibrium typical wavevector.

\subsection{Quantum-mechanical adiabaticity}
To be complete, we also consider the process where 
$\lambda$ is varied so slowly that there is adiabaticity in the many-body
quantum mechanical sense:
The adiabatic theorem of quantum mechanics~\cite{AdiabThmKato} 
implies that in the limit where $\lambda$ is changed infinitely slowly, 
the occupation probabilities of each eigenspace of the many-body Hamiltonian 
do not change with time, 
even in presence of level crossings~\cite{AdiabThmAvron}. 
We note that this may require macroscopically long evolution times
for a large system.
For an initial equilibrium state in the
canonical ensemble, the mean energy then varies with $\lambda$ as
\be
E^{\rm quant}_{\rm adiab}(\lambda)=
\sum_n \frac{e^{-\beta_0 E_n(\lambda_0)}}{Z_0}\,E_n(\lambda)
\label{eq:Ead1}
\ee
where the subscript $0$ refers to the initial state.
Taking the second order derivative of (\ref{eq:Ead1}) with respect
to $\lambda$ in $\lambda=\lambda_0$ gives
\be
\frac{d^2 E_{\rm adiab}^{\rm quant}}{d\lambda^2}
= \overline{\left(\frac{d^2E}{d\lambda^2}\right)}
<0.
\label{eq:d^2E_quant}
\ee
Note that the sign of the second order derivative of $E_{\rm adiab}^{\rm quant}$ remains negative
at all $\lambda$ provided one assumes that there is no level crossing in the many-body spectrum
when $\lambda$ is varied: $E_n(\lambda)-E_{n'}(\lambda)$ has the same sign
as $E_n(\lambda_0)-E_{n'}(\lambda_0)$ for all indices $n,n'$, which allows to conclude on the sign
with the same manipulation as the one having led to Eq.~(\ref{eq:manipulation}).

\noindent{\sl Thermodynamic vs quantum adiabaticity:}
The result of the isentropic transformation~(\ref{eq:d2us}) 
and the one of the adiabatic transformation in the quantum 
sense~(\ref{eq:d^2E_quant}) differ by the second term in the right hand 
side of~(\ref{eq:d2us}). A priori this term is extensive, and thus not negligible as compared to the first term. 
We have explicitly checked this expectation 
for the Bogoliubov model Hamiltonian of a weakly interacting Bose gas,
which is however not really relevant since this Bogoliubov model
corresponds to the peculiar case of an integrable dynamics.

For a quantum ergodic system we now show that the second term in the right hand side of (\ref{eq:d2us})
is negligible in the thermodynamic limit, as a consequence of the 
Eigenstate Thermalization Hypothesis  \cite{Deutsch1991,Srednicki1994,Srednicki1996,Srednicki1999}.
This Hypothesis  was tested numerically for several interacting quantum systems \cite{Olshanii,Rigol2009a,Rigol2009b}.
It states that, for a large system, the expectation value $\langle \psi_n|\hat{O}|\psi_n\rangle$ of a few-body observable $\hat{O}$ in a single
eigenstate $|\psi_n\rangle$ of energy $E_n$ can be identified with the microcanonical average $O_{\rm mc}(E_n)$ of $\hat{O}$
at that energy. Here the relevant operator $\hat{O}$ is the two-body observable (the so-called
{\sl contact operator}) such that $\frac{d}{d\lambda} E_n = \langle \psi_n| \hat{O}|\psi_n\rangle$.
In the canonical ensemble, the energy fluctuations scale as $\mathcal{V}^{1/2}$ where $\mathcal{V}$ is the system volume.
We can thus expand the microcanonical average around the mean energy $\bar{E}$:
\be
O_{\rm mc}(E) = O_{\rm mc}(\bar{E}) + (E-\bar{E}) O'_{\rm mc}(\bar{E}) + O(1)
\ee
To leading order, we then find that $\mbox{Cov}\!\left(E,\frac{dE}{d\lambda}\right)\sim O_{\rm mc}'(\bar{E})\mbox{Var}\, E$
and $\mbox{Var}\!\left(\frac{dE}{d\lambda}\right)\sim [O_{\rm mc}'(\bar{E})]^2\mbox{Var}\, E$,
so that the second term in the right hand side of ~(\ref{eq:d2us}) is $O(\mathcal{V}^{1/2})$
which is negligible as compared to the first term in that right hand side.
For the considered quantity, this shows the equivalence of the thermodynamic adiabaticity and of the quantum adiabaticity 
for a large system.

\noindent{\sl A microcanonical detour:} 
We now argue that the quantum adiabatic expression (\ref{eq:Ead1}) for the mean energy as a function of the slowly varying parameter
$\lambda$ can be obtained by a purely thermodynamic reasoning.
This implies that the exponentially long evolution times {\sl a priori} required to reach the quantum
adiabatic regime for a large system are actually not necessary to obtain (\ref{eq:Ead1}).
The first step is to realize that the initial canonical ensemble (for $\lambda=\lambda_0$) can be viewed
as a statistical mixture of microcanonical ensembles \cite{SinatraWitkowskaCastin}. These microcanonical ensembles correspond
to non-overlapping energy intervals of width $\Delta$, each interval contains many eigenstates,
but $\Delta$ is much smaller than the width of the probability distribution of the system energy in the canonical ensemble.
For further convenience, we take $\Delta \ll k_B T$.
One can label each energy interval by its central energy value, or more conveniently
by its entropy $S$.
If the eigenenergies $E_n(\lambda)$ are numbered in ascending order, the initial microcanonical ensemble of entropy
$S$ contains the eigenenergies with $n_1(S) \leq n <  n_2(S)$ and $S=k_B\ln[n_2(S)-n_1(S)]$. When $\lambda$ is slowly varied, the entropy
is conserved for our isolated system, and the microcanonical ensemble simply follows the evolution of the initial $n_2(S)-n_1(S)$ eigenstates,
which cannot cross for an ergodic system and remain bunched in energy space.
Furthermore, according to the Eigenstate Thermalization Hypothesis, the energy width $E_{n_2}-E_{n_1}$ remains close to its initial 
value $\Delta$: Each eigenenergy varies with a macroscopically large slope $dE_n/d\lambda$ but all the eigenenergies in the microcanonical
ensemble have essentially the same slope \footnote{One has $\frac{d}{d\lambda}(E_{n_2}-E_{n_1})= O_{\rm mc}(E_{n_2})
-O_{\rm mc}(E_{n_1})\simeq (E_{n_2}-E_{n_1})O'_{\rm mc}(E_{\rm mc}) = O(\Delta)$, where $O_{\rm mc}$ is the microcanonical expectation
value of the contact operator.}.
The mean microcanonical energy for this isentropic evolution is thus
\be
E_{\rm mc}(S,\lambda) = \frac{1}{n_2(S)-n_1(S)} \sum_{n=n_1(S)}^{n_2(S)-1} E_n (\lambda)
\label{eq:Emc}
\ee
Finally, we take the appropriate statistical mixture of the microcanonical ensembles
(so as to reconstruct the initial $\lambda=\lambda_0$ canonical ensemble): The microcanonical ensemble of entropy $S$ has
an initial central energy $E_{\rm mc}(S,\lambda_0)$, it is weighted in the statistical mixture
by the usual expression $P(S)=e^{S/k_B} e^{-\beta E_{\rm mc}(S,\lambda_0)}$.
Since $\Delta \ll k_B T$, one can identify $e^{-\beta E_{\rm mc}(S,\lambda_0)}$ with
$e^{-\beta E_n(\lambda_0)}$, for $n_1(S)\leq n< n_2(S)$. The corresponding
statistical average of (\ref{eq:Emc}) with the weight $P(S)$ gives (\ref{eq:Ead1}).

\section{Applications}\label{sec:appl}
In this Section, we apply some of the above relations in three dimensions, first to the two-body and three-body problems 
and then to the many-body problem. 
Except for the two-body case, we restrict to the infinite scattering length case $a=\infty$ in three dimensions.

\subsection{Two-body problem in a harmonic trap: Finite range corrections}
\label{subsec:appli_deux_corps}
Two particles interact with the compact-support potential $V(r_{12};b)$ of range $b$ and scattering length $a$
in an isotropic harmonic potential $U(\rr_i)=\frac{1}{2}m\omega^2 r_i^2$.
One separates out the center of mass, in an eigenstate of energy $E_{\rm cm}$.
The relative motion is taken with zero angular momentum; its wavefunction  $\psi(r)$ is an eigenstate of energy $E_{\rm rel}=E-E_{\rm cm}$
for a particle of mass $\mu=m/2$ in the potential $V(r;b)+\mu \omega^2 r^2/2$.
We take in this subsection $\hbar\omega$ as the unit of energy and $[\hbar/(\mu\omega)]^{1/2}$ as the unit of length. 
For $r\geq b$ the solution may be expressed in terms of the Whittaker function $W$, or equivalently of the Kummer function $U$, see \S 13 in \cite{Abramowitz}:
\bea
\label{eq:whit3d}
\frac{\psi(r)}{C_3}\!\!\! &\stackrel{3D}{=}&\!\!\! \frac{W_{\frac{E_{\rm rel}}{2},\frac{1}{4}}(r^2)}{r^{3/2}} =
e^{-\frac{r^2}{2}} U(\frac{3}{4}-\frac{E_{\rm rel}}{2},\frac{3}{2},r^2) \\
\frac{\psi(r)}{C_2}\!\!\! &\stackrel{2D}{=}&\!\!\! \frac{W_{\frac{E_{\rm rel}}{2},0}(r^2)}{r}= e^{-\frac{r^2}{2}} U(\frac{1-E_{\rm rel}}{2},1,r^2)
\label{eq:whit2d}
\eea
where the factors $C_2$ and $C_3$ ensure that $\psi$ is normalized to unity. 
The zero-range limit, where $V(r;b)$ is replaced by the Bethe-Peierls contact conditions at the origin, is exactly solvable; it gives eigenenergies 
$E_{\rm rel}^0$.  We give here the finite range corrections to the energy in terms of $r_e$. 

\noindent {\underline{\it Three dimensions:}} 
\\
Imposing the contact condition $\psi(r)=A[r^{-1}-a^{-1}]+O(r)$ to Eq.~(\ref{eq:whit3d}) gives an implicit equation for the spectrum in the zero-range limit,
obtained in \cite{Busch} with a different technique:
\be
f(E_{\rm rel}^0) = -\frac{1}{a} \ \ \ \mbox{with}\ \ \ f(E)\equiv -\frac{2\Gamma(\frac{3}{4}-\frac{E}{2})}{\Gamma(\frac{1}{4}-\frac{E}{2})}
\ee
We have calculated the finite range corrections up to order two in $b$ included, they remarkably involve only the effective range:
\be
E_{\rm rel} = E_{\rm rel}^0 + \frac{E_{\rm rel}^{0} r_e}{f'}+ \left(\frac{E_{\rm rel}^{0} r_e}{f'}\right)^2
\left(\frac{1}{E_{\rm rel}^{0}}-\frac{f''}{2 f'}\right) + O(b^3)
\label{eq:jusqua_re2}
\ee
where the first and second order derivatives $f'$ and $f''$ of $f(E)$ are taken in
$E=E_{\rm rel}^{0}$. To obtain this expansion, we have used the result of Appendix~\ref{app:piege_dans_interaction} 
that one can neglect, at this order,
the effect of the trapping potential for $r\leq b$, so that the wavefunction is proportional to the free space scattering
state at energy $E_{\rm rel}=\hbar^2 k^2/(2\mu)$, $\psi(r)=\mathcal{A}\chi(r)$. 
Such an approximation was already proposed in \cite{Bolda,Naidon,GaoPiege},
without analytical control on the resulting spectral error
\footnote{
We have employed two equivalent techniques.
The first one is to match in $r=b$ the logarithmic derivatives of Eq.~(\ref{eq:whit3d}) and of Eq.~(\ref{eq:norma_chi}) and to expand
their inverses up to order $b^4$ included. Due to Eq.~(\ref{eq:devuk_3D}) this involves only $r_e$.
The second one is to use relation (\ref{eq:appoz_3d}): The matching of $\mathcal{A}\chi$ with Eq.~(\ref{eq:whit3d}) in $r=b$
gives $\mathcal{A}/C_3=\frac{\pi^{1/2}}{\Gamma(\frac{3}{4}-\frac{E_{\rm rel}}{2})}[1+O(b^2)]$, and the normalization of $\psi$ to unity,
from relation 7.611(4) in \cite{Gradstein} together with the Smorodinski relation (\ref{eq:smorodinski}), 
gives $dE_{\rm rel}/dr_e$ up to order one in $b$ included, that one integrates to get the result.}.
We have checked that the term of Eq.~(\ref{eq:jusqua_re2}) linear in $r_e$ coincides with the prediction 
of [Tab.~V, Eq.~(1a)],
due to the fact that, from  relation 7.611(4) in \cite{Gradstein},
the normalization factor in the zero-range limit obeys  $(C_3^0)^2 2\pi^2 f'(E_{\rm rel}^0)/\Gamma^2(\frac{3}{4}-\frac{E_{\rm rel}^0}{2})=1$.

The term in Eq.~(\ref{eq:jusqua_re2}) linear in $r_e$ was already written explicitly in \cite{WernerThese}.
This corresponds to the first order perturbative use of the modified version of the zero-range model, as put forward in \cite{Efimov93}.
It can also be obtained by solving to first order in $r_e$ the self-consistent equation considered in \cite{Greene} 
obtained by replacing $a_0$ by $a_E$ [see Eq.~(5) of \cite{Greene}] into Eq.~(6) of \cite{Greene}. This self-consistent equation
was also introduced in \cite{Bolda}, and in \cite{GaoPiege} [see Eqs.~(11,12,30) of that reference]
with more elaborate forms for $a_E$. 
With our notations and units this self-consistent equation is simply
\be
f(E) = - u(k=\sqrt{2E})
\label{eq:autoc}
\ee
where $u(k)$ is related to the $s$-wave scattering amplitude by Eq.~(\ref{eq:introu}).
The self-consistent equation of \cite{Greene} corresponds to the choice $u(k)=\frac{1}{a}-\frac{1}{2} k^2 r_e$
in Eq.~(\ref{eq:autoc}). We have checked that solving that equation
to second order in $r_e$ then exactly gives the term of Eq.~(\ref{eq:jusqua_re2}) that is quadratic in $r_e$.
Our result of Appendix~\ref{app:piege_dans_interaction} shows that going to order three in $r_e$ with the self-consistent equation 
should not give the correct result, since one can then no longer neglect the effect of harmonic trapping
within the interaction range. This clarifies the status of that self-consistent equation.

To ascertain this statement, we have calculated the ground state relative energy up to third order included in $b$,
restricting for simplicity to an infinite scattering length, $1/a=0$
\footnote{The result is based on Appendix \ref{app:piege_dans_interaction}. The simplest calculation is as follows: One first neglects
the trapping potential for $r\leq b$, one matches the inverse of the logarithmic derivative of the scattering state
(\ref{eq:norma_chi}) for $r=b^-$ with the inverse of the logarithmic derivative of (\ref{eq:whit3d}) for $r=b^+$, and one expands
the resulting equation up to order $b^5$ included, using relations 13.1.2 and 13.1.3 in \cite{Abramowitz} for $r=b^+$. 
Then one includes the $r<b$ trapping effect by applying the usual
first order perturbation theory to the operator $\frac{1}{2}\mu \omega^2 r^2 \theta(b-r)$; at this order
the wavefunction for $r<b$ may be identified with the zero-energy scattering state $\phi(r)$.
An alternative, more complicated technique is to use $\psi^{(1)}$
of Appendix~\ref{app:piege_dans_interaction}. One finds that, up to order $b^4$ included, $\psi(b)/[-b\psi'(b)]=
u(1)/[-u'(1)] + f(1)/u(1)-f'(1)/u'(1)$, where we used the fact that $u(1)/[-u'(1)]=1$ to zeroth order in $b$
and $f(x)$ solves (\ref{eq:inhomo}).
Then from relations (\ref{eq:alphax},\ref{eq:betax}) and from the expression of $v(x)$ in terms of $u(y)$,
given above Eq.~(\ref{eq:estimations_du_chgt}), 
one finds $\psi(b)/[-b\psi'(b)]=u(1)/[-u'(1)] + \beta(1)/u^2(1)+ O(b^5)$. Matching this to the $r>b$ solution
gives (\ref{eq:devb3}).
}. We find
\begin{multline}
E_{\rm rel} = \frac{1}{2} + \frac{r_e}{2\pi^{1/2}} + \frac{2-\ln 2}{4\pi} r_e^2
+ \Big[\frac{(1-\ln 2)(2-\ln 2)}{4\pi^{3/2}}  \\
- \frac{\pi^2+12\ln^2 2}{192\pi^{3/2}} \Big] r_e^3 -\frac{\lambda_2 + \Lambda_2} {\pi^{1/2}}+O(b^4)
\label{eq:devb3}
\end{multline}
Here $\lambda_2$ is the coefficient of $k^4$ in the low-$k$ expansion of $u(k)$,
$u(k)=\frac{1}{a}-\frac{1}{2} k^2 r_e +\lambda_2 k^4+O(k^6)$,
it can be evaluated by a generalized Smorodinski relation \cite{smoro_en_prepa}.
On the contrary, $\Lambda_2$ is a new coefficient containing the effect of
the trapping potential within the interaction range. It  can be expressed in terms of 
the zero-energy free space scattering state $\phi(r)$,
normalized as in Eq.~(\ref{eq:normalisation_phi_tilde_3D}):
\be
\Lambda_2 = \int_0^{+\infty} dr\, r^2 [1-u_0^2(r)]
\ee
with $u_0(r)=r\phi(r)$. Although our derivation is for a compact support potential,
we expect that our result is applicable as long as $\lambda_2$ and $\Lambda_2$
are finite. For both quantities, this requires (for $1/a=0$) that the interaction potential drops faster
than $1/r^5$ \cite{smoro_en_prepa}.
Interestingly, if one expands the self-consistent Eq.~(\ref{eq:autoc}) up to order $b^3$ included,
one exactly recovers Eq.~(\ref{eq:devb3}), except for the term $\Lambda_2$. This was expected
from the fact that the derivation of (\ref{eq:autoc}) in \cite{GaoPiege} indeed neglects
the trapping potential within the interaction range.

\begin{figure}[t]
\includegraphics[width=0.9\linewidth,clip=]{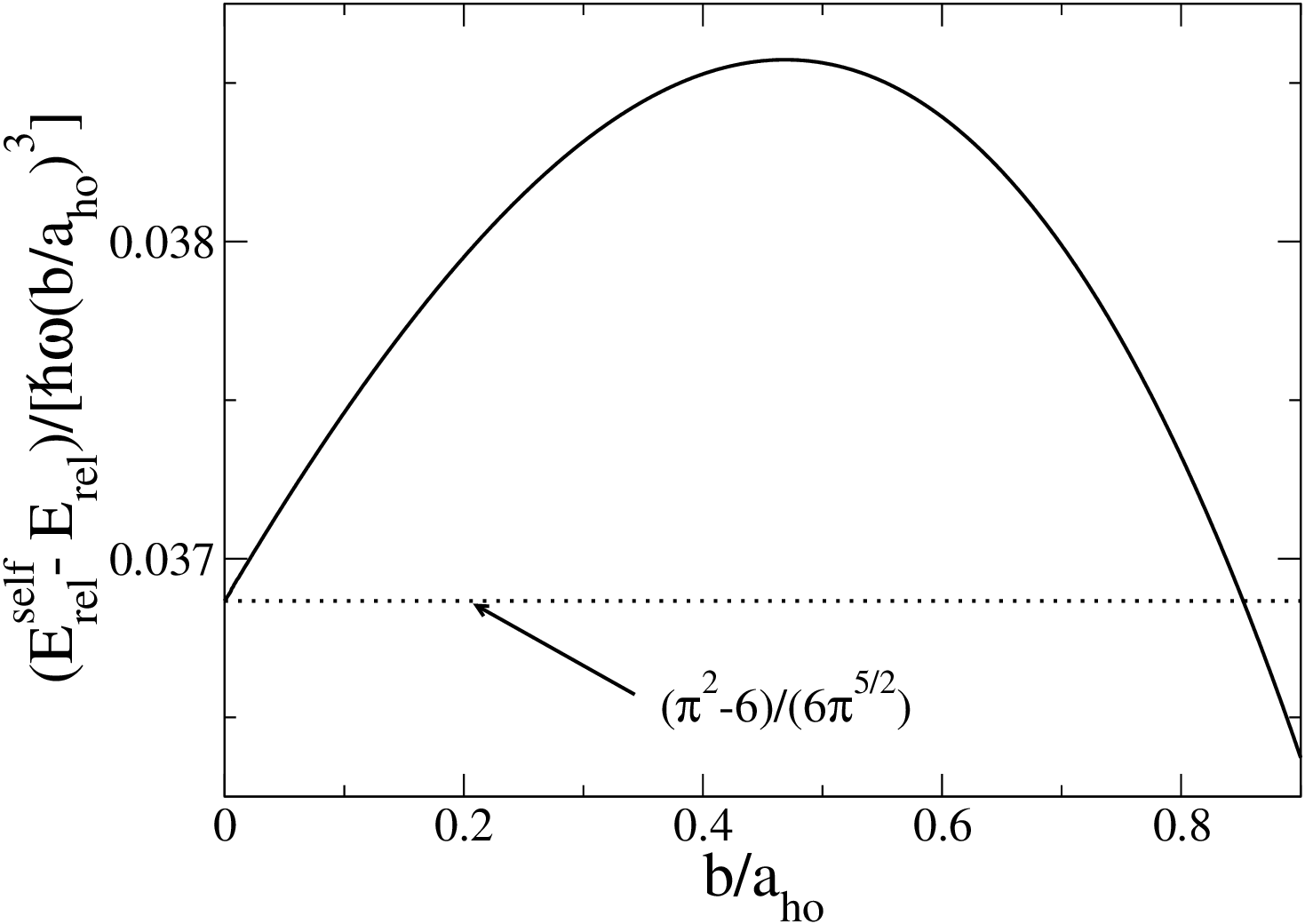}
\caption{For two opposite spin fermions interacting in $3D$ {\sl via} a potential of short range $b$ in an isotropic harmonic
trap, the self-consistent equation (\ref{eq:autoc}), derived e.g.\ in \cite{GaoPiege}, gives the eigenenergies
with an error of order $b^3$, due to the fact that it neglects the effect of the harmonic trap
within the interaction range, see Appendix~\ref{app:piege_dans_interaction}. This is illustrated with the ground state relative energy
for a square-well potential of infinite scattering length: The deviation (solid line) between the approximate energy $E_{\rm rel}^{\rm self}$
[solving Eq.~(\ref{eq:autoc})] and the exact one $E_{\rm rel}$ (calculated as in \cite{WernerThese}) vanishes
as $b^3$, with a coefficient given by Eq.~(\ref{eq:diff_autoc_exact}) (dotted line). $\mu$ is the reduced mass,
$\omega$ is the angular oscillation frequency in the trap and $a_{\rm ho}=[\hbar/(\mu \omega)]^{1/2}$.
\label{fig:autoc}}
\end{figure}

This discussion is illustrated for the particular case of the square-well potential (\ref{eq:puits_carre}) in Fig.\ref{fig:autoc}, 
with the exact spectrum obtained by matching the logarithmic derivative
of a Whittaker $M$ function for $r=b^-$  with the logarithmic derivative of a Whittaker $W$ function for $r=b^+$
as in Eqs.~(6.16,6.17,6.18) of \cite{WernerThese}
\footnote{In \cite{Pietro_r_e} a similar calculation was performed, except that the harmonic trap was neglected
within the interaction range.}. 
In this case, one finds $r_e=b$ \cite{YvanVarenna} and, remarkably, $\Lambda_2=-2 \lambda_2$ 
so that the difference between the ground state energy of (\ref{eq:autoc}) and the exact ground state energy obeys
\be
E_{\rm rel}^{\rm self} - E_{\rm rel} = \frac{\Lambda_2}{\pi^{1/2}} + O(b^4) = 
\left(\frac{1}{6}-\frac{1}{\pi^2}\right)\frac{b^3}{\pi^{1/2}} + O(b^4).
\label{eq:diff_autoc_exact}
\ee
Note that the case of two fermions with a square-well
interaction in a harmonic trap was numerically studied in \cite{Calarco}, for the $s$-wave and also for the $p$-wave case,
with the exact spectrum compared to the self-consistent equation (\ref{eq:autoc}) or to its $p$-wave equivalent.
No conclusion was given on the scaling with $b$ of the difference between the exact and the approximate spectrum.

\noindent {\underline{\it Two dimensions:}} 
\\
Imposing the contact condition $\psi(r)=A\ln(r/a)+O(r)$ to Eq.~(\ref{eq:whit2d}) gives an implicit equation for the spectrum in the zero-range limit
\cite{Busch,LeChapitre}:
\be
\psi\left(\frac{1-E_{\rm rel}^0}{2}\right)-2\psi(1)= -2 \ln a
\ee
where $\psi$ is the digamma function. We have obtained the finite range correction
\be
E_{\rm rel} = E_{\rm rel}^0 + \frac{4 r_e^2 E_{\rm rel}^0}{\psi'(\frac{1-E_{\rm rel}^0}{2})} + O(b^4\ln^4 b)
\label{eq:dev_expli_2d}
\ee
by neglecting the trapping potential for $r\leq b$ as justified by Appendix~\ref{app:piege_dans_interaction}, and by matching
in $r=b$ the scattering state $\mathcal{A}\chi(r)$ to Eq.~(\ref{eq:whit2d}). The bound on the error results in particular
from the statement that $\ldots$ in Eq.~(\ref{eq:devuk_2D}) are $O[(kb)^4\ln(a/b)]$, that one can e.g.\ check for the square-well potential. 
As expected, the value of $\partial E_{\rm rel}/\partial (r_e^2)$ in $r_e=0$ obtained from Eq.~(\ref{eq:dev_expli_2d}) coincides with
[Tab,~V, Eq.~(1b)], knowing that the normalization factor in the zero-range limit, according to relation 7.611(5) in \cite{Gradstein},
obeys $(C_2^0)^2 \pi \psi'(\frac{1-E_{\rm rel}^0}{2})/\Gamma^2(\frac{1-E_{\rm rel}^0}{2})=1$.

\subsection{Three-body problem: corrections to exactly solvable cases and comparison with numerics}\label{subsec:appl_3body}

In this Subsection, we use the known analytical expressions for the three-body wavefunctions to compute the corrections to the spectrum to first order in the inverse scattering length $1/a$ and in the effective range $r_e$. We shall consider not only spin-1/2 fermions, but also spinless bosons restricting to the universal stationary states~\cite{Werner3corpsPRL,Pethick3corps} which do not depend on the three-body parameter.

The problem of three identical spinless bosons~\cite{Werner3corpsPRL,Pethick3corps} 
or spin-1/2 fermions (say $N_\up=2$ and $N_\down=1$)~\cite{Werner3corpsPRL,TanScaling} 
is exactly solvable in the unitary limit in an isotropic harmonic trap $U(\rr)=\frac{1}{2}\,m\omega^2 r^2$.
Here we restrict to zero total angular momentum (see however the last line of Appendix~\ref{app:echelle})
with a center of mass in its ground state,
so that the normalization constants of the wavefunctions are also known analytically \cite{WernerThese}. Moreover we restrict to universal eigenstates~\footnote{For Efimovian eigenstates, computing the derivative of the energy with respect to the effective range would require to use a regularisation procedure similar to the one employed in free space in \cite{Efimov93,PlatterRangeCorrections}. However the derivative with respect to $1/a$ can be computed \cite{WernerThese}.}. The spectrum is then 
\be
E=E_{\rm cm}+(s+1+2q)\hbar\omega
\label{eq:echelle}
\ee
where $E_{\rm cm}$ is the energy of the center of mass motion,
$s$ belongs to the infinite set of real positive solutions of
\be
-s \cos\left(s\frac{\pi}{2}\right) + \eta\frac{4}{\sqrt{3}}\sin\left(s\frac{\pi}{6}\right)=0
\label{eq:s}
\ee
with $\eta=+2$ for bosons and $-1$ for fermions,
and
$q$ is a non-negative integer quantum number describing the degree of excitation of an exactly decoupled bosonic breathing mode
\cite{CRAS,WernerSym}.
We restrict to states with $q=0$. The case of a non-zero $q$ is treated in subsection
\ref{subsec:contactN}.

\paragraph{Derivative of the energy with respect to $1/a$.}

Injecting the expression of the regular part $A$ of the normalized wavefunction \cite{WernerThese} into 
[Tab.~II, Eqs.~(2a,4a)] or its bosonic version (Tab.~V, line~1 in \cite{CompanionBosons}) we obtain
\begin{equation}
\frac{\partial E}{\partial(-1/a)}\Big|_{a=\infty}\!\!\!\! = 
\frac{\sqrt{\frac{\hbar^3\omega}{m}}\Gamma(s+\frac{1}{2})\sqrt{2}s\sin\left(s\frac{\pi}{2}\right)/\Gamma(s+1)}
{
-\cos\left(s\frac{ \pi}{2 } \right) + s\frac{ \pi}{2 } \sin\left(s\frac{ \pi}{2 } \right)
+\eta\frac{2\pi }{3\sqrt{3} } \cos\left(s\frac{ \pi}{6 } \right)
}
\label{eq:dEda3corps}
\end{equation}
For the lowest fermionic state, this gives
$(\partial E/\partial(1/a))_{a=\infty}\simeq -1.1980\sqrt{\hbar^3\omega/m}$,
in agreement with the value  $-1.19(2)$ which we extracted from the numerical solution of a finite-range model
presented in Fig.~4a of \cite{StecherLong},
where the error bar comes from our simple way of extracting the derivative from the numerical data of \cite{StecherLong}.

\paragraph{Derivative of the energy with respect to the effective range.}

Using relation [Tab.~V, Eq.~(1a)], which holds not only for fermions but also for bosonic universal states, we obtain
\be
\left(\frac{\partial E}{\partial r_e}\right)_{a=\infty}\!\!\!\!\!=
\frac{\sqrt{\frac{\hbar m\omega^3}{8}} \Gamma(s-\frac{1}{2}) s 
(s^2-\frac{1}{2}) \sin(s\frac{\pi}{2})
/\Gamma(s+1)}
{-\cos(s\frac{\pi}{2})+s\frac{\pi}{2}\sin(s\frac{\pi}{2})
+\eta \,\frac{2\pi}{3\sqrt{3}}\cos(s\frac{\pi}{6})
}
 \label{eq:dEdre_3}
\ee
For bosons, this result was derived previously using the method of \cite{Efimov93} and found to agree with the numerical solution of a finite-range separable potential model for the lowest state \cite{WernerThese}.
For fermions, (\ref{eq:dEdre_3}) agrees with the numerical data from Fig. 3 of \cite{StecherLong} to $\sim0.3\%$ for the two lowest states and $5\%$ for the third lowest state
\footnote{Here we used the value of the effective range $r_e=1.435\,r_0$~\cite{ThogersenThese}
for the Gaussian interaction potential $V(r)=-V_0 e^{-r^2/r_0^2}$
with $V_0$ equal to the value where the first two-body bound state appears.};
(\ref{eq:dEdre_3}) also agrees to $3\%$
with the numerical data
from p.~21 of \cite{WernerThese} for the lowest state of a finite-range separable potential model. 
All these deviations are compatible with the estimated numerical accuracy.

\subsection{$N$-body problem in an isotropic trap: Non-zero $1/a$ and $r_e$ corrections}
\label{subsec:contactN}

We now generalize subsection \ref{subsec:appl_3body} to the case of an arbitrary number $N$ of spin-1/2 fermions 
(with an arbitrary spin configuration) at the unitary limit in an isotropic harmonic trap.
Although one cannot calculate $\partial E/\partial (1/a)$ and $\partial E/\partial r_e$, some useful information can be obtained
from the following remarkable property:
For any initial stationary state, and after an arbitrary change of the isotropic trap curvature, the system experiences 
an undamped breathing at frequency $2\omega$, $\omega$ being the single atom oscillation frequency in the final trapping potential \cite{CRAS}.
From this one can conclude that, in the case of a time independent trap, the system exhibits
a $SO(2,1)$ dynamical symmetry \cite{WernerSym}: The spectrum is a collection of semi-infinite
ladders indexed by the natural integer $q$. Another crucial consequence is that the eigenstate wavefunctions
are separable in $N$-body hyperspherical coordinates, with a know expression for the dependence with the hyperradius \cite{WernerSym}.
This implies that the functions $A_{ij}$ are also separable in $(N-1)$-body hyperspherical coordinates and that
their hyperradial dependence is also known.  As the eigenstates within a ladder have exactly the same hyperangular part,
one can relate the energy derivatives (with respect to $1/a$ or $r_e$) for step $q$ of a ladder 
to the derivative for the ground step of the {\sl same} ladder, as detailed
in Appendix~\ref{app:echelle}:
\begin{multline}
\label{eq:relat1}
\left[\frac{\partial E}{\partial (1/a)}\right]_{q} = \left[\frac{\partial E}{\partial (1/a)}\right]_{0} 
\frac{\Gamma(s+1)}{\Gamma(s+q+1)} \\
\times \sum_{k=0}^q \left[\frac{\Gamma(k+\frac{1}{2})}{\Gamma(k+1)\Gamma(\frac{1}{2})}\right]^2 
\frac{\Gamma(s+q-k+\frac{1}{2})\Gamma(q+1)}{\Gamma(s+\frac{1}{2})\Gamma(q-k+1)}
\end{multline}
with the eigenenergy of step $q$ is written as Eq.~(\ref{eq:echelle}), $s$ being now unknown for the general
$N$-body problem.
We have checked that this explicit result is consistent with the recursion relations derived
in \cite{Moroz}.
A similar type of result holds for the derivative with respect to $r_e$:
\begin{multline}
\label{eq:relat2}
\left[\frac{\partial E}{\partial r_e}\right]_{q} = \left[\frac{\partial E}{\partial r_e}\right]_{0}
\frac{\Gamma(s+1)}{\Gamma(s+q+1)} \\
\times \sum_{k=0}^q \left[\frac{\Gamma(k+\frac{3}{2})}{\Gamma(k+1)\Gamma(\frac{3}{2})}\right]^2 
\frac{\Gamma(s+q-k-\frac{1}{2})\Gamma(q+1)}{\Gamma(s-\frac{1}{2})\Gamma(q-k+1)}
\end{multline}
For non-zero $1/a$ or $r_e$, the level spacing is not constant within a ladder, the system will not respond to a trap change
by a monochromatic breathing mode. In a small system, a Fourier transform of the system response
can give access to the Bohr frequencies $(E_q-E_{q-1})/\hbar$, which would allow
an experimental test of Eqs.~(\ref{eq:relat1},\ref{eq:relat2}).  
In the large $N$ limit, for a system prepared in its ground state,
we now show that the main effects of non-zero $1/a$ or $r_e$ on the breathing
mode are a frequency change and a collapse.

Let us take the macroscopic limit of Eqs.~(\ref{eq:relat1},\ref{eq:relat2}) for a fixed $q$: 
Using Stirling's formula for $s\to +\infty$ we obtain
\bea
\label{eq:devq2unsura}
\frac{[\partial E/\partial(1/a)]_q}{[\partial E/\partial(1/a)]_0}&=&  1 -\frac{q}{4s} +\frac{q(9q+7)}{64 s^2} +\ldots\\
\frac{[\partial E/\partial r_e]_q}{[\partial E/\partial r_e]_0} &=& 1 + \frac{3q}{4s} -\frac{3q(5q+11)}{64s^2}
+\ldots
\label{eq:devq2re}
\eea
The first deviations from unity  are thus linear in $q$, and correspond to a shift of
the breathing mode frequency $\omega_{\rm breath}$ to the new value $2\omega + \delta\omega_{\rm breath}$,
that can be obtained to leading order in $1/a$ and $r_e$ from
\be
\frac{\partial \omega_{\rm breath}}{\partial (1/a)} = -\frac{\omega}{4 E_0} \frac{\partial E_0}{\partial (1/a)} 
\ \ \mbox{and}\ \ 
\frac{\partial \omega_{\rm breath}}{\partial r_e}  = \frac{3\omega}{4 E_0} \frac{\partial E_0}{\partial  r_e}
\label{eq:dbreath}
\ee
For a non-polarized gas (with the same number $N/2$ of particles in each spin state)
the local density approximation gives $4s \sim (3N)^{4/3} \xi^{1/2}$ \cite{TanScaling,BlumeUnivPRL} and it allows
to obtain the derivative of the energy with respect to $1/a$ \cite{WernerTarruellCastin} or to $r_e$ in terms of $\xi$,
$\zeta$ and $\zeta_e$, defined in Eqs.(\ref{eq:eq_d_etat},\ref{eq:def_zetae}), so that
\bea
\delta\omega_{\rm breath}= \frac{256\omega}{525\pi \xi^{5/4}} \left[\frac{\xi^{1/2}\zeta}{k_F a}+2\zeta_e k_F r_e\right]
\eea
where we have introduced the Fermi momentum $k_F$ of the unpolarized trapped ideal gas with the same atom number $N$
as the unitary gas, with $\hbar^2 k_F^2/(2m)=(3N)^{1/3}\hbar\omega$.
For $r_e=0$, we recover the superfluid hydrodynamic prediction of \cite{BulgacModes,CombescotLeyronasComment,YunStringari}.
We have checked that the change of the mode frequency due to finite range effects can also be obtained
from hydrodynamics \footnote{The hydrodynamic frequencies $\Omega$ are given by the
eigenvalue problem $-m\Omega^2 \delta \rho = \mbox{div}\, [\rho_0\, \nabla\, (\mu_{\rm hom}'[\rho_0]\, \delta \rho)]$
where $\delta \rho(\rr)$ is the infinitesimal deviation from the stationary density profile $\rho_0(\rr)$,
$\mu_{\rm hom}[\rho]$ is the ground state chemical potential of the homogeneous gas of density $\rho$ and the appex
$'$ indicates derivation.  For the equation of state $\mu_{\rm hom}[\rho]=A \rho^{2/3} + B \rho^{\gamma}$,  where $B$ is arbitrarily
small, we treat the term in $B$ to first order in perturbation theory around the breathing mode
to obtain $\Omega=2\omega + \omega \frac{96}{\pi} (\gamma-\frac{2}{3}) \frac{B}{\mu} \left(\frac{\mu}{A}\right)^{3\gamma/2}
\int_0^1 du u^2 (1-2u^2) (1-u^2)^{(3\gamma+1)/2}$ where $\mu=\omega N^{1/3} (2mA/\pi^{4/3})^{1/2}$ is the unperturbed
chemical potential of the trapped gas. To zeroth order in $B$,  scaling invariance gives
$\delta \rho^{(0)}(\rr) =\frac{d}{d\lambda}[\rho_0(\rr/\lambda)/\lambda^3]_{\lambda=1}$.
To use perturbation theory, we made the differential operator Hermitian with the change of function
$\delta f(\rr) = (\mu_{\rm hom}'[\rho_0(\rr)])^{1/2} \delta\rho(\rr)$. Hermiticity of the perturbation is guaranteed
(i.e.\ surface terms coming from the divergence theorem vanish) for $\gamma$ larger than $1/3$.
For finite-$r_e$ corrections, $\gamma=1$.
};
this change in typical experiments is of the order of $0.1\%$ for
lithium and $0.5\%$ for potassium, see subsection \ref{subsec:app_manips}.

Furthermore, due to the presence of $q^2$ terms in Eqs.~(\ref{eq:devq2unsura},\ref{eq:devq2re}), the Bohr frequencies
$(E_{q}-E_{q-1})/\hbar$ depend on the excitation degree $q$ of the mode: If many steps of the ground state ladder
are coherently populated, this can lead to a {\sl collapse}
of the breathing mode, which constitutes a mechanism for zero-temperature damping \cite{DalfovoMinnitiPitaevskii,CastinDumInstab}.
To coherently excite the breathing mode, we start with a ground state gas, with wavefunction $\psi_{\rm old}$, 
and we abruptly change at $t=0$ the trap frequency from $\omega_{\rm old}$ to $\omega =\lambda^2 \omega_{\rm old}$. For the unitary gas,
$\psi_{\rm old}$ is deduced from the $t=0^+$ ground state $\psi_0$ by a dilation with scaling factor $\lambda$,
\be
|\psi_{\rm old}\rangle = e^{-i \hat{D} \ln \lambda} |\psi_0\rangle
\ee
where $\hat{D}$ is the generator of the dilations \cite{LeChapitre,WernerSym}. Using the representation of $\hat{D}$ 
in terms of the bosonic operator $\hat{b}$ \cite{WernerSym}, that annihilates an elementary excitation of the breathing
mode ($\hat{b}|q\rangle = q^{1/2} |q-1\rangle$), and restricting to $|\epsilon|\ll 1$, where $\epsilon=\ln \lambda$,
one has
\be
\hat{D} \simeq -i s^{1/2} (\hat{b}^\dagger -\hat{b})
\ee
so that the trap change prepares the breathing mode in a Glauber coherent state with mean occupation number $\bar{q}= \epsilon^2 s$
and standard deviation $\Delta q=\bar{q}^{1/2}$. Similarly, the fluctuations of the squared radius of the gas
$\sum_i r_i^2/N$, that can be measured, are given by $-\frac{\hbar s^{1/2}}{m\omega} (\hat{b}+\hat{b}^\dagger)$ 
for small $\epsilon$.
In the large system limit, one can have $\bar{q}\gg 1$ so that $1\ll \Delta q \ll \bar{q}$. At times much shorter
than the revival time $2\pi\hbar/|\partial_q^2E_q|$, one then replaces the discrete sum over $q$ by an integral to obtain
\be
\left|\frac{\langle \hat{b}\rangle(t)}{\langle \hat{b}\rangle(0)}\right| = e^{-t^2/(2 t_c^2)} \ \ \ \mbox{with}\ \ \ \
t_c =  \frac{\hbar}{\Delta q\, \left|\partial_q^2 E_q\right|_{q=\bar{q}}}
\ee
For an unpolarized gas, using Eqs.~(\ref{eq:devq2unsura},\ref{eq:devq2re}) and the local density approximation, we obtain
the inverse collapse time due to non-zero $1/a$ or $r_e$:
\bea
(\omega t_c )^{-1} = \frac{64 |\epsilon|}{35\pi \xi (3N)^{2/3}} \left|\frac{3\zeta}{5 k_F a}+\frac{2\zeta_e k_F r_e}{3\xi^{1/2}}\right|
\eea 
For lithium experiments, $t_c$ is more than thousands of mode oscillation periods. To conclude with an exotic note,
we recall that the $q^2$ terms in Eqs.~(\ref{eq:devq2unsura},\ref{eq:devq2re})
lead to the formation of a Schr\"odinger-cat-like state for the breathing mode at half the revival time \cite{YurkeStoler}.

\subsection{Unitary Fermi gas: comparison with fixed-node Monte Carlo}
\label{sec:FNMC}

\begin{figure}[t]
\includegraphics[width=0.9\linewidth,clip=]{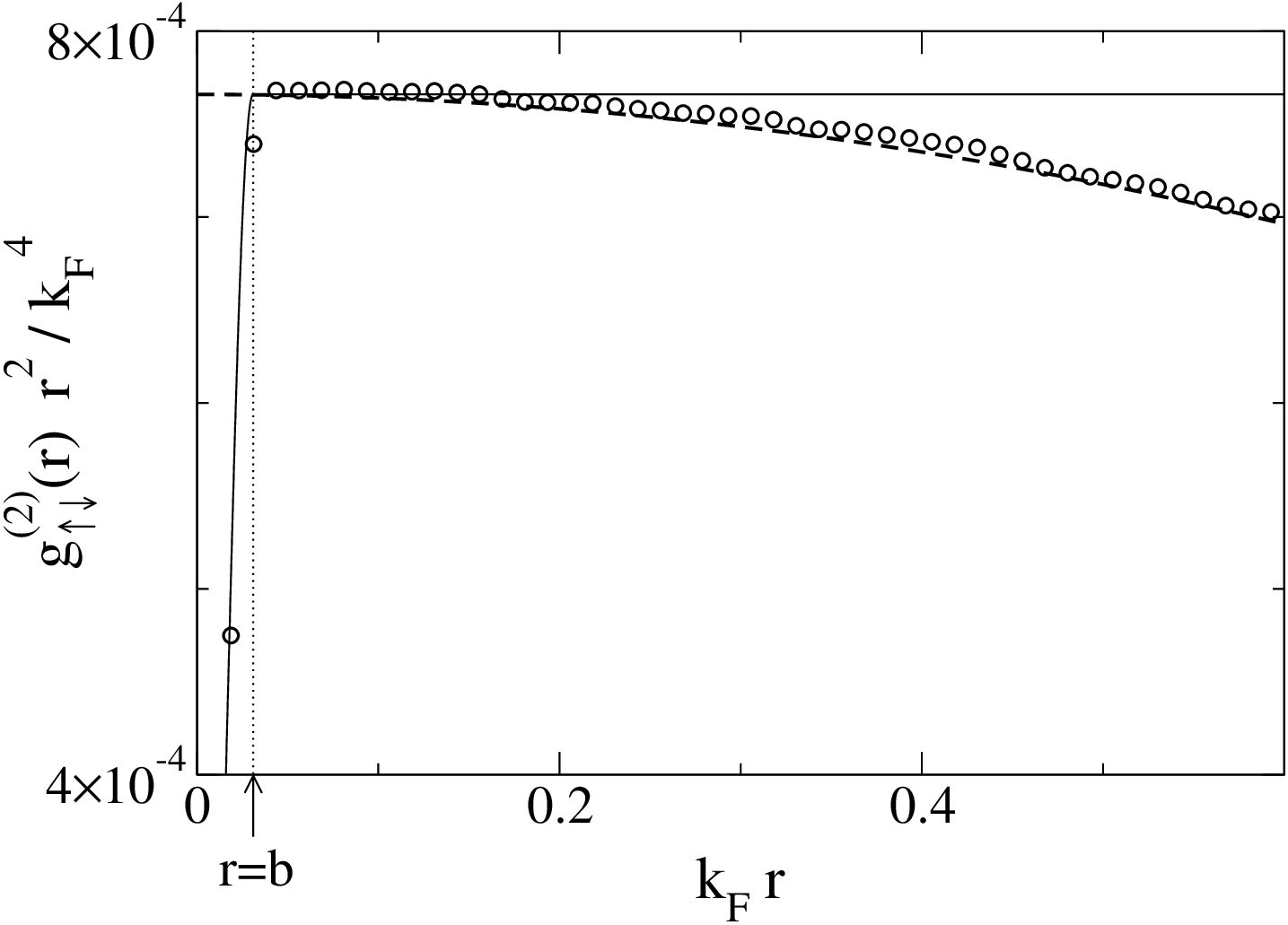}
\caption{Pair distribution function
$g_{\uparrow\downarrow}^{(2)}(r)=\langle
\hat{\psi}^\dagger_\uparrow(\rr)
\hat{\psi}^\dagger_\downarrow(\vn)
\hat{\psi}_\downarrow(\vn)
\hat{\psi}_\uparrow(\rr)
\rangle$ of the homogeneous non-polarized unitary gas at zero temperature. Circles: fixed-node Monte Carlo results from Ref.~\cite{LoboGiorgini_g2}. Solid line: analytic expression~(\ref{eq:g2_pour_MC}), where the value $\zeta=0.95$ was taken to fit the Monte Carlo results.
The arrow indicates the range $b$ of the square-well interaction potential.
Dashed line: analytic expression~(\ref{eq:g2jolie}), with $\zeta_e=0.12$ \cite{CarlsonAFQMC}.
\label{fig:g2}}
\end{figure}

\begin{figure}[t]
\includegraphics[width=0.9\linewidth,clip=]{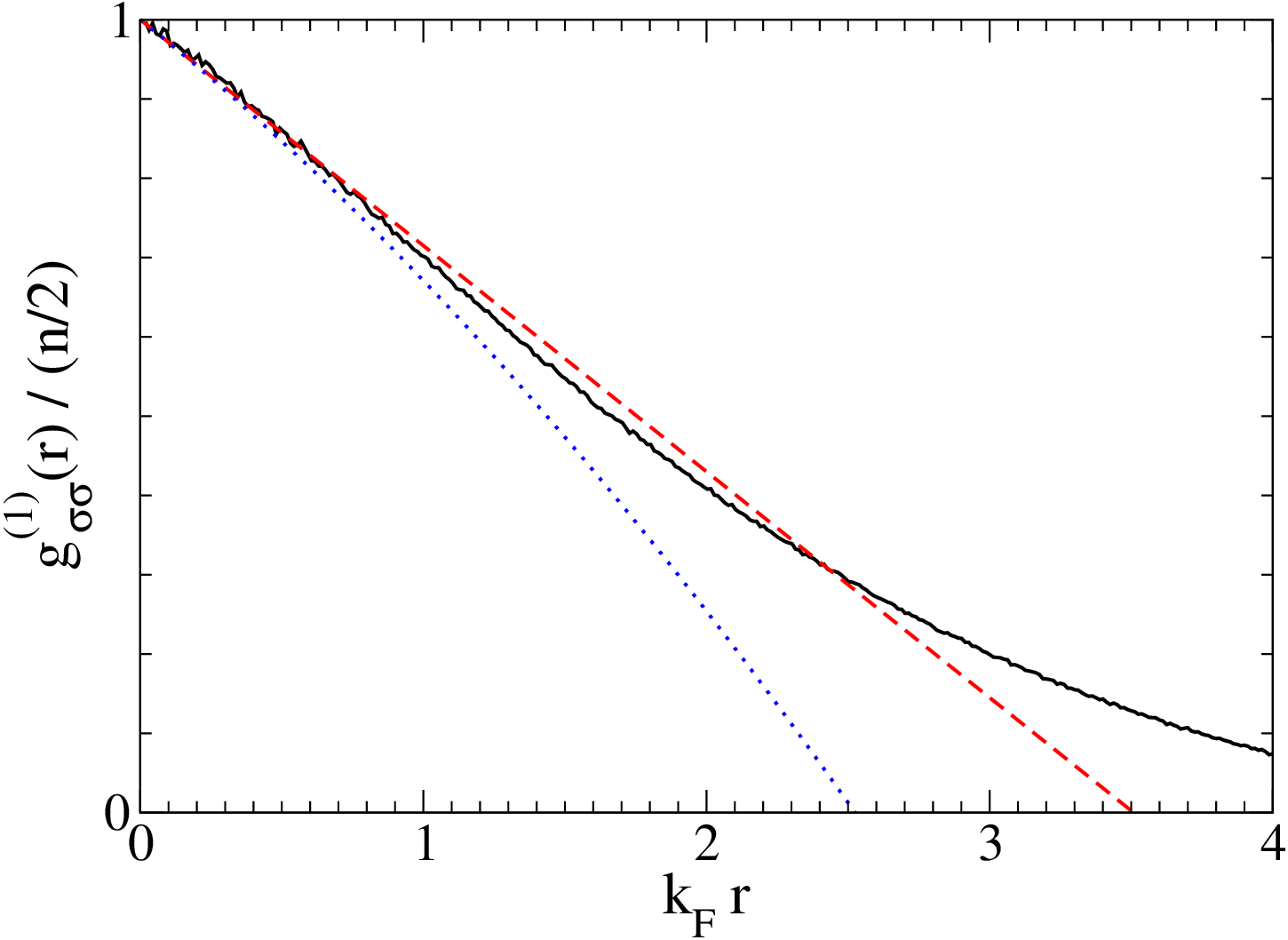}
\caption{(Color online) One-body density matrix $g^{(1)}_{\si\si}(r)=
\la\hat{\psi}^\dagger_\si(\rr)\hat{\psi}_\si(\vn)\ra$ of the homogeneous 
non-polarized unitary gas at zero temperature: comparison between the fixed-node Monte 
Carlo results from Ref.~\cite{Giorgini_nk}
(black solid line) and the
analytic expression~(\ref{eq:g1_pour_MC}) for the
small-$k_F r$ expansion of $g^{(1)}_{\sigma\sigma}$ up to first order 
(red dashed straight line) and second order (blue dotted parabola) 
where we took the value $\zeta=0.95$ extracted from the Monte 
Carlo data for $g^{(2)}_{\uparrow\downarrow}$, see Fig.~\ref{fig:g2}.
\label{fig:g1} }
\end{figure}

For the homogeneous non-polarized unitary gas (i.e. the spin-1/2 Fermi gas in $3D$ with $a=\infty$ and $N_\up=N_\down$) at zero temperature, we can compare our analytical expressions for the short-distance behavior of the one-body density matrix $g^{(1)}_{\sigma\sigma}$ and the pair distribution function $g^{(2)}_{\uparrow\downarrow}$ to the fixed-node Monte Carlo results in~\cite{Giorgini,Giorgini_nk,LoboGiorgini_g2}. In this case, $g^{(1)}_{\si\si}({\bf R}-\rr/2,{\bf R}+\rr/2)$ and
$g^{(2)}_{\up\down}({\bf R}-\rr/2,{\bf R}+\rr/2)$ depend only on $r$ and not on $\si$, ${\bf R}$ and  the direction of $\rr$.
Expanding the energy to first order in $1/(k_F a)$ around the unitary limit yields:
\be
E=E_{\rm ideal}\left(\xi - \frac{\zeta}{k_F a}+\dots\right)
\label{eq:eq_d_etat}
\ee
where $E_{\rm ideal}$ is the ground state energy of the ideal gas, $\xi$ and $\zeta$ are universal dimensionless numbers, and the Fermi wavevector is related to the density through $k_F=(3\pi^2 n)^{1/3}$. 
Expressing $C$ in terms of $\zeta$ thanks to [Tab.~II, Eqs.~(2a,4a)] and Eq.~(\ref{eq:eq_d_etat}), and inserting this into~[Tab.~II, Eq.~(7a)], we get
\be
g^{(1)}_{\si\si}(r)\simeq\frac{n}{2}\left[ 1 - \frac{3\zeta}{10} k_F r - \frac{\xi}{10} (k_F r)^2 + \dots\right].
\label{eq:g1_pour_MC}
\ee
For a finite interaction range $b$, this expression is valid for $b\ll r \ll k_F^{-1}$ 
\footnote{For a finite-range potential one has $g^{(1)}_{\si\si}(r)=n/2-r^2 m E_{\rm kin}/(3\hbar^2 \mathcal{V})+\dots$ where $\mathcal{V}$ is the volume; the kinetic energy diverges in the zero-range limit as $E_{\rm kin}\sim -E_{\rm int}$, thus $E_{\rm kin}\sim-C/(4\pi)^2 \int d^3r\,V(r)|\phi(r)|^2$ from 
[Tab.~IV, Eq.~(2a)], so that $E_{\rm kin}\sim C\pi\hbar^2/(32 m b)$ for the square-well interaction. This behavior of $g^{(1)}(r)$ only holds at very short distance $r\ll b$ and is below the resolution of the Monte Carlo data.}.
[Tab.~IV, Eq.~(4a)] yields
\be
g^{(2)}_{\up\down}(r)\underset{k_F r\ll1}{\simeq}\frac{\zeta}{40\pi^3}k_F^4 |\phi(r)|^2.
\label{eq:g2_pour_MC}
\ee
The interaction potential used in the Monte Carlo simulations~\cite{Giorgini,Giorgini_nk,LoboGiorgini_g2} is a square-well:
\be
V(r)=-\left(\frac{\pi}{2}\right)^2 \frac{\hbar^2}{m b^2}\, \theta(b-r) 
\label{eq:puits_carre}
\ee
The corresponding zero-energy scattering state is
\be
\phi(r)=\frac{\sin\left(\frac{\pi r}{2 b}\right)}{r}\ \ \mbox{for}\ \ r<b, \ \phi(r)=\frac{1}{r} \ \ \mbox{for}\ \ r>b
\ee
and the range $b$ was taken such that $n b^3=10^{-6}$ i.e. $k_F b=0.0309367\dots$. Thus we can assume that we are in the zero-range limit $k_F b\ll1$, so that (\ref{eq:g1_pour_MC},\ref{eq:g2_pour_MC}) are applicable.

Figure \ref{fig:g2} shows that the expression (\ref{eq:g2_pour_MC}) for $g^{(2)}_{\uparrow\downarrow}$ fits well the Monte Carlo data of \cite{LoboGiorgini_g2} if one adjusts the value of $\zeta$ to $0.95$. This value is close to the value $\zeta\simeq1.0$ extracted from (\ref{eq:eq_d_etat}) and the $E(1/a)$-data of~\cite{Giorgini}. 

Using $\zeta=0.95$ we can compare the expression (\ref{eq:g1_pour_MC}) for $g^{(1)}_{\sigma\sigma}$ with  Monte Carlo data of~\cite{Giorgini_nk}  without adjustable parameters.
Figure \ref{fig:g1} shows that the first order derivatives agree, while the second order derivatives are compatible within the statistical noise. This provides an interesting check of the numerical results, even though any wavefunction satisfying the contact condition~[Tab. I, Eq. (1a)] leads to $g^{(1)}_{\sigma\sigma}$ and $g^{(2)}_{\uparrow\downarrow}$ functions satisfying [Tab.~II, Eqs.~(3a,6a)] with values of $C$ compatible with each other. 

A more interesting check is provided by our expression [Tab.~V, Eq.(3a)] for the subleading term in the short range behavior
of $g^{(2)}_{\uparrow\downarrow}(r)$, which here reduces to
\be
g^{(2)}_{\uparrow\downarrow}(r) = \frac{\zeta}{40\pi^3} \frac{k_F^4}{r^2} - \frac{\zeta_e}{20\pi^3} k_F^6 + O(r)
\label{eq:g2jolie}
\ee
where $\zeta_e$ is defined in Eq.~(\ref{eq:def_zetae}). Remarkably, this expression is consistent with the fixed node Monte Carlo 
results of \cite{LoboGiorgini_g2} if one uses the value of $\zeta_e$ of \cite{CarlsonAFQMC}, see Fig.~\ref{fig:g2}.

\subsection{Finite-range correction in simulations and experiments} 
\label{subsec:app_manips}

 We recall that, as we have seen in Section~\ref{sec:re}, the finite-range corrections to eigenenergies
are, to leading order, of the form
$(\partial E/\partial r_e)\,r_e$ for continuous-space models
or (\ref{eq:dEdRe}) for lattice models,
where the coefficients $\partial E/\partial r_e$, and $\partial E/\partial R_e$ for lattice models, are model-independent.
This can be used in practice by extracting the values of these coefficients from numerical simulations,
done with some convenient continuous-space or lattice models
(usually a dramatic simplification of the atomic physics reality);
then, knowing the value of $r_e$ in an experiment,
one can compute the finite-range corrections present in the measurements,
assuming that the universality of finite range corrections, derived in section
\ref{sec:re} for compact support potentials, also applies for multichannel $O(1/r^6)$ models.
The value of $r_e$ is predicted in Ref.~\cite{GaoFeshbach} to be
\begin{multline}
r_e = -2\,R_*\,\left( 1-\frac{a_{\rm bg}}{a}\right)^2
\\
+ \frac{4\pi\,b}{3\, \Gamma^2(1/4)}\,
\left[
\left( \frac{\Gamma^2(1/4)}{2\pi} - \frac{b}{a} \right)^2 + \frac{b^2}{a^2}
\right]
\label{eq:re_Gao}
\end{multline}
where $b$ is the van~der~Waals length $b=(m C_6/\hbar^2)^{1/4}$,
$a_{\rm bg}$ is the background scattering length
and $R_*$ is the so-called Feshbach length~\cite{PetrovBosons}.
We recall that the magnetic-field dependence of $a$ close to a Feshbach resonance reads
$a(B) = a_{\rm bg} [ 1 - \Delta B / (B-B_0)]$
where $B_0$ is the resonance location
and $\Delta B$ is the resonance width,
and that
$R_* = \hbar^2 / (m a_{\rm bg} \mu_b \Delta B)$
where $\mu_b$ is the effective magnetic moment of the closed-channel molecule.
We note that the $a$-dependent terms in the second term of~(\ref{eq:re_Gao})
are $O(b^2)$ and thus do not contribute to the leading-order correction in $b$.
In contrast, the $a$-dependence of the first term of ~(\ref{eq:re_Gao}) can be significant
since $a_{\rm bg}$ can be much larger than $b$
(this is indeed the case for $^6{\rm Li}$)~\footnote{The general structure of Eq.~(\ref{eq:re_Gao}) already appeared for a simple separable two-channel model~\cite{WernerTarruellCastin} with exactly the same expression for the first term,
which explains why the $a$-dependence is correctly reproduced by the simple expression of~\cite{WernerTarruellCastin},
as observed in~\cite{NaidonCRAS} by comparison with a coupled-channel calculation,
provided that the separable-potential range in~\cite{WernerTarruellCastin} was adjusted to reproduce the correct value of $r_e$ at resonance.}.
A key assumption of Ref.~\cite{GaoFeshbach} is that the open-channel interaction potential is well approximated by $-C_6/r^6$ down to interatomic distances $r\ll b$.
This assumption is well satisfied for alkali atoms~\cite{GaoFeshbach,Gao}.
Although we have not calculated the off-shell length $\rho_e$ explicitly, we have checked that it is finite for a $-C_6/r^6$ potential
\cite{smoro_en_prepa}.

As an illustration, we estimate the finite-range corrections to the non-polarized unitary gas energy in typical experiments.
Similarly to (\ref{eq:eq_d_etat}), we have the expansion 
\be
E=E_{\rm ideal}\left(\xi + \zeta_e k_F r_e+\dots\right)
\label{eq:def_zetae}
\ee
where $E$ and $E_{\rm ideal}$ are the ground state energies of the homogeneous Fermi gas (of fixed density $n=k_F^3/(3\pi^2)$)
for $1/a=0$ and $a=0$ respectively.
The value of $\zeta_e$ was estimated both from fixed-node Monte Carlo and Auxiliary Field Quantum Monte Carlo to be $\zeta_e=0.12(3)$
~\cite{CarlsonAFQMC}~\footnote{As discussed around Eq.~(\ref{eq:dEdRe}), one has to take into account not only $r_e$ but also $R_e$ for lattice models,
which was not done in~\cite{CarlsonAFQMC}.}.
The value of $r_e$ as given by Eq.~(\ref{eq:re_Gao}) is $4.7\,{\rm nm}$
for the $B_0\simeq 834\,{\rm G}$ resonance of $^6{\rm Li}$ (in accordance with~\cite{Strinati})
and $6.7\,{\rm nm}$ for the $B_0\simeq 202.1\,{\rm G}$ resonance of $^{40}{\rm K}$.
The typical value of $1/k_F$ is $\simeq 400\,{\rm nm}$ in~\cite{ZwierleinEOS}, while $1/k_F$ at
the trap center is $\simeq 250\,{\rm nm}$ in~\cite{HuletClosedChannel}
and $\simeq 100\,{\rm nm}$ in~\cite{JinPotentialEnergy},
which respectively leads to a finite range correction to the homogeneous gas energy:
\be
\frac{\delta E}{E} \simeq 0.4\%, 0.6\%\ \mbox{and}\ 2\%.
\ee
In the case of lithium, this type of analysis was used in \cite{ZwierleinEOS} to estimate
the resulting experimental uncertainty on $\xi$.

\section{Conclusion} \label{sec:conclusion}

We derived relations between various observables for $N$ spin-1/2 fermions in an external potential
with zero-range or short-range interactions, in continuous space or on a lattice, in two or three dimensions.
Some of our results generalize the ones of
\cite{Olshanii_nk, TanLargeMomentum, TanEnergetics, ZhangLeggettUniv, TanSimple,CombescotC}:
Large-momentum behavior of the momentum distribution,
short-distance behavior of the pair distribution function and of the one-body density matrix, derivative of the energy with respect to the scattering length or to time, norm of the regular part of the wavefunction (defined through the behavior of the wavefunction when two particles approach each other),
and, in the case of finite-range interactions, interaction energy, 
are all related to the same quantity $C$; 
and
the difference between the total energy and the trapping potential energy is
related to $C$ and to a functional of the momentum distribution (which is also equal to 
 the second order term in the short-distance expansion of the one-body density matrix).
We also obtained entirely new relations:
The second order derivative of the energy with respect to the inverse scattering length (or to the logarithm of the scattering length in two dimensions) is related to the regular part of the wavefunctions, and is negative at fixed entropy;
and
the derivative of the energy with respect to the effective range  $r_e$ 
of the interaction potential (or to $r_e^2$ in $2D$) is also related to the regular part,
to the subleading short distance behavior of the pair distribution function,
and to the subleading $1/k^6$ tail of the momentum distribution.
We have found unexpected subtleties in the validity condition of the derived expression of this derivative
in $2D$: Our expression for $\partial E/\partial(r_e^2)$  applies because, 
for the class of interaction potentials that we have specified, the effective
range squared $r_e^2$ is much larger than the true range squared $b^2$, than the length squared $\rho_e^2$
characterizing the low-energy $s$-wave off-shell $T$-matrix, and than the length squared $R_1^2$
characterizing the low energy $p$-wave scattering amplitude,  by logarithmic factors that diverge in the zero-range limit.
In $3D$, for lattice models, our expression for $\partial E/\partial r_e$ applies only
for magic dispersion relations where an extra parameter $R_e$ quantifying the breaking
of Galilean invariance (as predicted in \cite{zhenyaNJP}) vanishes; also, the magic dispersion relation
should not have cusps at the border of the first Brillouin zone otherwise
the so-called Juillet effect compromises the validity of our $\partial E/\partial r_e$ expression for
finite size systems.
We have explicitly constructed such a magic relation, that may be useful
to reduce lattice discretization effects in Quantum Monte Carlo simulations.
We also considered models with a momentum cut-off used in Quantum Monte Carlo calculations,
either in continuous space \cite{zhenyas_crossover} or on a lattice \cite{bulgacQMC,BulgacCrossover,BulgacPG,BulgacPG2}: 
Surprisingly, in the infinite cut-off limit,
the breaking of Galilean invariance survives and one does not exactly recover the unitary gas.

Applications of general relations were presented in three dimensions.
For two particles in an isotropic harmonic trap, finite-interaction-range corrections
were obtained, and were found to be universal up to order $r_e^2$ included in $3D$; in particular, this clarifies
analytically the validity of some approximation and self-consistent equation introduced in \cite{Bolda,Naidon,Greene,GaoPiege} 
that neglect the effect of the trapping potential within the interaction range.
For the universal states of three particles with an infinite scattering length in an isotropic harmonic trap,
the derivatives of the energy with respect to the inverse scattering length
and with respect to the effective range
were computed analytically and found to agree with available numerics.
For the unitary gas in an isotropic harmonic trap, which has a $SO(2,1)$ dynamical symmetry
and an undamped breathing mode of frequency $2\omega$,
we have determined the relative finite-$1/a$ and finite range energy corrections within each $SO(2,1)$ ladder, which
allows in the large-$N$ limit to obtain the frequency shift and the collapse time of the breathing mode.
For the bulk unitary Fermi gas, existing fixed-node Monte Carlo data  were checked to satisfy exact relations. 
Also, the finite-interaction-range correction to the unitary gas energy
expected from our results to be (to leading order) model-independent and thus extractable from Quantum Monte Carlo results,
was estimated for typical experiments: This quantifies one of the experimental uncertainties on the Bertsch parameter $\xi$.

The relations obtained here may be used in various other contexts. For example, 
the result [Tab.~II, Eqs.~(11a,11b)] on the sign of the second order derivative of $E$ at constant entropy is relevant to adiabatic ramp experiments~\cite{CarrCastin,GrimmCrossover,JinPotentialEnergy,thomas_entropie_PRL,thomas_entropie_JLTP},
and the relation [Tab.~III, Eq.~(8a)] allows to directly compute $C$ using determinantal diagrammatic Monte Carlo
\cite{Goulko_C} and bold diagrammatic Monte Carlo~\cite{VanHouckePrepa,FelixIntTalk,VanHouckePrepaC}. $C$ is directly related to the closed-channel fraction in a two-channel model \cite{BraatenLong,WernerTarruellCastin}, which allowed to extract it \cite{WernerTarruellCastin} from the experimental photoassociation measurements in \cite{HuletClosedChannel}. 
$C$ was measured from the tail of the momentum distribution \cite{JinC}. For the homogeneous gas
$C$ was extracted from measurements of the equation of state \cite{SylEOS}.
$C$ also plays an important role in the theory of radiofrequency spectra \cite{ZwergerRF,BaymRF,ZhangLeggettUniv,StrinatiRF,RanderiaRF,ZwergerRFLong} and in finite-$a$ virial theorems \cite{TanViriel,Braaten,WernerViriel}, as
verified experimentally \cite{JinC}.
$C$ was also extracted from the momentum tail of the static structure factor $S(k)$, which is the Fourier transform
of the spin-independent pair distribution function $\langle \hat{n}(\rr) \hat{n}(\mathbf{0})\rangle$ 
and was measured by Bragg spectroscopy \cite{AustraliensC,AustraliensT}.
In principle one can also measure {\sl via} $S(k)$ the parameter $\zeta_e$ quantifying the finite range correction to the unitary
gas energy, from the relation
\be
\frac{\partial E}{\partial r_e}=-\frac{\pi\hbar^2}{m} \int \frac{d^3k}{(2\pi)^3} \left[S(k)-\frac{C}{4k}\right]
\ee  
resulting from [Tab.~V, Eq.~(3a)]. This procedure is not hampered by the small value of $k_F r_e$ in present experiments,
contrarily to the extraction of $\zeta_e$ from a direct measurement of the gas relative energy correction
$\propto \zeta_e k_F r_e \lesssim 10^{-2}$.

We can think of several generalizations of
the relations presented here.
All relations can be extended to the case of periodic boundary conditions.
The techniques used here can be applied to the
 one-dimensional case to generalize the relations of \cite{Olshanii_nk}.
For two-channel or multi-channel models one may derive relations other than the ones of \cite{BraatenLong,WernerTarruellCastin,ZhangLeggettUniv}.
Generalization of the present relations to arbitrary mixtures of atomic species,
and to situations (such as indistinguishable bosons) where the Efimov effect takes place,
was given in \cite{CompanionBosons}.

\acknowledgments
We thank E.~Burovski, J.~Dalibard, B.~Derrida, V.~Efimov, O.~Goulko, R.~Ignat, O.~Juillet, D.~Lee, S.~Nascimb\`ene, M.~Olshanii, 
N.~Prokof'ev, B.~Svistunov, S.~Tan, for useful discussions, as well as
S.~Giorgini and J.~von~Stecher  for sending numerical data 
from~\cite{Giorgini_nk,LoboGiorgini_g2,StecherLong,ThogersenThese
}.
The idea of using the range corrections present in Monte Carlo calculations to estimate the range corrections in cold atom experiments was given by W.~Ketterle during the Enrico Fermi School in 2006.
The question of a possible link between $\partial E/\partial r_e$ and the $1/k^6$ subleading
tail of $n_{\sigma}(\kk)$ was asked by L. Platter during a talk given by F.W. at the INT workshop 10-46W in Seattle.
The work of F.W. at UMass was supported by NSF under Grants No.~PHY-0653183 and No.~PHY-1005543.
Our group at ENS is a member of IFRAF. We acknowledge support from ERC Project 
FERLODIM N.228177.

\section*{Note}
[Tab.~II, Eq.~(4b)], as well as [Tab.~II, Eq.~(12b)], were obtained independently by Tan 
\cite{TanUnpublished}
using the formalism of~\cite{TanSimple}.
After our preprint \cite{50pages} appeared, some of our $2D$ relations were tested in \cite{Giorgini2D}
and some of them were rederived in \cite{Moelmer}.
 
\appendix

\section{Two-body scattering for the lattice model}

\label{app:2body}
For the lattice model defined in Sec.~\ref{sec:models:lattice},
we recall that $\phi(\rr)$ denotes the
zero-energy two-body scattering state with the normalization~(\ref{eq:normalisation_phi_tilde_3D},\ref{eq:normalisation_phi_tilde_2D}).
In this Appendix we derive the relation (\ref{eq:g0_3D},\ref{eq:g0_2D}) between the coupling constant $g_0$ and the scattering length, as well as the expressions (\ref{eq:phi0_vs_g0},\ref{eq:phi0_vs_g0_2D},\ref{eq:phi_tilde_3D},\ref{eq:phi_tilde_2D}) of $\phi(\vn)$.
Some of the calculation resemble the ones in~\cite{MoraCastin, YvanHouchesLowDShort}.

We consider a low-energy scattering state $\Phi_\qq(\rr)$ of wavevector $q\ll b^{-1}$ and energy $E=2\epsq\simeq\hbar^2q^2/m$, i.e. the solution of the two-body Schr\"odinger equation (with the center of mass at rest):
\be
(H_0+V)|\Phi_\qq\ra=E |\Phi_\qq\ra
\label{eq:schro_2corps}
\ee
where $H_0=\int_D d^dk/(2\pi)^d\,2\epsk|\kk\ra\la\kk|$ and $V=g_0|\rr=\vn\ra\la\rr=\vn|$, with the asymptotic behavior
\bea
\Phi_\qq(\rr)&\underset{r\to\infty}{=}&e^{i\qq\cdot\rr}+f_\qq \frac{e^{iqr}}{r}+\ldots\ \ \ \ {\rm in}\ 3D
\label{eq:phiq_asymp_3D}
\\
\Phi_\qq(\rr)&\underset{r\to\infty}{=}&e^{i\qq\cdot\rr}+f_\qq \sqrt{\frac{2}{i\pi q r}}e^{iqr}+\ldots\ \ {\rm in}\ 2D.
\label{eq:phiq_asymp_2D}
\eea
Here $f_\qq$ is the scattering amplitude , which in the present case
is independent of the direction of $\rr$ as we will see.
Note that,  in $2D$, the present definition corresponds to the convention
(\ref{eq:def_fk_a_2D}), it differs e.g.\ from~\cite{ShlyapHoucheslowDShort} by a factor $1/(4i)$.
Also $\sqrt{i}\equiv e^{i\pi/4}$.
We then have the well-known expression
\be
|\Phi_\qq\ra=(1+G V)|\qq\ra
\label{eq:phiq_GV}
\ee
where $G\equiv(E+i0^+-H)^{-1}$. Since
$G=G_0+G_0 V G$,
with $G_0\equiv(E+i0^+-H_0)^{-1}$, Eq.~(\ref{eq:phiq_GV}) 
is equivalent to
\be
|\Phi_\qq\ra=(1+G_0 T)|\qq\ra
\label{eq:phiq_G0T}
\ee
where the $T$-matrix is
$T=V+VGV$.
Indeed, (\ref{eq:phiq_GV}) clearly solves (\ref{eq:schro_2corps}), and one can check [using the fact that $\la \rr |G_0|\rr=\vn\ra$ behaves for $r\to\infty$ as
$-m/(4\pi\hbar^2)\,e^{iqr}/r$ in $3D$ and
$-(m/\hbar^2)\sqrt{i/(8\pi q r)}e^{iqr}$ in $2D$] that
(\ref{eq:phiq_G0T}) satisfies (\ref{eq:phiq_asymp_3D},\ref{eq:phiq_asymp_2D}) with
\bea
f_\qq&=&-\frac{m}{4\pi\hbar^2}b^3\la\rr=\vn|T|\qq\ra\ \ \ \ {\rm in}\ 3D
\label{eq:f_T_3D}
\\
f_\qq&=&\frac{m}{4i\hbar^2}b^2\la\rr=\vn|T|\qq\ra\ \ \ \ {\rm in}\ 2D.
\label{eq:f_T_2D}
\eea
Using $T=V+VGV$ and $G=G_0+G_0 V G$ one gets
\be
\la\rr=\vn|T|\qq\ra=b^{-d}\left[\frac{1}{g_0}-\int_D \frac{d^d k}{(2\pi)^d}\,\frac{1}{E+i0^+-2\epsk}\right]^{-1}.
\ee
In $3D$ the scattering length in defined by $\ds f_\qq\underset{q\to0}{\rightarrow}-a$, which gives the relation (\ref{eq:g0_3D}) between $a$ and $g_0$. In $2D$, 
\be
f_\qq\underset{q\to0}{=}\frac{i\pi/2}{\ln(qae^\gamma/2)-i\pi/2+o(1)}
\label{eq:fq_2D_lowE}
\ee
where $a$ is by definition the $2D$ scattering length.
Identifying the inverse of the right-hand-sides of Eqs.~(\ref{eq:f_T_2D}) and (\ref{eq:fq_2D_lowE}) and taking the real part gives the desired (\ref{eq:g0_2D}). We note that Eqs.~(\ref{eq:fq_2D_lowE},\ref{eq:g0_2D}) remain true if $q\to0$ is replaced by the limit $b\to0$ taken for fixed $a$.

To derive (\ref{eq:phi0_vs_g0},\ref{eq:phi0_vs_g0_2D}) we start from
$V|\Phi_\qq\ra = T |\qq\ra$,
which directly follows from (\ref{eq:phiq_GV}).
Applying $\la\rr=\vn|$ on the left and using (\ref{eq:f_T_3D},\ref{eq:f_T_2D}) yields
\bea
g_0 \Phi_\qq(\vn)&=&-\frac{4\pi\hbar^2}{m}f_\qq\ \ \ {\rm in}\ 3D
\\
g_0 \Phi_\qq(\vn)&=&\frac{4i\hbar^2}{m}f_\qq\ \ \ {\rm in}\ 2D.
\label{eq:phiq0_f_2D}
\eea
In $3D$, we simply have $\ds\phi=-a^{-1}\underset{q\to0}{\lim}\Phi_\qq$~\footnote{In the case of an infinite scattering length, one has to take a finite $a$ so that this expression makes sense, and only then take the limit $|a|\to\infty$ (this comes from the fact that the scattering amplitude at zero energy is infinite in this case).}, and the result (\ref{eq:phi0_vs_g0}) follows. 
In $2D$, the situation is a bit more tricky because $\underset{q\to0}{\lim}\Phi_\qq(\vn)=0$. We thus start with $q>0$, and we will take the limit $q\to0$ later on. 
At finite $q$, we define
$\phi_\qq(\rr)$ as being proportional to $\Phi_\qq(\rr)$, and normalized by imposing the same condition 
(\ref{eq:normalisation_phi_tilde_2D}) than at zero energy, but only for $b\ll r\ll q^{-1}$.
Inserting (\ref{eq:fq_2D_lowE}) into (\ref{eq:phiq0_f_2D}) gives an expression for $\Phi_\qq(\vn)$.
To deduce the value of $\phi(\vn)$, it remains to calculate the $\rr$-independent ratio $\phi_\qq(\rr)/\Phi_\qq(\rr)$. But for $r\gg b$ we can replace $\phi_\qq(\rr)$ and $\Phi_\qq(\rr)$ by their values within the zero-range model (since we also have $b\ll q^{-1}$) which we denote by $\phi_\qq^{\rm ZR}(\rr)$ and $\Phi_\qq^{\rm ZR}(\rr)$.
The two-body Schr\"odinger equation
\be
-\frac{\hbar^2}{m}\Delta\Phi_\qq^{\rm ZR} = E\,\Phi_\qq^{\rm ZR},\ \forall r>0
\ee
implies that
\be
\Phi_\qq^{\rm ZR}(\rr)=e^{i\qq\cdot\rr}+\mathcal{N} H_0^{(1)}(q r)
\ee
where $\mathcal{N}$ is a constant and $H_0^{(1)}$ is an outgoing Hankel function.
The contact condition 
\be
\exists A/\ \Phi_\qq^{\rm ZR}(\rr)\underset{r\to0}{=} A\ln(r/a)+O(r)
\ee
together with the known short-$r$ expansion of the Hankel function~\cite{Lebedev} then gives
\be
A=\frac{-1}{\ln(q a e^\gamma/2)-i\pi/2}.
\ee
Of course we also have $\Phi_\qq^{\rm ZR}/\phi_\qq^{\rm ZR}=A$, which gives (\ref{eq:phi0_vs_g0_2D}).

Finally, Eqs.~(\ref{eq:phi_tilde_3D},\ref{eq:phi_tilde_2D}) are obtained from (\ref{eq:phi0_vs_g0},\ref{eq:phi0_vs_g0_2D})
using the relations
$d(m/(4\pi\hbar^2a))/d(1/g_0) = 1$ in $3D$
and
$d(1/g_0)/d(\ln a)=-m/(2\pi\hbar^2)$ in $2D$,
which are direct consequences of
the relations (\ref{eq:g0_3D},\ref{eq:g0_2D}) between $g_0$ and $a$.

\section{Derivation of a lemma} \label{app:lemme}
In this Appendix, we derive the lemma~(\ref{eq:lemme_3D}) in three dimensions, as well as its two-dimensional version~(\ref{eq:lemme_2D}).

\noindent{\underline{\it Three dimensions:}}
\\
By definition we have
\begin{multline}
\langle \psi_1, H \psi_2 \rangle - \langle H \psi_1, \psi_2 \rangle = -\frac{\hbar^2}{2 m} \int' d^3 r_1 \ldots d^3 r_N \\
 \sum_{i=1}^N \left[ \psi_1^* \Delta_{\rr_i} \psi_2 - \psi_2 \Delta_{\rr_i} \psi_1^* \right].
\end{multline}
Here the notation $\int'$ means that the integral is restricted to the set where none of the particle positions coincide~\footnote{In other words, the Dirac distributions originating from the action of the Laplacian onto the $1/r_{ij}$ divergences can be ignored.}.
We rewrite this as:
\begin{multline}
\langle \psi_1, H \psi_2 \rangle - \langle H \psi_1, \psi_2 \rangle = -\frac{\hbar^2}{2 m} \sum_{i=1}^N \int' \Big( \prod_{k\neq i} d^3 r_k \Big) \\
\lim_{\epsilon\to0}
 \int_{\{\rr_i / \forall j\neq i, r_{ij}>\epsilon \}} d^3 r_i
\left[ \psi_1^* \Delta_{\rr_i} \psi_2 - \psi_2 \Delta_{\rr_i} \psi_1^* \right].
\label{eq:Toto_echange}
\end{multline}
We note that this step is not trivial to justify mathematically.
The order of integration has been changed and the limit $\epsilon\to 0$ has been exchanged with the integral over $\rr_i$.
We expect that this is valid in the presently considered case of equal mass fermions, and more generally provided the wavefunctions are sufficiently regular in the limit where several particles tend to each other.

Since the integrand is the divergence of $\psi_1^* \nabla_{\rr_i} \psi_2 - \psi_2 \nabla_{\rr_i} \psi_1^*$,
the divergence theorem gives
\begin{multline}
\langle \psi_1, H \psi_2 \rangle - \langle H \psi_1, \psi_2 \rangle = \frac{\hbar^2}{2 m} \sum_{i=1}^N \int' \Big( \prod_{k\neq i} d^3 r_k \Big) \\
\lim_{\epsilon\to 0}
 \sum_{j, j\neq i} \ \oiint_{S_\epsilon(\rr_j)} \left[
 \psi_1^* \nabla_{\rr_i} \psi_2 - \psi_2 \nabla_{\rr_i} \psi_1^*
 \right] \cdot \mathbf{dS}
 \label{eq:ostro}
\end{multline}
where the surface integral is for $\rr_i$ belonging to  the sphere $S_\epsilon(\rr_j)$ of center $\rr_j$ and radius $\epsilon$, and the vector area $\mathbf{dS}$ points out of the sphere.
We then expand the integrand by using the contact condition, in the limit $r_{ij}=\epsilon\to0$ taken for fixed $\rr_j$ and fixed $(\rr_k)_{k\neq i,j}$. Using $\RR_{ij}=\rr_j + \epsilon \uu /2$ with 
$\uu\equiv(\rr_i-\rr_j)/r_{ij}$ we get
\bea
\!\!\! \psi_n \!\!\!\! & \underset{\epsilon\to0}{=}&\!\!\!\! \left(\frac{1}{\epsilon}
 -\frac{1}{a_n} \right)
  A_{ij}^{(n)}+ \frac{1}{2} \uu \cdot \nabla_{\RR_{ij}} A_{ij}^{(n)}
 +O(\epsilon)
 \label{eq:psi_n_3D}
 \\
 \!\!\! \nabla_{\rr_i} \psi_n\!\!\!\! & \underset{\epsilon\to0}{=}& \!\!\!\!
 -\frac{\uu}{\epsilon^2} A_{ij}^{(n)} \!\!\!
 +\!\frac{1}{2\epsilon}\! \left[
 \nabla_{\RR_{ij}} A_{ij}^{(n)}\!\!
 -\! \uu (
  \uu \cdot\! \nabla_{\RR_{ij}} A_{ij}^{(n)}) \right]\!\!+\! O(1) \nonumber \\
&&
  \label{eq:grad_psi_n_3D}
 \eea
where $n$ equals $1$ or $2$, and the functions $A_{ij}^{(n)}$ and $\nabla_{\RR_{ij}} A_{ij}^{(n)}$ are taken at $\left( \rr_j , (\rr_k)_{k\neq i,j} \right)$.
This simply gives
\begin{multline}
\oiint_{S_\epsilon(\rr_j)} \left[
 \psi_1^* \nabla_{\rr_i} \psi_2 - \psi_2 \nabla_{\rr_i} \psi_1^*
 \right] \cdot \mathbf{dS} \underset{\epsilon\to0}{=} 4\pi \left( \frac{1}{a_1}-\frac{1}{a_2}\right) \\
\times A_{ij}^{(1)\,*} A_{ij}^{(2)} +O(\epsilon)
 \label{eq:int_surface_3D}
\end{multline}
because the leading order term cancels and most angular integrals vanish.
Inserting this into (\ref{eq:ostro}) gives the desired lemma (\ref{eq:lemme_3D}).

\noindent{\underline{\it Two dimensions:}}
\\
The derivation is analogous to the $3D$ case.
In (\ref{eq:ostro}), the double integral on the sphere of course has to be replaced by a simple integral on the circle.
Instead of (\ref{eq:psi_n_3D},\ref{eq:grad_psi_n_3D}),
we now obtain,
from
the $2D$ contact condition~[Tab.~I, Eq.~(1b)],
\bea
\psi_n & \underset{\epsilon\to0}{=}&  
\ln( \epsilon/a_n)\ 
A_{ij}^{(n)}
 +O(\epsilon \ln \epsilon)
 \label{eq:psi_n_2D}
 \\
 \nabla_{\rr_i} \psi_n & \underset{\epsilon\to0}{=}&
 \frac{\uu}{\epsilon} A_{ij}^{(n)}
  +O(\ln \epsilon),
  \label{eq:grad_psi_n_2D}
 \eea
which gives
\begin{multline}
\oint_{S_\epsilon(\rr_j)} \left[
 \psi_1^* \nabla_{\rr_i} \psi_2 - \psi_2 \nabla_{\rr_i} \psi_1^*
 \right] \cdot \mathbf{dS} \underset{\epsilon\to0}{=} 2\pi 
 \ln(a_2/a_1) \\
  \times A_{ij}^{(1)\,*} A_{ij}^{(2)} +O(\epsilon \ln^2\epsilon)
 \label{eq:int_surface_2D}
\end{multline}
and yields the lemma (\ref{eq:lemme_2D}).

\section{Zero-range limit of the lattice model's contact}
\label{app:C_b}

In this appendix, we show that our definition~[Tab.~III, Eqs.~(1a,1b)] of the contact operator $\hat{C}$ within the lattice model agrees in the zero-range limit $b\to0$ with the way~[Tab.~II, Eq.~(1)] $C$ is usually defined within the zero-range model.

\vspace{-6mm}
\subsection{Stationary state} 

\vspace{-4mm}
Let us first 
consider an eigenstate $|\psi\ra$ of the zero-range model with an energy $E$.
Let $|\psi_b\ra$ denote the eigenstate of the lattice model which tends to $|\psi\ra$ when $b\to0$, and $E_b$ the corresponding eigenenergy.
Then, 
$C_b\equiv\la\psi_b|\hat{C}|\psi_b\ra$
tends to 
the contact $C$ of the state $\psi$
[defined in~Tab.~II, Eq.~(1)]
when $b\to0$.
Indeed,
$C$ is related to $dE/d(-1/a)$ by~[Tab.~II, Eq.~(4a)];
$C_b$ is related to $dE_b/d(-1/a)$ by~[Tab.~II, Eq.~(4a)];
and the function $E_b(1/a)$ should tend smoothly to $E(1/a)$
when $b\to0$.

\vspace{-6mm}
\subsection{Arbitrary pure state}
\label{app:C_b_2}

\vspace{-4mm}
We now consider
any pure state $|\psi\ra$
satisfying the contact condition~[Tab. I, Eq. (1a)].
We will show that 
$C_b\equiv\la\psi_b|\hat{C}|\psi_b\ra$
tends to 
the contact $C$ of the state $|\psi\ra$
[defined in~Tab.~II, Eq.~(1)]
when $b\to0$,
where $|\psi_b\ra$ is defined as follows:
Writing $|\psi\ra$ as a linear combination $\sum_n c^{(n)} |\psi^{(n)}\ra$ of the zero-range model's eigenstates $|\psi^{(n)}\ra$,
we define the linear combination $|\psi_b\ra\equiv\sum_n c^{(n)} |\psi_b^{(n)}\ra$ of the lattice-model's eigenstates $|\psi_b^{(n)}\ra$.

We consider only the $3D$ case, the derivation being almost identical in $2D$.
 Let $A$ and $A^{(n)}$ denote the regular parts of $\psi$ and $\psi^{(n)}$ [defined by the contact condition~Tab.~I, Eq.~(1a)], and
$A_b$ and $A_b^{(n)}$ denote the regular parts of $\psi_b$ and $\psi_b^{(n)}$ [defined by (\ref{eq:def_A_reseau})]. Linearity
immediately gives
$A=\sum_n c^{(n)} A^{(n)}$ and $A_b=\sum_n c^{(n)} A_b^{(n)}$,
as well as
$C_b=\sum_{n,m} (c^{(n)}_b)^* c^{(m)}_b \la\psi^{(n)}_b|\hat{C}|\psi^{(m)}_b\ra$. Expressing $\hat{C}$ in terms of $H_{\rm int}$ thanks to~[Tab.~III, Eq.~(2)],
and using the lemma~(\ref{eq:lemme_W})
as well as~(\ref{eq:phi0_vs_g0}),
we get
$ \la\psi^{(n)}_b|\hat{C}|\psi^{(m)}_b\ra=(4\pi)^2\,(A^{(n)}_b,A^{(m)}_b)$.
When $b\to0$,
we expect that
this last quantity tends to $(4\pi)^2\,(A^{(n)},A^{(m)})$
because $A_b^{(n)}\to A^{(n)}$
[see~(\ref{eq:def_A_reseau}) and the discussion thereafter]. Thus $C_b$ indeed tends to $C$.

\section{Spectral effect of the trapping potential
within the interaction range}
\label{app:piege_dans_interaction}

The motivation of this Appendix is to justify the fact that, in Eq.~(\ref{eq:espc}) and in its equivalent form in $2D$
for a $N$-body problem, we have neglected the effect
of the trapping potential within the interaction range. In the case of an isotropic harmonic trap,
the exact form of Eq.~(\ref{eq:espc}) contains the external potential term $\frac{1}{4}m\omega^2 r_{ij}^2$.
This issue is thus mappable to the two-body problem in a trap with a finite range interaction,
which was the object of numerous studies in $3D$ \cite{Greene,Bolda,Naidon,GaoPiege} that have however not analytically quantified the effect 
of the trapping potential within the interaction range.
After elimination
of the center of mass motion and restriction to a zero angular momentum,
one faces the $3D$ or $2D$ eigenvalue problem
\be
E \psi(r) = -\frac{\hbar^2}{m} \Delta \psi(r) +
\left[\frac{1}{4} m \omega^2 r^2 + V(r;b)\right] \psi(r)
\ee
with the conditions that $\psi$ diverges neither in $r=0$ nor
at infinity.
The rotationally invariant compact support potential $V(r;b)$
of range $b$ is of the minimal depth ensuring a fixed scattering length $a$
(as discussed in subsection \ref{subsec:wwlftdpov}). 
In the limit $b\to 0$, where $E$ converges to a finite value,
we show that neglecting the effect of the trapping
potential {\sl within} the interaction range $r\leq b$,
as done in subsection \ref{subsec:dotef}, introduces 
on the eigenenergy $E$ an error $O(b^3)$ in $3D$ and $O[b^4 \ln^2(a/b)]$
in $2D$, which thus does not affect the results [Tab.~V, Eqs.~(1a,1b)].

The starting point is the Hellmann-Feynman theorem, with $\psi$ real
and normalized to unity:
\be
\frac{dE}{db}= \int d^dr\, \psi^2(r)\, \partial_b V(r;b).
\label{eq:hfapp}
\ee
To reexpress this integral in a more operational way, we introduce
the solution $\tilde{\psi}(r)$ of Schr\"odinger's equation with
the same eigenvalue $E$ but for the interaction potential
$V(r;\tilde{b})$ of a different range $\tilde{b}$. This solution
$\tilde{\psi}(r)$ remains finite in $r=0$ but it diverges at infinity
and cannot be $L^2$-normalized. In what follows we take a convenient normalization
of $\tilde{\psi}$ such that $\lim_{\tilde{b}\to b}\tilde{\psi}=\psi$.

We multiply Schr\"odinger's equation for $\psi$ (respectively $\tilde{\psi}$)
by $\tilde{\psi}$ (respectively $\psi$) and we integrate the difference
of the two resulting equations over the domain $r<R$. Using the divergence
theorem, the Wronskian $W(R)$ appears, 
\be
W(r) \equiv \tilde{\psi}(r) \psi'(r) -\psi(r) \tilde{\psi}'(r).
\ee
For $r > b, \tilde{b}$, the Wronskian satisfies the differential
equation $W'(r)=-\frac{d-1}{r^{d-1}} W(r)$, so that, for large $R$,
$W(R)= \frac{w}{R^{d-1}}$ and
\be
w = \frac{m}{\hbar^2} \int_0^{+\infty}\!\!\!  
dr\, r^{d-1} [V(r;b)-V(r;\tilde{b})]
\tilde{\psi}(r)\psi(r).
\ee
Turning back to the Hellmann-Feynman formula (\ref{eq:hfapp}), we obtain
the exact relation
\be
\frac{dE}{db} = \frac{2(d-1)\pi\hbar^2}{m} \lim_{\tilde{b}\to b} 
\frac{w}{b-\tilde{b}}
\label{eq:relation_exacte}
\ee
It remains to calculate $w$ treating perturbatively the trapping potential
within the interaction range.

To zeroth order, one neglects the trapping potential for $r\leq b$ [or $r\leq \tilde{b}$ for
$\tilde{\psi}$], so that $\psi^{(0)}(r)= \mathcal{A} \chi(r)$,
where $\chi$ is the scattering state of energy $E$ for $V(r;b)$. 
Taking for simplicity $E>0$, we set $E=\hbar^2 k^2/m$, $k>0$, and $\chi$
is normalized as in Eqs.~(\ref{eq:norma_chi},\ref{eq:asympt2d}).
Note that $\mathcal{A}$ is then fully specified by the continuous matching of
$\psi^{(0)}$ in $r=b$ to the outer solution in the trapping potential
(that can be expressed in terms of Whittaker functions, see subsection
\ref{subsec:appli_deux_corps}) and by the fact that 
$\psi$ is normalized to unity.
We also have $\tilde{\psi}^{(0)}(r)=\mathcal{A} \tilde{\chi}(r)$ for $r\leq \tilde{b}$,
where $\tilde{\chi}$ is the scattering state of energy $E$ for $V(r;\tilde{b})$
and the same prefactor $\mathcal{A}$ was taken for convenience.
The zeroth-order Wronskian $W^{(0)}$ can then be calculated explicitly,
in particular using relations 8.477(1) and 8.473(4,5) of \cite{Gradstein}.
We use Eqs.~(\ref{eq:devuk_3D},\ref{eq:devuk_2D}), 
with $\ldots= O[(kb)^4 \ln(a/b)]$ in (\ref{eq:devuk_2D}) [as we have checked for the square well],  to obtain
\bea
\label{eq:appoz_3d}
\left(\frac{dE}{db}\right)^{(0)} & \stackrel{3D}{=} &  
2\pi E \mathcal{A}^2\frac{dr_e}{db} + 
O(b^2) \\
\left(\frac{dE}{db}\right)^{(0)} & \stackrel{2D}{=} &  
\pi E \mathcal{A}^2 \frac{d}{db}(r_e^2) + O[b^3\ln (a/b)]
\eea
We have checked that the $b\to 0$ limit of these relations coincide
with [Tab.~V, Eqs.~(1a,1b)].

To first order, we treat the trapping potential perturbatively within
the interaction range. We rescale the distance by $b$,
so that $\psi^{(1)}(r)=f(x)$,  and $\chi(r)= \mathcal{N}
u(x)$, where $x=r/b$ and the function $u(x)$ is normalized by the condition
$u(0)=1$. 
The function $f$ solves the inhomogeneous Schr\"odinger equation:
\begin{multline}
\!\!\!\!\!\! f''(x) + \frac{d-1}{x} f'(x) + \left[k^2 b^2\! -\!\frac{m b^2}{\hbar^2} V(bx;b)
\right]\!
f(x) = \mathcal{F} x^2 u(x) \\
\mbox{with}\ \mathcal{F} = \frac{1}{4} \mathcal{A} \mathcal{N} \frac{m^2\omega^2}{\hbar^2} b^4 
\label{eq:inhomo}
\end{multline}
The function $u(x)$ is a solution of the corresponding homogeneous equation.
A second solution $v(x)$ can be constructed, that
diverges for $x\to 0$. It is of the form
$v(x)=-\frac{u(x)}{x} +Z_3(x)$ with $Z_3(x)=O(x)$ for $x\to 0$ in $3D$,
and $v(x)=u(x)\ln x + Z_2(x)$ with $Z_2(x)=O(x^2)$ for $x\to 0$ in $2D$.
More precisely, one has $Z_d(x)=u(x) \int_0^{x} dy\, y^{1-d} [-1+1/u^2(y)]$.
Since the expression between square brackets in the left-hand side
of Eq.~(\ref{eq:inhomo}) is $O(1)$, $u(x)$ and $Z_d(x)$
are $O(1)$ for $x\leq 1$. A first consequence is
that the factor $\mathcal{N}$ scales as $1/b$
in $3D$ and as $\ln(a/b)$ in $2D$
\footnote{
This also results from the fact that $u(1)$ is not particularly close to zero:
For $1/a=0$ in $3D$, $u(1)/u'(1)=-1$.}.
A second consequence is that, both in two and three dimensions,
\be
\psi^{(1)}(b) \ \mbox{and}\  b\, \psi^{(1)'}\!(b) = O(\mathcal{F}).
\label{eq:estimations_du_chgt}
\ee
This can be seen with the method of variation of constants, where one
sets $(f(x),f'(x))=\alpha(x) (u(x),u'(x)) + \beta(x) (v(x),v'(x))$, with the boundary conditions 
$\alpha(0)=0$ (so that $\psi^{(1)}$ does not duplicates the zeroth order solution)
and $\beta(0)=0$ (so that $\psi^{(1)}$ does not diverge in $r=0$). This leads to
\bea
\label{eq:alphax}
\alpha(x) &=& -\mathcal{F} \int_0^x dy\, y^{d+1} u(y) v(y) \\
\beta(x) &=& \mathcal{F} \int_0^x dy\, y^{d+1} u^2(y) 
\label{eq:betax}
\eea
Similar results hold for $\tilde{\psi}^{(1)}$.  From Eq.~(\ref{eq:estimations_du_chgt}) and its counterpart
for $\tilde{\psi}^{(1)}(\tilde{b})$, $\tilde{\psi}^{(1)'}(\tilde{b})$, we can estimate the variation
of the Wronskian $W(R)$ for $R$ close to $b,\tilde{b}$, and thus the variation $w^{(1)}$ of $w$ due to the trapping
potential. Dividing by $b-\tilde{b}$ and taking the limit $\tilde{b}\to b$
as in Eq.~(\ref{eq:relation_exacte}) amounts to taking a derivative with respect to $\tilde{b}$, which gives
an additional factor $O(1/b)$. Finally the error $\delta E$ introduced on the eigenenergy by the neglection of the
trapping potential within the interaction range is bounded in the zero range limit $b\to 0$ as
\bea
\delta E & \stackrel{3D}{=} &  O(m\omega^2 b^3 \mathcal{A}^2) \\
\delta E & \stackrel{2D}{=} &  O[m\omega^2 b^4 \mathcal{A}^2 \ln^2(a/b)]
\eea
where the factor $\mathcal{A}$ converges to a finite, energy-dependent value for $b\to 0$.

\section{Low-energy $T$-matrix parameters in $2D$}
\label{app:param2d}

We derive the hierarchy (\ref{eq:hier1},\ref{eq:hier2},\ref{eq:hier3}) for a $2D$ non-positive minimal-depth
potential of finite range $b$, $V(r)=\frac{\hbar^2 k_0^2}{m} v(r/b)$, for $b\to 0$ and $k_0$ adjusted to have
a constant $s$-wave scattering length $a$. The key point is then that $k_0b\to 0$ (differently from $3D$).

In the $s$-wave channel, we write the zero-energy scattering wavefunction as $\psi(r)=f(x)$, with $x=r/b$.
The function $f$ solves $f''(x)+\frac{1}{x} f'(x)= (k_0b)^2 v(x) f(x)$ and it is normalized as $f(0)=1$. 
We expand $f(x)$ in powers of $(k_0b)^2$. To zeroth order, $f_0(x)=1$. To first order,
$f_1''+\frac{1}{x}f_1'=(k_0 b)^2 v(x)$, with $f_1(0)=0$. This is integrated with 
the method of variation of constants, $f_1(x)=\alpha(x)+\beta(x)\ln x$ and $f_1'(x)=\beta(x)/x$:
\bea
\alpha(x) &=& - (k_0b)^2 \int _0^x dy\, y\, v(y) \ln y \\
\beta(x) &=&  (k_0b)^2 \int _0^x dy\, y\, v(y).
\eea
Expressing that $f_1(x)\simeq \beta(+\infty) \ln(r/a)$ at infinity gives
\be
-\frac{1}{\ln(a/b)} \simeq \frac{\beta(+\infty)}{1+\alpha(+\infty)} \simeq \frac{m}{\hbar^2} 
\int_0^{+\infty} dr\, r\, V(r)
\label{eq:betainfini}
\ee
and further using Eq.~(\ref{eq:rhoe_2D}) leads to
\be
\frac{1}{2}\rho_e^2 \sim b^2 \int_0^{+\infty} \!\!\!\! dx\, 
x\left[\frac{\beta(x)-\beta(+\infty)}{\beta(+\infty)}\ln x
+\frac{\alpha(x)-\alpha(+\infty)}{\beta(+\infty)}\right]
\label{eq:rhoe_interm}
\ee
Integration by parts then gives Eq.~(\ref{eq:hier2}). 
Using Eqs.~(\ref{eq:smorodinski2d},\ref{eq:betainfini},\ref{eq:rhoe_interm}) and
realizing that $\phi(r)+\ln (r/a)=\frac{2}{\beta(+\infty)}+O(1)$ for $b\to 0$ 
with $0<r/b\leq 1$ fixed, gives
Eq.~(\ref{eq:hier1}).
Reproducing this perturbative expansion with the same $v(x)$ in the $l$-wave, one
gets
\be
R_l^{2l} \underset{b\to 0}{\sim} \left(\frac{b^l}{2^l l!}\right)^2\frac{1}{\ln(a/b)} 
\frac{\int_0^{\infty} dx\, x^{2l+1}v(x)}{\int_0^{\infty} dx\, x\, v(x)}
\ee
This relation for $l=1$, combined with Eq.~(\ref{eq:hier2}), gives Eq.~(\ref{eq:hier3}).

\section{Some maths for the Juillet effect}
\label{app:R1}

Here, in the context of the Juillet effect for lattice models,  we justify the expansion 
(\ref{eq:devR1}). The quantity $R_1$ defined in Eq.~(\ref{eq:R1}) may be expressed
in terms of the difference between an integral and a $3D$ Riemann sum.
We are then guided by the following type of results: {\sl If $f(\xx)$ is a $C^\infty$
function inside the cube $B=[-1/2,1/2]^3$, then for $\varepsilon=1/(2N+1)$, with
the integer $N\to +\infty$:
\be
\int_B d^3x \, f(\xx) -\varepsilon^3 \sum_{\nn}
f(\varepsilon\nn)
= \frac{\varepsilon^2}{24} \int_B d^3x \,\Delta f(\xx) + O(\varepsilon^4)
\label{eq:lemme_riemann}
\ee
where $\Delta f$ is the Laplacian of $f$ and the sum over $\nn$ ranges over $\{-N,\ldots,N\}^3$.} 
To show this lemma, we introduce the short-hand notation $S[f]$ for the left-hand side of (\ref{eq:lemme_riemann}) 
and we pave $B$ with little cubes of volume $\varepsilon^3$ and of centers $\varepsilon\nn$:
\be
S[f]= \sum_{\nn} \varepsilon^3 \int_{B} d^3 x [f(\varepsilon\nn+\varepsilon\xx)-f(\varepsilon\nn)].
\label{eq:Rfinterm}
\ee
Then we use the fourth-order Taylor-Lagrange formula for $f$ restricted to the line connecting
$\varepsilon \nn$ to $\varepsilon \nn + \varepsilon \xx$:
$f(\varepsilon \nn + \varepsilon \xx)-f(\varepsilon \nn) = 
\frac{\varepsilon^2}{2} \sum_{i,j} x_i x_j \partial_i\partial_j f(\varepsilon\nn)
+\mbox{odd} + O(\varepsilon^4)$
where ``odd" stands for terms that are linear and cubic in the components of $\xx$, and $O(\varepsilon^4)$
results from the fact that
the fourth-order derivatives of $f$ are uniformly bounded on $B$. Integration over $\xx$ inside the cube $B$
eliminates the odd terms, and the $i\neq j$ quadratic terms, so that
\be
S[f]=\frac{\varepsilon^5}{24} \sum_{\nn} \left[\Delta f(\varepsilon\nn) +O(\varepsilon^2)\right].
\label{eq:Rftrans}
\ee
A Riemann sum thus deviates from the integral by $O(\varepsilon^2)$, for a $C^\infty$ integrand. 
Applying this conclusion to Eq.~(\ref{eq:Rftrans}), where $\Delta f$ is $C^\infty$, we obtain
the desired Eq.~(\ref{eq:lemme_riemann}). This result
is however not immediate to apply to the quantity $R_1$  because the integrand of $R_1$ is singular in $\kk=\mathbf{0}$.
We thus use several steps.

We first consider the quantity $R_1$ for a quadratic dispersion relation
that is cut in a smooth way: One twice replaces $1/(2\epsilon_{\kk})$ in Eq.~(\ref{eq:R1})
by $\phi(\kk b/2\pi)/(\hbar^2 k^2/m)$ where $\phi(\xx)$
is a $C^\infty$ rotationally invariant function, equal to $1$ in $\xx=\mathbf{0}$,
and  of compact support included
inside $B\equiv [-1/2,1/2]^3$ (which allows to replace the set $D$ by $\mathbb{R}^3$ in the integration
and in the summation). After the change of variable $\kk = 2\pi \xx/L$,
we decompose $\mathbb{R}^3$ as a collection of cubes of size unity (as in \cite{boite}), to obtain
\begin{multline}
\!\!\!\!\! \frac{h^2 L}{m} R_1^\phi=\!\!\!\!\sum_{\nn\in\mathbb{Z}^{3*}}\! \int_B \!\!\! d^3x\! \left[
\frac{\phi(\varepsilon \nn + \varepsilon \xx)}{(\nn+\xx)^2} -\frac{\phi(\varepsilon \nn)}{n^2}
\right]
+\! \int_B \!\! d^3x \frac{\phi(\varepsilon \xx)}{x^2}
\label{eq:R1phi}
\end{multline}
with $h=2\pi\hbar$ is Planck's constant and $\varepsilon\equiv b/L$ is the small parameter.
As shown in \cite{boite}, the right-hand side of Eq.~(\ref{eq:R1phi}) has a finite limit when $\varepsilon\to 0$,
here called $\mathcal{C}\simeq 8.91363$,
that one can obtain by taking $\varepsilon$ to zero inside the sum and the integral, which amounts to replacing
$\phi$ by unity. The deviation of Eq.~(\ref{eq:R1phi}) from its $\varepsilon\to 0$
limit can thus be exactly written as $\{S[f]+\varepsilon^3 f(\mathbf{0})\}/\varepsilon$, with
$S[f]=\varepsilon^3 \sum_{\nn\in\mathbb{Z}^{3}}\int_B d^3x [f(\varepsilon \nn + \varepsilon \xx)-f(\varepsilon \nn)]$,
that we treat as we did for Eq.~(\ref{eq:Rfinterm}).
Here $f(\xx)=[\phi(\xx)-1]/x^2$ (extended by continuity to $\xx=\mathbf{0}$)
is a $C^\infty$ function since $\phi$ is rotationally invariant. 
In the fourth-order Taylor-Lagrange formula, $O(\varepsilon^4)$ is replaced with the more accurate
$O(\frac{\varepsilon^4}{(1+\varepsilon^2n^2)^3})$, 
due to the fact that the fourth order derivatives of $f(\xx)$ are uniformly bounded and decrease as $1/x^6$ at infinity.
The integral of the Laplacian of $f$ appears as in Eq.~(\ref{eq:lemme_riemann}), except that is in
integrated over the whole $\mathbb{R}^3$ space, which gives zero.
We finally obtain
\be
\frac{h^2 L}{m} R_1^\phi \underset{b\to 0}{=} \mathcal{C} + \left(\frac{b}{L}\right)^2 \lim_{\xx\to\mathbf{0}}
\frac{\phi(\xx)-1}{x^2} + O(b/L)^3.
\label{eq:resR1phi}
\ee

Turning back to the lattice model, we now evaluate how $R_1$ deviates from
its $b\to 0$ limit for the uncut parabolic dispersion relation $\kk\to \hbar^2 k^2/(2m)$. The difference between
the smoothly-cut $R_1^{\phi}$ and the uncut $R_1^{\rm parab}$ (times $h^2 L/m$) 
is now of the form $\varepsilon^2 f(\mathbf{0})$ plus
$\frac{1}{\varepsilon}$ times the difference $S[f]$ between an integral and a Riemann
sum, with $f(\xx)=[\phi(\xx)-1]/x^2$ as before is $C^\infty$. We then use the result (\ref{eq:lemme_riemann}),
the key point being that the integration domain is $B$ (rather than the whole space), so that the integral
of the Laplacian of $f$ over $B$ gives a non-zero surface contribution, equals to the flux of the gradient of $f$
through the surface of $B$. 
This leads to Eq.~(\ref{eq:devR1}) for the particular case of the parabolic dispersion relation.
The surface term can be evaluated explicitly, as in subsection \ref{subsec:wwlftdpov}, from the integral
evaluated in polar coordinates:
\be
\int_{[-1,1]^2} \frac{dx dy}{(1+x^2+y^2)^2} = \sqrt{8} \arcsin \frac{1}{\sqrt{3}}
\label{eq:une_integrale}
\ee

Finally, we consider a general dispersion relation (\ref{eq:gdr}), with $\eta_\xx=\frac{1}{2} x^2 +O(x^4)$
for $\xx\to\mathbf{0}$. One can consider the difference between the corresponding $R_1$ and
$R_1^{\rm parab}$. The corresponding function $f(\xx)=
1/(2\eta_\xx)-1/x^2$ is then not $C^\infty$ in $\xx=\mathbf{0}$.
E.g.\ for the Hubbard model, $\eta_\xx=[3-\sum_i \cos 2\pi x_i]/(2\pi)^2$
is not rotationally invariant and $f(\xx)$ behaves as $\sum_i x_i^4/x^4$ at low $x$, its $\xx\to\mathbf{0}$
limit depends on the direction of $\xx$. This limiting behavior is however scaling invariant,
a feature that holds for a general dispersion relation.
The $n^{\rm th}$ order derivatives of $f$ are then $O(1/x^n)$ for $\xx\to\mathbf{0}$. For this class of functions,
we introduce $S^*[f]$ defined as $S[f]$ in Eq.~(\ref{eq:lemme_riemann}) except that
one excludes the term $\nn=\mathbf{0}$ in the sum. This implies
that in the equivalent of Eq.~(\ref{eq:Rfinterm}), there is an isolated contribution,
the integral of $f$ over $\varepsilon B$, which is $O(\varepsilon^3)$ and negligible.  
Then reproducing the analysis with the fourth-order Taylor-Lagrange formula we obtain
\be
S^*[f]= \frac{\varepsilon^2}{24} \int_B d^3x \,\Delta f(\xx) + O(\varepsilon^3).
\ee
As $\frac{h^2 L}{m} (R_1-R_1^{\rm parab})=S^*[f]/\varepsilon$, we obtain Eq.~(\ref{eq:devR1}).

\section{Isentropic derivatives of the mean energy 
in the canonical ensemble}
\label{app:adiab}

One considers a system with a Hamiltonian $H(\lambda)$ depending 
on some parameter
$\lambda$, and at thermal equilibrium in the canonical ensemble
at temperature $T$, with a density operator $\rho=\exp(-\beta H)/Z$.
In terms of the partition function 
$Z(T,\lambda) = \mbox{Tr}\, e^{-\beta H(\lambda)}$, with $\beta=
1/(k_B T)$, one
has the usual relations for the free energy $F$, the mean energy $\bar{E}
=\mbox{Tr}(\rho H)$
and the entropy $S=-k_B \mbox{Tr}(\rho\ln\rho)$:
\bea
\label{eq:def}
F(T,\lambda) &=& -k_B T \ln Z(T,\lambda) \\
\label{eq:utile}
F(T,\lambda) &=& \bar{E}(T,\lambda)-T S(T,\lambda) \\
\partial_T F(T,\lambda) &=& -S(T,\lambda).
\eea
One now varies $\lambda$ for a fixed entropy $S$. The temperature
is thus a function $T(\lambda)$ of $\lambda$ such that
$S(T(\lambda),\lambda) = \mbox{constant}$.
The derivatives of the mean energy
for fixed entropy are
$(\frac{d\bar{E}}{d\lambda})_S \equiv  \frac{d}{d\lambda}
[\bar{E}(T(\lambda),\lambda)]$ and
$(\frac{d^2\bar{E}}{d\lambda^2})_S \equiv  \frac{d^2}{d\lambda^2}
[\bar{E}(T(\lambda),\lambda)]$.
Writing (\ref{eq:utile}) for $T=T(\lambda)$ and
taking the first order and the second order derivatives
of the resulting equation with respect to $\lambda$,
one finds
\bea
\label{eq:deriv1}
\left(\frac{d\bar{E}}{d\lambda}\right)_S \!\!\!\! &=& \!\!
\partial_\lambda F(T(\lambda),\lambda) \\
\label{eq:deriv2}
\left(\frac{d^2\bar{E}}{d\lambda^2}\right)_S \!\!\!\! &=& \!\!
\partial_\lambda^2F(T(\lambda),\lambda)
-\frac{\left[\partial_T\partial_\lambda F (T(\lambda),\lambda)\right]^2}
{\partial_T^2 F (T(\lambda),\lambda)}
\eea
It remains to use (\ref{eq:def}) to obtain a microscopic
expression of the above partial derivatives of $F$, from
the partition function expressed as a sum
$Z=\sum_n e^{-\beta E_n}$  over the eigenstates $n$ of the
Hamiltonian: 
\bea
\label{eq:dl}
\partial_\lambda F(T,\lambda) &=& \overline
{\frac{dE}{d\lambda}} \\
\label{eq:dldl}
\partial_\lambda^2 F(T,\lambda) &= &
\overline{\frac{d^2E}{d\lambda^2}}
-\beta\, \mbox{Var}\!\left(\frac{dE}{d\lambda}\right) \\
\label{eq:dtdt}
\partial_T^2 F(T,\lambda) &=& -\frac{\mbox{Var}E}{k_B T^3} \\
\label{eq:dtdl}
\partial_T \partial_\lambda F(T,\lambda) &= &
\frac{\mbox{Cov}(E,dE/d\lambda)}{k_B T^2}.
\eea
Here the expectation value $\overline{(\ldots)}$ stands for a sum
over the eigenenergies with the canonical probability weights, 
and $\mbox{Var}$ and $\mbox{Cov}$ are the corresponding variance and 
covariance,  e.g.\ 
$\mbox{Cov}(E,dE/d\lambda) \!\! \equiv \!\!
\sum_n E_n \frac{dE_n}{d\lambda}\frac{e^{-\beta E_n}}{Z} 
-\overline{E}\  \overline{\frac{dE}{d\lambda}}$.
Insertion of (\ref{eq:dl}) into (\ref{eq:deriv1})
gives (\ref{eq:relation_T}).
Insertion of (\ref{eq:dldl},\ref{eq:dtdt},\ref{eq:dtdl})
into (\ref{eq:deriv2}) gives (\ref{eq:d2us}).

\section{Non-zero $1/a$ and $r_e$ corrections within a ladder of the trapped unitary gas}
\label{app:echelle}

For $N$ spin-1/2 fermions at the unitary limit in an isotropic harmonic trap, there is separability of the wavefunction
in internal hyperspherical coordinates \cite{WernerSym}:
\be
\psi(\rr_1,\ldots,\rr_N) = \psi_{\rm cm}(\CC) R^{-(3N-5)/2} F(R) \Phi(\Omega)
\label{eq:separee}
\ee
where $\CC$ is the center-of-mass location of the $N$ fermions, $R$ is the hyperradius and $\Omega$ is a set
of $3N-4$ hyperangles constructed from the Jacobi coordinates
(see e.g.\ \cite{LeChapitre}).  One has the general formulas
$\CC=\sum_{i=1}^{N} m_i \rr_i/M$ and $R^2 =\sum_{i=1}^{N} m_i (\rr_i-\CC)^2/\bar{m}$, where $M=\sum_{i=1}^{N} m_i$ is the total mass,
$\bar{m}$ an arbitrary mass unit, and $m_i$ is the mass of particle $i$ (here equal to $m$).
We shall not need the expression of the hyperangles.
Eq.~(\ref{eq:separee}) is due to the separability of the non-interacting Hamiltonian in a harmonic trap,
and to the fact that the Bethe-Peierls contact condition do not break this separability for $1/a=0$.
One finds that $\Phi(\Omega)$ is an eigenstate of the Laplacian on the unit sphere
of dimension $3N-4$, with contact conditions. Corresponding eigenvalues are conveniently written as $(\frac{3N-5}{2})^2-s^2$, $s>0$. 
In the $N$-body case, $s$ is not known analytically.
On the contrary, $F(R)$ solves a simple $2D$
Schr\"odinger-like equation
\begin{multline}
(E-E_{\rm cm})  F(R) =  -\frac{\hbar^2}{2\bar{m}} \left[ F''(R) +\frac{1}{R} F'(R)\right]  \\
+\left(\frac{\hbar^2 s^2}{2\bar{m} R^2} + \frac{1}{2} 
\bar{m}\omega^2 R^2\right) F(R)
\label{eq:eqpF}
\end{multline}
This leads to a spectrum of the form (\ref{eq:echelle}), with eigenfunctions expressed in terms of generalized Laguerre
polynomials multiplied by a Gaussian \cite{WernerSym}.

To derive Eqs.~(\ref{eq:relat1},\ref{eq:relat2}), one uses the fact that this separability extends to the functions
$A_{ij}(\RR_{ij},(\rr_k)_{k\neq i,j})$. One takes the limit $r_{ij}\to 0$ for a fixed $\RR_{ij}$ in
(\ref{eq:separee}): $\Phi(\Omega)$ diverges as $R/r_{ij}$ (since it depends on the hyperangles only),
$\CC$ and $R$ respectively tend to the center-of-mass position $\check{\CC}$ and the hyperradius $\check{R}$
of a fictitious system of $N-1$ particles of total mass $M=Nm$, composed of a particle of position $\RR_{ij}$ and mass $2m$,
and $N-2$ fermions of positions $\rr_k$, $k\neq i,j$ and mass $m$ \footnote{If the first Jacobi coordinates of the $N$ particles
are chosen to be $\propto \rr_{ij}$, the other ones tend to the Jacobi coordinates of the fictitious system.}. We thus obtain the form
\be
A_{ij}(\RR_{ij},(\rr_k)_{k\neq i,j}) = \psi_{\rm cm}(\check{C}) \check{R}^{-(3N-7)/2} F(\check{R}) \check{\Phi}(\check{\Omega})
\label{eq:separa_Aij}
\ee
It remains to express the Hamiltonian [Tab.~V, Eq.~(2)] of the fictitious system 
in terms of its center-of-mass $\check{\CC}$ and hyperspherical coordinates $(\check{R},\check{\Omega})$:
$\mathcal{H}_{ij}= -\frac{\hbar^2}{2M}\Delta_{\check{C}} + \frac{1}{2} M \omega^2 \check{C}^2
-\frac{\hbar^2}{2\bar{m}}[\partial_{\check{R}}^2+\frac{3N-7}{\check{R}}\partial_{\check{R}}
+\frac{1}{\check{R}^2}\Delta_{\check{\Omega}}] + \frac{1}{2} \bar{m} \omega^2 \check{R}^2$.
In the integral over $\check{R}$, we use the fact that $F$ solves (\ref{eq:eqpF}) and we integrate
by parts to obtain for $s>1/2$
\footnote{Note that $F(R)$ scales as $R^s$ for $R\to 0$. Also, each term of the sum over $i<j$ gives the same contribution, due to the fermionic
antisymmetry, and we have dropped this sum and the $ij$ indices for simplicity.}:
\bea
(A,A) = \int_0^{\infty} \!\!\!\! d\check{R}\, F^2(\check{R}) \!\int\!\! d\check{\Omega}\, \check{\Phi}^2(\check{\Omega}) \ \ \ \ \ \
\ \ \ \ \ \ \ \ \ \ \ \ \ \ \ \ \ \ \ \ \ \   && \\
\!\!\! (A,(\mathcal{H}\!-\!E)A) =\!\! \int_0^{\infty}\!\!\!\!\!\!  d\check{R} \, \frac{\hbar^2 F^2(\check{R})}{2\bar{m}R^2}  \!\!
\int \!\! d\check{\Omega} \check{\Phi}(\check{\Omega}) [\Lambda\! -\!\Delta_{\check{\Omega}}]
\check{\Phi}(\check{\Omega})\ \   &&
\eea
with $\Lambda=(\frac{3N-8}{2})^2+\frac{1}{4}-s^2$. Within a given $SO(2,1)$ energy ladder, $\check{\Phi}$ is fixed,
only $F$ depends on the quantum number $q$. The normalization of $\psi$ to unity imposes that
$\int_0^{+\infty} dR \, R F^2(R)$ is also fixed within a ladder. From known integrals involving the Laguerre polynomials,
see e.g.\ Eq.~(F7) in \cite{Tignone}, one gets Eqs.~(\ref{eq:relat1},\ref{eq:relat2}). Another byproduct is for $N=3$,
where $\check{\Phi}(\check{\Omega})$ is a spherical harmonic of spin $l$: This leads to
$\partial_{r_e}E/\partial_{(-1/a)}E=\frac{m\omega}{4\hbar} \frac{\Gamma(s-1/2)}{\Gamma(s+1/2)}[s^2-\frac{1}{2}-l(l+1)]$.



\end{document}